\documentclass{article}
\usepackage{microtype,graphicx,booktabs}
\usepackage{hyperref}
\providecommand{\doi}[1]{doi: {\urlstyle{rm}\href{https://doi.org/#1}{\nolinkurl{#1}}}}
\newcommand{\theHalgorithm}{\arabic{algorithm}}
\usepackage[preprint]{tbpfpreprint}
\makeatletter
\renewcommand{\Notice@String}{}
\renewcommand{\printAffiliationsAndNotice}[1]{%
  \global\tbpf@noticeprintedtrue
  {\let\thefootnote\relax\footnotetext{Department of Statistics, Texas A\&M University. Correspondence to: Ning Ning \textless\href{mailto:patning@tamu.edu}{patning@tamu.edu}\textgreater.}}}
\makeatother
\usepackage{amsmath,amssymb,amsthm,mathtools,longtable,xcolor,needspace} 
\definecolor{tbgreen}{RGB}{0,105,60}
\newtheorem{theorem}{Theorem}
\newtheorem{proposition}[theorem]{Proposition}
\newtheorem{corollary}[theorem]{Corollary}
\newtheorem{lemma}[theorem]{Lemma}
\theoremstyle{remark}
\newtheorem{remark}[theorem]{Remark}
\theoremstyle{plain}
\newcommand{\statementlabel}[1]{\label{#1}}
\usepackage{placeins} 
\newcommand{\appendixsection}[1]{\clearpage\section{#1}} 

\newcommand{\E}{\mathbb E}
\newcommand{\Var}{\operatorname{Var}}
\newcommand{\cK}{\mathcal K}
\newcommand{\cX}{\mathcal X}

\tbpftitlerunning{Twisted Block Particle Filters}
\RequirePackage{color}\definecolor{RED}{rgb}{1,0,0}\definecolor{BLUE}{rgb}{0,0,1} 
\DeclareOldFontCommand{\sf}{\normalfont\sffamily}{\mathsf} 
\providecommand{\DIFaddtex}[1]{{\protect\color{black} #1}} 
\providecommand{\DIFaddbegin}{\color{black}} 
\providecommand{\DIFaddend}{\color{black}} 
\providecommand{\DIFaddFL}[1]{\DIFadd{#1}} 
\providecommand{\DIFaddbeginFL}{\color{black}} 
\providecommand{\DIFaddendFL}{\color{black}} 
\providecommand{\DIFadd}[1]{\texorpdfstring{\DIFaddtex{#1}}{#1}} 
\RequirePackage{listings} 
\RequirePackage{color} 
\lstdefinelanguage{DIFcode}{ 
  moredelim=[il][\color{black}\scriptsize]{\%DIF\ <\ }, 
  moredelim=[il][\color{blue}\sffamily]{\%DIF\ >\ } 
} 
\lstdefinestyle{DIFverbatimstyle}{ 
	language=DIFcode, 
	basicstyle=\ttfamily, 
	columns=fullflexible, 
	keepspaces=true 
} 
\lstnewenvironment{DIFverbatim}{\lstset{style=DIFverbatimstyle}}{} 
\lstnewenvironment{DIFverbatim*}{\lstset{style=DIFverbatimstyle,showspaces=true}}{} 

\usepackage{etoolbox}
\AtBeginEnvironment{thebibliography}{%
 \let\revisionbibitem\bibitem
 \renewcommand{\bibitem}[2][]{\color{black}\revisionbibitem[#1]{#2}%
  \ifstrequal{#2}{lin2013}{\color{black}}{}%
  \ifstrequal{#2}{alaluhtala2016}{\color{black}}{}%
  \ifstrequal{#2}{billingsley1999}{\color{black}}{}%
  \ifstrequal{#2}{doucet2009}{\color{black}}{}%
  \ifstrequal{#2}{doucet2006}{\color{black}}{}%
  \ifstrequal{#2}{andrieu2010}{\color{black}}{}%
  \ifstrequal{#2}{heng2020}{\color{black}}{}%
  \ifstrequal{#2}{gordon1993}{\color{black}}{}%
  \ifstrequal{#2}{arulampalam2002}{\color{black}}{}%
  \ifstrequal{#2}{bon2022}{\color{black}}{}%
  \ifstrequal{#2}{xue2025}{\color{black}}{}}}
\AtEndEnvironment{thebibliography}{\color{black}}
\begin{document}
\twocolumn[
\tbpftitle{Twisted Block Particle Filters: A Local Particle-System\texorpdfstring{\\}{ }Change of Measure for High-Dimensional State-Space Models}
\begin{tbpfauthorlist}
\tbpfauthor{Ning Ning}{}
\end{tbpfauthorlist}
\tbpfkeywords{Sequential Monte Carlo, block particle filters, twisting}
\vskip 0.3in
]
\printAffiliationsAndNotice{}
\begin{abstract}
Block particle filters mitigate weight degeneracy in high-dimensional state-space models, but their normalizer estimates can remain highly variable. We introduce a twisted block particle filter (TBPF) that incorporates predictive guidance through an exact change of measure at blockwise resampling. The method guide-tilts one uniformly selected offspring per block, leaves the state-transition kernel unchanged, and computes the resampling correction using finite empirical guide averages. We prove that this correction exactly preserves the finite-particle block particle filter normalizer expectation, separating the effect of guided sampling from the approximation introduced by spatial blocking. Our analysis provides exact conditional local moments and a one-step conditional central limit result. We further derive explicit bounds on the correction accumulated across blocks and time, establish fixed-horizon consistency of the block filter and corrected normalizer under suitable regularity and guide-range conditions, and identify conditional criteria for guide perturbations that reduce variance. Experiments on coupled linear Gaussian and stochastic-volatility models with up to 256 latent dimensions show lower measured log-normalizer variance. For stochastic volatility with a five-step guide, the median reduction is $38\%$. An application to a 100-dimensional financial portfolio panel extends the evaluation to real data.
\end{abstract}

\section{Introduction}
\label{sec:introduction}

Sequential Monte Carlo (SMC), or particle filtering, provides a flexible simulation-based framework for inference in nonlinear and non-Gaussian state-space models. By propagating, weighting, and resampling an interacting particle population, particle filters (PFs) approximate filtering distributions and, in standard formulations, provide unbiased estimates of normalizing constants or marginal likelihoods \citep{gordon1993,doucet2001,delmoral2004,doucet2009}. These properties make particle filters useful not only for state estimation but also as computational building blocks for likelihood-based parameter inference and particle Markov chain Monte Carlo (MCMC) \citep{andrieu2010}. Their flexibility, however, comes with a well-known difficulty: when the state and observation dimensions are large, global importance weights can become highly variable, leading to rapid weight degeneracy and a particle requirement that grows prohibitively with dimension \citep{bengtsson2008,snyder2008}.

A major response to this difficulty is \emph{localization}. For graphical or spatially structured state-space models, block particle filters (BPFs) replace a global selection step by lower-dimensional blockwise updates and then recombine the selected blocks. Under suitable locality and mixing conditions, the resulting local filtering error can depend on block size rather than total model dimension, at the price of a deterministic blocking bias near interactions across block boundaries \citep{rebeschini2015}. This principle has subsequently been developed in several directions, including practical high-dimensional filtering and parameter learning \citep{park2020,ning2023}, blocked or divide-and-conquer SMC constructions \citep{crucinio2024}, and bagged filters with spatiotemporally localized weights \citep{ionides2023bagged}. In particular, \citet{ning2023} developed the iterated BPF for parameter learning and established dimension-stable local error bounds for graphical state-space models. \citet{ionides2024} extended iterated BPF to coupled systems with shared and unit-specific parameters. 

A different line of work addresses particle degeneracy by changing \emph{how a fixed target is simulated}. Auxiliary and look-ahead particle filters use predictive information to steer resampling toward particles likely to explain future observations \citep{pitt1999,lin2013}. Twisted particle filters (TPFs) formalize this idea through a change of measure on the particle system. \citet{whiteley2014} characterized an essentially optimal particle-system transition, under a long-time relative-variance criterion, through a positive eigenfunction or ``twisting'' function. Subsequent work developed practical approximations to twisting functions and connections with auxiliary particle filtering, controlled SMC, and Monte Carlo approximations of otherwise intractable twists \citep{alaluhtala2016,guarniero2017,heng2020,bon2022}. More recent work has connected twisting with continuous-time control and learned path-space guidance \citep{luwang2024}, while online controlled-SMC methods continue this direction for sequential real-time inference \citep{xue2025}. 

Blocking and twisting therefore attack different sources of difficulty. Blocking localizes selection and thereby changes the approximation target; twisting changes the sampling law used to represent a specified target and introduces the corresponding correction. This distinction raises a natural question that, to our knowledge, has not been addressed in the form considered here: \emph{can particle-system twisting be localized to the resampling stage of a block particle filter, while leaving the propagation kernel unchanged, in a way that is exactly correct for the finite-particle BPF baseline and whose conditional sampling behavior can be quantified?}

We introduce the \emph{twisted block particle filter} (TBPF), which localizes particle-system twisting to BPF resampling while leaving propagation unchanged. This construction separates the effect of guidance from the approximation error introduced by blocking. Our contributions are:

\begin{figure*}[t]
\centering
\includegraphics[width=\textwidth]{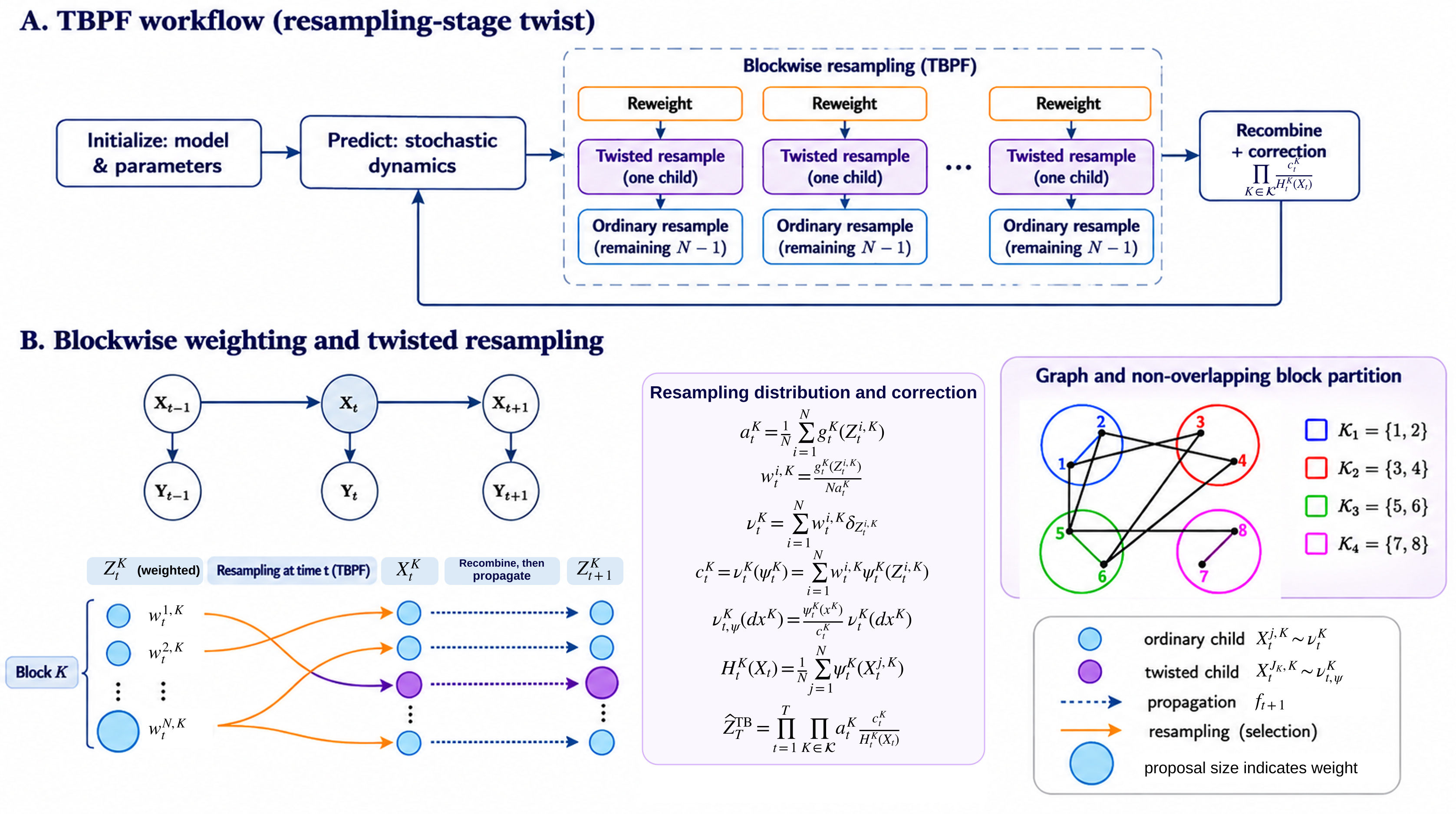}
\caption{Schematic of the resampling-stage TBPF. (A) Propagation is followed by blockwise weighting, selection of one guide-tilted child and $N-1$ ordinary children per block, recombination, and correction. (B) Particle selection within a block and an example graph partition. The weights, offspring law, and normalizer are specified by \eqref{eq:weights}, \eqref{eq:twist}, and \eqref{eq:normalizer}, respectively.}
\label{fig:tbpf-workflow}
\end{figure*}

\begin{itemize}
    \item \textbf{A localized particle-system twist for block filtering.}
    We construct a resampling-stage TBPF that guide-tilts one uniformly selected offspring per block while leaving the state-transition kernel unchanged. Figure~\ref{fig:tbpf-workflow} illustrates the blockwise selection, recombination, and correction. The construction is implementable using empirical guide averages and reduces to the BPF when the guides are constant.

    \item \textbf{An exact finite-particle change-of-measure correction.}
    On an augmented particle-system history, we derive the Radon--Nikodym correction between TBPF and the BPF and prove exact preservation of baseline expectations. In particular, the corrected TBPF normalizer has exactly the same expectation as the finite-particle BPF normalizer. This result carefully separates finite-particle block-normalizer preservation from exact-model likelihood unbiasedness.

    \item \textbf{Local moment, stability, and consistency theory.}
    For a filter with $N$ particles, we obtain conditional moment identities for the modified offspring law and show that replacing one child per block introduces an $O(N^{-1})$ shift in local conditional means without changing the leading $N^{-1/2}$ resampling fluctuation scale. We also control the second moment of the accumulated correction and establish fixed-horizon consistency for the block filtering recursion and, under additional guide-range conditions, for the corrected block normalizer.

    \item \textbf{A broad empirical study.}
    We compare PF, BPF, TPF, and TBPF on coupled linear Gaussian (LG) and stochastic-volatility (SV) models over dimensions $64$, $128$, and $256$, and we examine a 100-dimensional portfolio panel under an SV working model. The experiments show that localized twisting can substantially reduce measured normalizer variability in many matched BPF--TBPF comparisons.
\end{itemize}

\section{Model and Twisted Block Algorithm}\label{sec:model}
\subsection{Baseline block filter}\label{sec:baseline}
Let $V$ be the vertex set of a finite graph, with $d=|V|$. For each $v\in V$, let $\cX^v$ be a measurable local state space and $N(v)\subseteq V$ its transition neighborhood. For $J\subseteq V$, write $\cX^J=\prod_{v\in J}\cX^v$ and $x^J=(x^v)_{v\in J}$, and set $\cX=\cX^V$. Fix a time horizon $T\ge1$. For each $t=1,\ldots,T$ and $v\in V$, let $f_t^v(x^{N(v)},dz^v)$ be a Markov kernel from $\cX^{N(v)}$ to $\cX^v$. Fix observations $y_{1:T}=(y_1,\ldots,y_T)$. For $t=1,\ldots,T$, let $g_t^v:\cX^v\to(0,\infty)$ be the measurable local likelihood at time $t$. The prescribed kernel $f_t$ propagates the state from time $t-1$ to $t$ and may vary with $t$. Its sitewise factors and the likelihood factorize as
\begin{equation}
\begin{gathered}
 f_t(x,dz)=\prod_{v\in V}f_t^v(x^{N(v)},dz^v),\\
 g_t(z)=\prod_{v\in V}g_t^v(z^v)>0.
\end{gathered}
\end{equation}
For a probability measure $\mu$ on $\cX$, write $\mu(h)=\int h(x)\,\mu(dx)$ for an integrable measurable function $h$. Define the time-$t$ prediction operator $P_t$ and, whenever $0<\mu(g_t)<\infty$, the observation-update operator $C_t$ by
\[
\begin{aligned}
 (P_t\mu)(A)&=\int_{\cX} f_t(x,A)\,\mu(dx),\\
 (C_t\mu)(A)&=\frac{\int_A g_t(z)\,\mu(dz)}{\mu(g_t)},
\end{aligned}
\]
for each measurable $A\subseteq\cX$. Equivalently, $P_t\mu=\mu f_t$ and $C_t\mu=g_t\mu/\mu(g_t)$. Let $\pi_0$ be the initial state law. The exact filtering law at time $t$ is $\pi_t=C_tP_t\pi_{t-1}$.

For a partition $\cK$ of $V$ into nonempty disjoint blocks, let $\mu^K$ denote the marginal of $\mu$ on $\cX^K$ and define $B\mu=\bigotimes_{K\in\cK}\mu^K$. Starting from a  law $\pi_0=\bigotimes_{K\in\cK}\pi_0^K$ that is a product across partition blocks, define
\begin{equation}\label{eq:block}
 \widetilde\pi_t=C_tBP_t\widetilde\pi_{t-1}\qquad\text{\normalfont and}\qquad \widetilde\pi_0=\pi_0.
\end{equation}
Thus $B$ is the approximation step: after prediction, it replaces the joint law by the product of its block marginals before the observation correction is applied. The block likelihood is $g_t^K=\prod_{v\in K}g_t^v$.

The finite-particle BPF mirrors this recursion as follows. Let $N\ge1$ be the number of particles and write $X_t^{1:N}=(X_t^1,\ldots,X_t^N)\in\cX^N$, with $X_t^{i,K}$ the block-$K$ coordinates of particle $i$. Unspecified sums over $i,j$, over $K$, and over $t$ range over $\{1,\ldots,N\}$, $\cK$, and $\{1,\ldots,T\}$, respectively; products use the same ranges. Let $\delta_x$ denote the unit point mass at $x$. Given particles $X_{t-1}^{1:N}$, draw independent proposals $Z_t^i\sim f_t(X_{t-1}^i,\cdot)$. For each block $K\in \cK$, define
\begin{align}\label{eq:weights}
 a_{t,N}^{K}=N^{-1}\sum_i g_t^K(Z_t^{i,K}),\quad
 w_t^{i,K}=\frac{g_t^K(Z_t^{i,K})}{N a_{t,N}^{K}},\\
 \nu_t^K=\sum_iw_t^{i,K}\delta_{Z_t^{i,K}}\quad\text{and}\quad
 \widehat Z_T^B=\prod_{t=1}^T\prod_{K\in\cK}a_{t,N}^{K}.\nonumber
\end{align}
BPF independently draws $N$ children from each $\nu_t^K$, then concatenates the $i$th children across blocks. The normalizer estimator $\widehat Z_T^B$ in \eqref{eq:weights} is evaluated along this BPF history. With one block this is the ordinary bootstrap likelihood estimate; with several interacting blocks it is generally an approximate likelihood functional.

\subsection{One distinguished child per block}\label{sec:algorithm}
Let $\mathcal G_t$ be the sigma-field containing the past, current proposal pool, and any guide-construction randomness, before current offspring are generated. For each $K$, choose a positive finite guide $\psi_t^K:\cX^K\to(0,\infty)$ using only $\mathcal G_t$. For a candidate offspring array $x=(x^1,\ldots,x^N)\in\cX^N$, define
\begin{equation}
\begin{gathered}
 c_t^K=\nu_t^K(\psi_t^K),\quad H_t^K(x)=N^{-1}\sum_j\psi_t^K(x^{j,K}).
\end{gathered}
\end{equation}
The quantity $c_t^K$ is the guide average under the weighted proposal law. The empirical guide average $H_t^K(x)$ instead depends on the candidate offspring array $x$ and hence on the resampling outcome.   {The ordinary and twisted conditional } full-array   {laws are
}\[
{R_t(dx\mid\mathcal G_t)=\bigotimes_{K\in\cK}(\nu_t^K)^{\otimes N}(dx^{1:N,K}).
}\]
 \begin{equation}\label{eq:twist}
 \widetilde R_t(dx\mid\mathcal G_t)
 =R_t(dx\mid\mathcal G_t)\prod_K\frac{H_t^K(x)}{c_t^K}.
\end{equation}
For a probability vector $w=(w_1,\ldots,w_N)$, $\operatorname{Cat}(w)$ assigns probability $w_i$ to label $i$. In Algorithm~\ref{alg:tbpf}, $J_t^K$ is the distinguished child label, $A_t^{j,K}$ is the proposal label selected for child $j$ in block $K$, and $L_t$ is the accumulated log normalizer through time $t$. We write $w_t^{1:N,K}=(w_t^{1,K},\ldots,w_t^{N,K})$ and abbreviate the current offspring array $X_t^{1:N}$ by $X_t$. For each block this law draws a uniform child label, samples the other children independently from $\nu_t^K$, and then samples the selected child independently from   \[{\nu_{t,\psi}^K=\psi_t^K\nu_t^K/c_t^K.}\]  Execute blocks independently given $\mathcal G_t$.   {Under $R_t$, $\E[H_t^K\mid\mathcal G_t]=c_t^K$; conditional block independence makes the product density integrate to one. Sampling and correction use the same realized, finite, strictly positive guide values, frozen before drawing offspring.
} 

\begin{algorithm}[t]
\caption{Resampling-stage TBPF}\label{alg:tbpf}
\begin{algorithmic}[1]
\setlength{\itemsep}{2pt}
\STATE Independently (ind) draw $X_0^1,\ldots,X_0^N\sim\pi_0$; set $L_0\gets0$.
\FOR{$t=1,\ldots,T$}
\STATE Ind draw $Z_t^i\sim f_t(X_{t-1}^i,\cdot)$.
\STATE Compute $a_{t,N}^{K},w_t^{i,K},\nu_t^K$ by \eqref{eq:weights}.
\STATE Freeze $s_i^K=\psi_t^K(Z_t^{i,K})>0$ for every block; set $s_i^K=1$ at $t=T$.
\STATE Store weighted output $\widehat\pi_t^{TB}=\bigotimes_K\nu_t^K$.
\FOR{each block $K$, ind.}
\STATE Set $c_t^K=\sum_iw_t^{i,K}s_i^K$.
\STATE Draw $J_t^K\sim\operatorname{Uniform}\{1,\ldots,N\}$.
\STATE Ind draw  $A_t^{j,K}\sim\operatorname{Cat}(w_t^{1:N,K})$ for $j\ne J_t^K$.
\STATE Ind draw  $A_t^{J_t^K,K}\sim\operatorname{Cat}((w_t^{i,K}s_i^K/c_t^K)_{i=1}^N)$.
\STATE Set $X_t^{j,K}=Z_t^{A_t^{j,K},K}$ for every $j$.
\STATE Set $H_t^K=N^{-1}\sum_js_{A_t^{j,K}}^K$.
\ENDFOR
\STATE Concatenate block children to obtain $X_t^{1:N}$.
\STATE $L_t\gets L_{t-1}+\sum_K(\log a_{t,N}^{K}+\log c_t^K-\log H_t^K)$.
\ENDFOR
\STATE Return $\widehat Z_T^{TB}=e^{L_T}$ and weighted outputs.
\end{algorithmic}
\end{algorithm}

The correction is evaluated on the same offspring used in the next propagation step:
\begin{equation}\label{eq:normalizer}
 \widehat Z_T^{TB}=\prod_{t,K}a_{t,N}^{K}\frac{c_t^K}{H_t^K(X_t)}.
\end{equation}
Figure~\ref{fig:tbpf-workflow} illustrates the blockwise resampling and recombination in Algorithm~\ref{alg:tbpf}.
Set $\psi_T^K=1$ because terminal resampling cannot improve an estimate that no longer depends on the children. Constant guides at every time recover BPF. The weighted output $\widehat\pi_t^{TB}$ supplies experimental filtering means; the recombined empirical measure $\eta_t^N=N^{-1}\sum_i\delta_{X_t^i}$ is a distinct output.

Write $\E_{TB}$ and $\E_B$ for expectations under the twisted and baseline particle-system laws, and $\Var_{TB},\Var_B$ for the corresponding variances. For a probability measure $\mu$ and square-integrable $u$, set $\Var_\mu(u)=\mu(u^2)-\mu(u)^2$. A block test $u:\cX^K\to\mathbb R$ is also viewed as the full-state function $x\mapsto u(x^K)$; $\|u\|_\infty$ denotes its supremum norm. For a probability law $\rho$, write $\operatorname{Cov}_\rho(a,b)=\rho(ab)-\rho(a)\rho(b)$ when these integrals exist.
If guides are randomized, the reference history also records their construction variables. Both laws use the same conditional guide-generation rule; the baseline particle marginal remains BPF. Any retained resampling marks are attached using the conditional extension in Theorem~\ref{supp:thm:rn}. This distinction is needed because a realized correction need not be measurable from the particle arrays alone (Theorem~\ref{supp:thm:rn}).

\begin{theorem}[Exact preservation of the baseline]\label{thm:rn}
Let $\mathbb P_B,\mathbb P_{TB}$ be the baseline and twisted laws on the augmented history that records the proposal/offspring arrays and all retained guide randomness. Use the common guide-generation kernels and baseline auxiliary extension specified in Theorem~\ref{supp:thm:rn}. For positive finite frozen guides,
\begin{gather}
 D_T=\prod_{t,K}\frac{c_t^K}{H_t^K(X_t)}
 =\frac{d\mathbb P_B}{d\mathbb P_{TB}},\\
\text{\normalfont and}\quad \E_{TB}[D_TF]=\E_B[F]
\end{gather}
for nonnegative measurable $F$, and for $\E_B|F|<\infty$. The particle-array marginal of $\mathbb P_B$ is BPF. In particular,
  {$\E_{TB}\widehat Z_T^{TB}=\E_B\widehat Z_T^B$}.
\end{theorem}
The full statement and proof appear in Theorem~\ref{supp:thm:rn}.
This yields exact-likelihood unbiasedness for one block and for independent subsystems aligned with the partition.

\section{Local   {Sampling Effects } and   {Correction Stability} }\label{sec:accuracy}
\label{sec:consistency}
When taking $N\to\infty$, $\nu_{t,N}^K$ denotes the weighted proposal law $\nu_t^K$ with its particle-count dependence explicit. The notation $\mathcal N(m,\Sigma)$ denotes a Gaussian law with mean $m$ and covariance $\Sigma$.
\begin{theorem}[Conditional local moments]\label{thm:moments}
Fix $t,K$ and a bounded block test  \DIFaddbegin \DIFadd{$u(x^K)$ that may be a random function with evaluation jointly measurable in }\DIFaddend $\mathcal G_t$ \DIFaddbegin \DIFadd{and the block state; its realized function is fixed before offspring are drawn}\DIFaddend. Use the positive finite frozen guides of Algorithm~\ref{alg:tbpf}. Conditional on $\mathcal G_t$,
 \DIFaddbegin \begin{gather}
 \DIFadd{\E_{TB}[\eta_t^N(u)\mid\mathcal G_t]=\nu_t^K(u)+\frac{\nu_{t,\psi}^K(u)-\nu_t^K(u)}N,}\\
 \DIFadd{\Var_{TB}(\eta_t^N(u)\mid\mathcal G_t)=\frac{(N-1)\Var_{\nu_t^K}(u)+\Var_{\nu_{t,\psi}^K}(u)}{N^2}.\nonumber
}\end{gather}\DIFaddend 
For $|u|\le1$, the conditional mean squared error relative to $\nu_t^K(u)$ is at most $N^{-1}+4N^{-2}$. For uniformly bounded pool-measurable tests $u_N$, if $\Var_{\nu_{t,N}^K}(u_N)\to\sigma^2$ in probability for deterministic $\sigma^2$, the conditional characteristic functions of $\sqrt N(\eta_t^N(u_N)-\nu_{t,N}^K(u_N))$ converge in probability to those of $\mathcal N(0,\sigma^2)$.
\end{theorem}
The full statement and proof appear in Theorem~\ref{supp:thm:moments}.
Thus the local shift is $O(N^{-1})$ and does not change the leading resampling covariance. At a common pool, replacing one ordinary child per block gives the pathwise bound $2q(J)\|u\|_\infty/N$ for a test depending on a set $J$ meeting $q(J)$ blocks. 

\begin{proposition}[Correction stability]\label{prop:correction}
For the positive finite frozen guides of Algorithm~\ref{alg:tbpf}, let $r_t^K$ be the ratio of the largest to the smallest guide value on the block proposal support. For $C_t^K=c_t^K/H_t^K(X_t)$,
 \DIFaddbegin \begin{equation}
\DIFadd{\begin{aligned}
 \E_{TB}[C_t^K\mid\mathcal G_t]&=1,\\
 \E_{TB}[(C_t^K)^2\mid\mathcal G_t]&\le1+\frac{(\sqrt{r_t^K}-1)^2}{N}.
\end{aligned}
}\end{equation}\DIFaddend 
If deterministic bounds  \DIFaddbegin \DIFadd{$r_t^K\le \rho_t^K<\infty$ }\DIFaddend hold almost surely across histories and pools, then
 \DIFaddbegin \begin{equation}\DIFadd{\statementlabel{eq:global}
 \E_{TB}D_T^2\le
 \exp\left\{\frac{1}N\sum_{t,K}(\sqrt{\rho_t^K}-1)^2\right\}.
}\end{equation}\DIFaddend 
\end{proposition}
\vspace{-0.7cm}

The full statement and proof appear in Proposition~\ref{supp:prop:second}.
Fixed-range clipping is  \DIFaddbegin \DIFadd{sufficient. If $\rho_t^K\le\rho$, the exponent is at most $T|\cK|(\sqrt\rho-1)^2/N$. }\DIFaddend The bound also shows why a small conditional-mean shift does not imply small global importance fluctuations: correction factors accumulate over both space and time.

\begin{figure*}[!t]
	\centering
	\DIFaddbeginFL \includegraphics[width=\textwidth,height=.58\textheight,keepaspectratio]{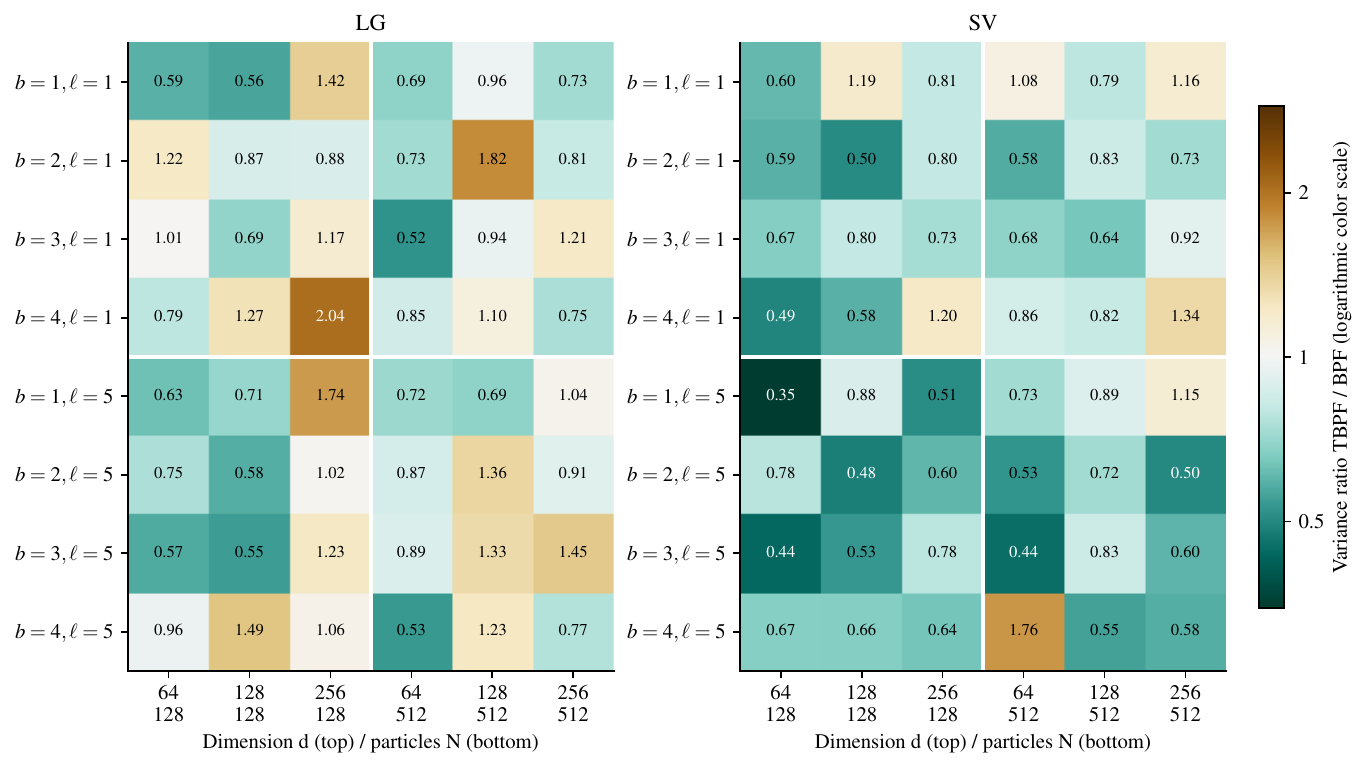}
	\DIFaddendFL \caption{Effect of block size $b$ and guide horizon $\ell$ on log-normalizer variance. Every cell reports TBPF variance divided by BPF variance at the same model, dimension, particle count, and block size. Lower ratios favor TBPF, with values below one indicating a reduction. Color is logarithmic and centered at one, with a shared scale across both models. All 96 fixed-configuration comparisons are shown, including configurations where twisting increases variance. }
	\label{fig:block_horizon_heatmap}
\end{figure*}

\DIFaddbegin \paragraph{\DIFadd{When does a guide reduce variance?}}
\DIFadd{Let $\mathcal Z_T=\prod_{t,K}a_{t,N}^{K}$ be the same history functional under both laws. If both estimators have finite second moments, then
}\begin{equation}\DIFadd{\label{eq:variance-identity}
 \Var_{TB}(D_T\mathcal Z_T)-\Var_B(\mathcal Z_T)=\E_B[(D_T-1)\mathcal Z_T^2].
}\end{equation}
\DIFadd{Thus improvement requires alignment between the correction and the downstream squared payoff; bounding $\E_{TB}D_T^2$ alone is insufficient. Equal means also imply the same difference in squared-error risk relative to any common deterministic reference. 
}

\begin{proposition}[Conditional guide improvement]\label{prop:guide-improvement}
Condition on a fixed block pool with weighted law $\nu$ and offspring law $R=\nu^{\otimes N}$. Let $X=(X^1,\ldots,X^N)$ and $U$ have baseline joint law $P_0(dx,du)=R(dx)\mathsf M(x,du)$, where $\mathsf M$ is the continuation kernel. Fix a nonnegative measurable payoff $W$ with $\mathcal A(x):=\int W(x,u)^2\mathsf M(x,du)$ integrable under $R$, and a bounded measurable block function $h$. For a real parameter $\theta$ satisfying $|\theta|\|h\|_\infty<1$, use $\psi_\theta=1+\theta h$, holding the pool, $\mathsf M$, $W$, and $h$ fixed. Set $m_h=\nu(h)$ and $\bar h(x)=N^{-1}\sum_jh(x^j)$, and define
\[
\begin{gathered}
 \Lambda_\theta(x)=\frac{1+\theta\bar h(x)}{1+\theta m_h},\quad
 Q_\theta(dx,du)=\Lambda_\theta(x)P_0(dx,du),\\
 \mathcal V(\theta)=\Var_{Q_\theta}\!\left(\frac{W(X,U)}{\Lambda_\theta(X)}\right).
\end{gathered}
\]
Then $\mathcal V'(0)=-\operatorname{Cov}_R(\mathcal A,\bar h)$. Some positive admissible $\theta$ reduces variance if and only if this covariance is positive; in that case, all sufficiently small positive $\theta$ do so.
\end{proposition}
\DIFadd{For an exact example with one block, take $N=2$ and an equal-weight pool $\{0,1\}$ before the terminal step. Use identity propagation and terminal likelihood $g_{\mathrm{ex}}(0)=1$, $g_{\mathrm{ex}}(1)=3$. For the guide $\psi_\beta=g_{\mathrm{ex}}^\beta$, $\beta\in\mathbb R$, the corrected continuation payoff has variance
}\[
 \mathcal V_{\mathrm{ex}}(\beta)=\frac{(3^\beta-3)^2}{8\cdot3^\beta}.
\]
\DIFadd{The constant guide $\beta=0$ recovers BPF variance $1/2$; the aligned guide $\beta=1$ gives zero variance. Improvement holds exactly for $0<\beta<2$. Appendix~\ref{supp:pool-example} gives the payoff, correction, and full enumeration.
}

\DIFadd{For log variance, additionally assume $W>0$ almost surely under $P_0$ and $\E_{P_0}[(\log W)^2]<\infty$ in Proposition~\ref{prop:guide-improvement}. Set $Y_{\log}=\log W$, $\mu_{\log}=\E_{P_0}Y_{\log}$, and $\mathcal V_{\log}(\theta)=\Var_{Q_\theta}[Y_{\log}-\log\Lambda_\theta(X)]$. Then
}\begin{equation}\label{eq:log-guide-improvement}
 \begin{aligned}
 \mathcal V_{\log}'(0)
 &=\E_{P_0}\!\big[(\bar h(X)-m_h)\\
 &\qquad\times\{(Y_{\log}-\mu_{\log})^2-2(Y_{\log}-\mu_{\log})\}\big].
 \end{aligned}
\end{equation}
\DIFadd{A negative derivative gives strict log-variance reduction for all sufficiently small positive $\theta$. Appendix~\ref{supp:variance-guides} proves both criteria and bounds the finite-step change in $\mathcal V$.
}

\DIFaddend \begin{theorem}[Fixed-horizon consistency]\label{thm:consistency}
Fix $d,T,\cK$ and the observations, and let $N\to\infty$. Suppose component spaces are Polish, each $f_t$ is Feller, likelihood factors are bounded continuous and strictly positive, and initial particles are independent and identically distributed (i.i.d.) from  \DIFaddbegin \DIFadd{$\pi_0=\bigotimes_{K\in\cK}\pi_0^K$}\DIFaddend. For any positive finite predictable guides, including $N$-dependent guides, for each  \DIFaddbegin \DIFadd{$1\le t\le T$,
}\[
 \DIFadd{\eta_t^N,\ \widehat\pi_t^{TB}\xrightarrow{\mathrm{weak},\mathbb P}\widetilde\pi_t\quad(t\ge1).
}\]\DIFaddend 
If the deterministic range bounds of Proposition~\ref{prop:correction} also hold uniformly in $N$, then $D_T\to1$ in $L^2$ and the corrected normalizer satisfies
\begin{equation}
 \widehat Z_T^{TB}\to
 Z_T^{\mathrm{block}}=\prod_{t,K}(P_t\widetilde\pi_{t-1})^K(g_t^K)\quad\text{\normalfont in }L^2.
\end{equation}
\end{theorem}
 \DIFaddbegin 

\DIFaddend Here $\xrightarrow{\mathrm{weak},\mathbb P}$ means weak convergence of random probability measures in probability. For random scalars, convergence in $L^2$ means convergence in mean square under the particle-system law.
\DIFaddbegin \DIFadd{At $t=0$, $\eta_0^N$ converges weakly in probability to $\pi_0$. }\DIFaddend The full statement and proof appear in Theorem~\ref{supp:thm:consistent}. \DIFaddbegin \DIFadd{The guarantee is for fixed $d,T,\cK$ and bounded continuous tests. Coordinate-mean root mean squared error (RMSE) is evaluated empirically.
}\DIFaddend

\section{Numerical and Empirical Experiments}\label{sec:experiments}
In the continuous-state simulations and real-data analysis, we compare PF, BPF, TPF, and TBPF. The implemented TPF twists the ancestor selection and propagation of one child using a Gaussian-shaped full-state guide. PF and TPF  \DIFaddbegin \DIFadd{estimate the model likelihood unbiasedly; }\DIFaddend BPF and TBPF  \DIFaddbegin \DIFadd{share the BPF normalizer expectation at the stated $N$ and partition}\DIFaddend.
\DIFaddbegin\DIFadd{Here $K\in\cK$ is a block, $b=\max_{K\in\cK}|K|$ is the maximum number of sites per block, and $|\cK|$ is the number of blocks.}\DIFaddend

\subsection{Models, guides, comparators, and design}\label{sec:design}
\paragraph{Models.}
Following the LG and SV examples of \citet{whiteley2014}, we consider spatially coupled state-space models on a periodic ring. For sites $v\in V$, the latent process evolves according to
\begin{equation}
 X_t^v=0.7X_{t-1}^v+0.1\left(X_{t-1}^{v-1}+X_{t-1}^{v+1}\right)
       +\sigma\epsilon_t^v,
\end{equation}
where $V=\{1,\ldots,d\}$ and spatial indices are interpreted periodically. Here $X_t^v$ is the latent state, $Y_t^v$ below is its observation, and $\sigma$ is the state-noise scale. The symbols $\epsilon_t^v,e_t^v$ denote state and observation noises. We initialize $X_0\sim\mathcal N(0,I_d)$, where $I_d$ is the $d\times d$ identity matrix. All state and observation noises are mutually independent standard Gaussian variables across sites and times.
The two observation models are
\[
\begin{aligned}
 \text{LG:}\quad &Y_t^v=X_t^v+e_t^v, &&\sigma=1,\\
 \text{SV:}\quad &Y_t^v=\exp(X_t^v/2)e_t^v, &&\sigma=0.3.
\end{aligned}
\]

\DIFaddbegin \paragraph{\DIFadd{Information timing.}}
\DIFadd{At time $t$, TBPF resampling uses observations through $\min(T,t+\ell)$; its weighted output is affected by preceding twists and can depend on observations through $\min(T,t+\ell-1)$. Thus for $\ell=1$ the weighted output uses observations through $t$, while subsequent resampling uses $y_{t+1}$. TPF uses the arrival window $y_{t:t+\ell-1}$.  Here ``predictable'' means frozen before drawing offspring, not causal in observation time.
}

\DIFaddend \paragraph{Experimental design.}
We consider dimensions $d\in\{64,128,256\}$, particle counts $N\in\{128,512\}$, and horizon $T=80$. For blocked methods, the maximum block size is $b\in\{1,2,3,4\}$; when $b=3$, the final contiguous block may be smaller. Twisted methods use guide horizons $\ell\in\{1,5\}$. For each model and dimension, we generate one fixed dataset and use 32 Monte Carlo replications of each applicable procedure. \DIFaddbegin \DIFadd{Runs share seeds across methods. Appendix~\ref{supp:independent} examines data sensitivity at a fixed configuration using two independently simulated datasets per model and dimension.}\DIFaddend

\paragraph{Guides.}
For TBPF, the block guide targets the look-ahead likelihood
\begin{equation}\label{eq:lookahead-guide}
 \psi_t^K(x^K)\approx
 p(y_{t+1:t+\ell}^K\mid X_t^K=x^K).
\end{equation}
Here $y_{t+1:t+\ell}^K$ denotes the next $\ell$ block observations, truncated at $T$. We approximate this likelihood using isolated scalar dynamics with autoregression $0.9$, then combine site messages within each block. The resulting positive guide enters both the tilted law $\nu_{t,\psi}^K=\psi_t^K\nu_t^K/c_t^K$ and the normalizer correction $c_t^K/H_t^K(X_t)$. The coupled transition and observation likelihood remain unchanged.
For TPF, the full-state guide at arrival time $t+1$ has the quadratic form
\[
 \psi_{t+1}(z)=\exp\!\left\{-\frac{1}{2}\sum_{v\in V}
 a_{t+1}^v(z^v-m_{t+1}^v)^2\right\},
\]
where $a_{t+1}^v>0$ and $m_{t+1}^v$ are guide precisions and centers. For particles $Z_t^i$ with normalized weights $w_t^i$, the distinguished parent's weight is proportional to $w_t^i(f_{t+1}\psi_{t+1})(Z_t^i)$, where $(f_{t+1}\psi_{t+1})(x)=\int f_{t+1}(x,dz)\psi_{t+1}(z)$. Gaussian integration and tilted sampling are exact for this chosen guide. TPF and TBPF therefore differ in guide families as well as blocking. The LG recursions, SV approximations, and clipping rules are detailed in Appendices~\ref{supp:subsec:guide-design} and~\ref{supp:subsec:tpf}. 

\paragraph{Ablation studies and sensitivity analyses.}
We isolate the resampling twist by comparing BPF and TBPF at the same model, dataset, particle count, and partition. \DIFaddbegin \DIFadd{Under these matched conditions, both methods have the same normalizer expectation.}\DIFaddend Within each twisted method, the guide horizons $\ell=1$ and $\ell=5$ examine sensitivity to look-ahead. Varying $b\in\{1,2,3,4\}$ examines block-size sensitivity, although it also changes the blocked target.

\paragraph{Evaluation metrics.}
LG filtering RMSE is computed against the exact Kalman filtering mean, whereas SV filtering RMSE is computed against the simulated latent state. We report \DIFaddbegin \DIFadd{observed }\DIFaddend log-normalizer variance\DIFaddbegin \DIFadd{. Appendix~\ref{supp:revision-validation} reports paired bootstrap diagnostics and ordinary-scale concentration results. }\DIFaddend

\begin{figure*}[!t]
	\centering
	\DIFaddbeginFL \includegraphics[width=0.9\textwidth,height=.58\textheight,keepaspectratio]{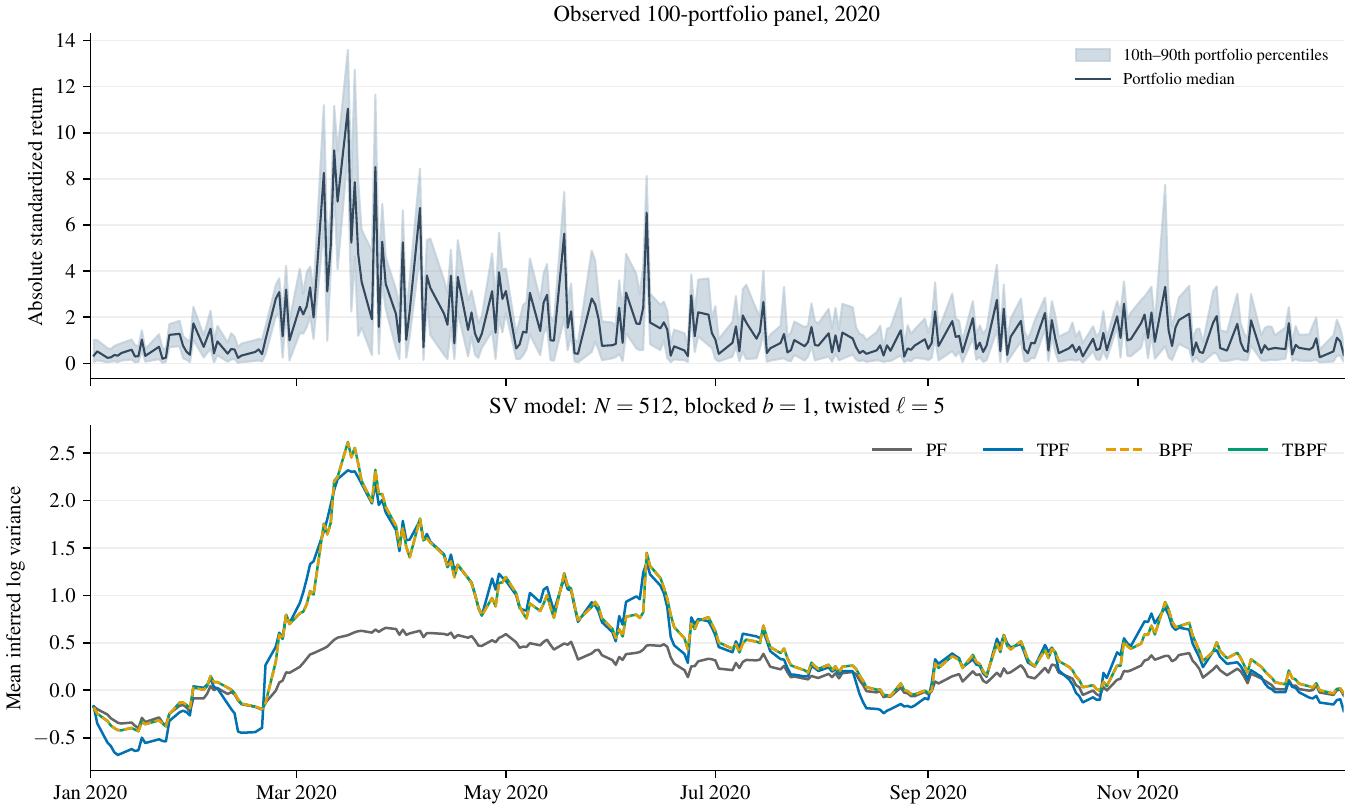}
	\DIFaddendFL \caption{Real 100-portfolio data and model-based inference during 2020. Top: cross-sectional median and 10th--90th percentiles of absolute returns standardized using 2018--2019 moments. The shaded band is cross-sectional dispersion, not a confidence interval. Bottom: SV empirical filtered log variance, averaged over all 100 coordinates and 32 runs, for PF, TPF, BPF, and TBPF at $N=512$, with $b=1$ for blocked methods and $\ell=5$ for twisted methods. Higher values indicate larger absolute returns (top) or higher inferred log variance (bottom).}
	\label{fig:real100_series}
\end{figure*}

\subsection{Continuous-state simulation results}\label{sec:simulations}
\paragraph{Ablation study: resampling twist.}
Figure~\ref{fig:block_horizon_heatmap} displays all 96 fixed-configuration variance ratios across block sizes and guide horizons. For each model, data sequence, dimension, particle count $N$, block size $b$, and guide horizon $\ell$, the ratio is calculated as
 \DIFaddbegin \begin{gather*}
 \DIFadd{r_V=\widehat{\Var}_r(\log\widehat Z_{T,r}^{TB})\slash\widehat{\Var}_r(\log\widehat Z_{T,r}^{B}),
}\end{gather*}\DIFaddend 
where  \DIFaddbegin \DIFadd{$r$ indexes the filter run at fixed observations and configuration, and $\widehat{\Var}_r$ denotes }\DIFaddend sample variance across 32  \DIFaddbegin \DIFadd{runs}\DIFaddend, and BPF and TBPF use the same partition. Lower ratios favor TBPF; for example, $r_V=0.6$ indicates a 40\% reduction in measured log-normalizer variance. The heatmap retains every tested block-size and horizon combination. 

TBPF has lower measured variance in 29/48 LG and 41/48 SV comparisons, totaling 70/96. The size of the reduction depends on block size and guide horizon. Reductions are more frequent in SV, particularly at $\ell=5$, where 22 of 24 settings favor TBPF. Table~\ref{tab:matched} summarizes these comparisons: the median variance ratio is below one in all four model/horizon groups and equals 0.620 for SV at $\ell=5$.

\begin{table}[!t]\color{black}
\caption{\color{black}Resampling-twist ablation study: same-partition TBPF/BPF comparisons from the simulation runs. Each row contains 24 fixed $(d,N,b)$ settings. The guide horizon $\ell$ is the number of future block observations used to construct the TBPF guide in \eqref{eq:lookahead-guide}. $r_V$ is the TBPF/BPF log-variance ratio: lower ratios favor TBPF, with values below one indicating a reduction. Med.\ denotes the median. Higher counts of observed reductions favor TBPF.}
\label{tab:matched}
\centering\small
\begin{tabular}{lcrr}
\toprule
Model & $\ell$ & $r_V<1$ & Med.\ $r_V$\\\midrule
LG & 1 & 15/24 & 0.876\\
LG & 5 & 14/24 & 0.899\\
SV & 1 & 19/24 & 0.795\\
SV & 5 & 22/24 & 0.620\\
\bottomrule\end{tabular}
\end{table}

\begin{table}[!t]\color{black}
\caption{\color{black}Four-method, horizon-matched point-estimate minima: TBPF is compared with PF, BPF at the same block size, and TPF at the same horizon. These counts do not select a minimum over block sizes or horizons. Higher counts favor TBPF.}
\label{tab:summary}
\centering
\begin{tabular}{llr}
\toprule Model & Metric & TBPF minima\\\midrule
LG & Log variance & 28/48\\
LG & Filtering RMSE & 39/48\\
SV & Log variance & 41/48\\
SV & Filtering RMSE & 33/48\\
\bottomrule
\end{tabular}
\end{table}

\paragraph{Cross-method benchmarks.}
The preceding comparison isolates TBPF relative to BPF at the same partition. We next compare all four method families (Table~\ref{tab:summary}). TBPF has the smallest log-estimate variance in 28/48 LG and 41/48 SV horizon-matched comparisons, totaling 69/96. Appendix~\ref{supp:subsec:validation} gives additional simulation results and method comparisons, including log-normalizer variance and filtering RMSE at both particle counts in Figures~\ref{supp:fig:variance_dimension} and~\ref{supp:fig:filtering_accuracy}.

\DIFaddbegin\DIFadd{Separately minimizing filtering RMSE within BPF and TBPF at each model, dimension and particle count gives differences below 0.6\% in LG and 0.2\% in SV, relative to BPF. These are state-estimation errors, not log-normalizer variances. The selected LG configurations use $b=1$; SV uses $b=2$--$4$. TBPF has the smallest measured filtering RMSE in 72/96 horizon-matched four-method comparisons.}\DIFaddend


\subsection{Real 100-portfolio panel}\label{sec:real}
\DIFaddbegin\DIFadd{As a computational stress test, we analyze all 100 daily value-weighted size/book-to-market portfolios from the Kenneth French Data Library \citep{FrenchData}, retaining their published order. The August 2026 Center for Research in Security Prices (CRSP) release supplies 503 training days in 2018--2019 and 253 evaluation days in 2020, with no missing observations. Training means and sample standard deviations standardize returns; the largest cross-sectional median absolute standardized return is 11.04 on 16 March 2020.}

\DIFadd{The fixed working SV model has self/neighbor coefficients $0.7,0.1,0.1$ and innovation scale $0.3$ on a source-order ring. All filters start from $\mathcal N(0,I_{100})$, without training-period warm-up; only preprocessing moments are fitted. Both particle counts, all block sizes and horizons, and 32 replications give 960 runs. Figure~\ref{fig:real100_series} uses a fixed illustrative configuration. Table~\ref{tab:real} compares both horizons at fixed partition. Separately selected minima, both at $b=1$, give observed reductions of 20.1\% and 56.3\% for $N=128$ and $512$. These descriptive minima do not validate a tuning rule. Appendix~\ref{supp:sec:real_data} provides preprocessing and model details and separately selected method minima.}\DIFaddend

\begin{table}[t]\color{black}\centering\small
\caption{\color{black}Fixed-partition real-panel comparisons at $b=1$. Ratios are TBPF/BPF observed log variance with paired-seed 95\% bootstrap intervals (32 runs, 4,000 resamples). Lower ratios favor TBPF.}\label{tab:real}
\begin{tabular}{rrrr}\toprule $N$ & $\ell$ & Variance ratio & 95\% interval\\\midrule
128 & 1 & 0.950 & [0.471,1.923]\\
128 & 5 & 0.799 & [0.435,1.526]\\
512 & 1 & 0.437 & [0.210,0.833]\\
512 & 5 & 0.527 & [0.245,1.064]\\
\bottomrule\end{tabular}\end{table}

\section{Discussion}
 \DIFaddbegin \DIFadd{TBPF preserves the BPF normalizer expectation by an exact finite-pool change of resampling law. Under stated fixed-setting assumptions, its corrected normalizer converges to $Z_T^{\mathrm{block}}$. Guide alignment can either reduce or inflate ordinary variance; the conditional criterion identifies a local improvement direction without claiming global optimality. The empirical evidence is conditional on the tested datasets: }\DIFaddend 70\DIFaddbegin\DIFadd{/}\DIFaddend{}96  \DIFaddbegin \DIFadd{matched comparisons show lower observed log-normalizer variance. The independent-data check gives 3/6 LG and 6/6 SV observed reductions, with wide conditional intervals.}

%

\DIFaddend \label{disclosures-end}
\clearpage
 \DIFaddbegin \bibliographystyle{tbpfpreprint}
\bibliography{references_ms7}
\DIFaddend \clearpage
\onecolumn
\appendix
\section*{Guide to the appendices}\label{supplement-start}
The appendices provide  \DIFaddbegin \DIFadd{related literature}\DIFaddend, proofs, experimental details,  \DIFaddbegin \DIFadd{diagnostic calculations, and notation. }\DIFaddend Select a section or an indented subsection below to navigate directly to it.
\begingroup\small
\begin{itemize}
\item \hyperref[app:background]{Appendix~\ref*{app:background}: Methodological context and related literature}
\begin{itemize}
\item \hyperref[app:bg:pf]{\ref*{app:bg:pf}: Particle filtering, likelihood estimation, and degeneracy}
\item \hyperref[app:bg:block]{\ref*{app:bg:block}: High-dimensional and localized particle filtering}
\item \hyperref[app:bg:guided]{\ref*{app:bg:guided}: Look-ahead, auxiliary, and guided PFs}
\item \hyperref[app:bg:twisting]{\ref*{app:bg:twisting}: TPFs and controlled SMC}
\item \hyperref[app:bg:positioning]{\ref*{app:bg:positioning}: Relationship of TBPF to the preceding literature}
\end{itemize}
\item \hyperref[supp:nav:a-tractable-twisted-block-resampling-law]{Appendix~\ref*{supp:nav:a-tractable-twisted-block-resampling-law}: A tractable twisted block resampling law}
\item \hyperref[supp:nav:local-sampling-and-global-correction-behavior]{Appendix~\ref*{supp:nav:local-sampling-and-global-correction-behavior}: Local sampling and global correction behavior}
 \begin{itemize}
\DIFaddbegin
\item \hyperref[supp:variance-guides]{\ref*{supp:variance-guides}: Variance comparison and conditional guide quality}
\item \hyperref[supp:pool-example]{\ref*{supp:pool-example}: An exact finite-pool guide-strength example}
\item \hyperref[supp:tv]{\ref*{supp:tv}: A full-history total-variation consequence}
\DIFaddend \end{itemize}
\item \DIFaddbegin \hyperref[supp:sec:theory]{Appendix~\ref*{supp:sec:theory}: Fixed-horizon consistency}
\item \DIFaddend \hyperref[supp:sec:experiments]{Appendix~\ref*{supp:sec:experiments}: Detailed experimental models and comparisons}
\begin{itemize}
\item \hyperref[supp:subsec:models]{\ref*{supp:subsec:models}: Coupled models and design}
\item \hyperref[supp:subsec:guide-design]{\ref*{supp:subsec:guide-design}: Guide design}
\item \hyperref[supp:subsec:tpf]{\ref*{supp:subsec:tpf}: TPF comparator}
\item \hyperref[supp:subsec:validation]{\ref*{supp:subsec:validation}: Simulation results and method comparisons}
\item  \DIFaddbegin \hyperref[supp:nav:which-method-works-best]{\ref*{supp:nav:which-method-works-best}: Retrospective minima over the tested configurations}
\DIFaddend \item \hyperref[supp:nav:graphical-comparison]{\ref*{supp:nav:graphical-comparison}: Graphical comparison}
\item \hyperref[supp:sec:real_data]{\ref*{supp:sec:real_data}: Real data: a 100-dimensional portfolio panel}
\DIFaddbegin \item \hyperref[supp:revision-validation]{\ref*{supp:revision-validation}: Gaussian references, uncertainty, and implementation diagnostics}
\item \hyperref[supp:independent]{\ref*{supp:independent}: Independent-data fixed-configuration check}
\DIFaddend \end{itemize}
\item \hyperref[supp:app:notation]{Appendix~\ref*{supp:app:notation}: Notation reference}
\end{itemize}
\endgroup
\numberwithin{equation}{section}
\numberwithin{theorem}{section}
\setcounter{figure}{0}
\setcounter{table}{0}
\setcounter{algorithm}{0}
\renewcommand{\thefigure}{S\arabic{figure}}
\renewcommand{\thetable}{S\arabic{table}}
\renewcommand{\thealgorithm}{S\arabic{algorithm}}
\renewcommand{\theHalgorithm}{supp.\arabic{algorithm}}

\appendixsection{Methodological context and related literature}\label{app:background}
Spatial blocking and particle-system twisting address different difficulties in SMC. Blocking replaces a global update with local updates and changes the approximation target. Twisting changes the simulation law for a specified target and compensates through a change-of-measure correction. TBPF combines these operations by retaining BPF propagation and guide-tilting one uniformly selected offspring per block. Its correction uses finite empirical guide averages over the proposal and offspring pools.

 \subsection{Particle filtering, likelihood estimation, and degeneracy}
\label{app:bg:pf}

PFs originated as simulation-based recursive filters for nonlinear and non-Gaussian state-space models \citep{gordon1993,doucet2001,arulampalam2002,delmoral2004}. In a bootstrap PF, particles are propagated through the state transition, weighted by the observation density, and resampled. More general SMC formulations permit alternative proposals, resampling schemes, and Feynman--Kac representations \citep{delmoral2004,doucet2009}. Besides approximating filtering expectations, PFs provide normalizing-constant estimates with an important unbiasedness property in standard Feynman--Kac settings. This property underlies pseudo-marginal and particle MCMC methodology, where an unbiased likelihood estimate can be substituted for an unavailable exact likelihood inside a valid Monte Carlo procedure \citep{andrieu2010}.

The basic PF nevertheless scales poorly in high dimension. When an observation likelihood factors across many coordinates, a global importance weight is a product of many local factors; its variability can therefore grow rapidly with dimension. Empirical and theoretical analyses demonstrate severe weight collapse unless particle numbers increase very rapidly \citep{bengtsson2008,snyder2008}. This phenomenon motivates two broad remedies relevant here. One is to \emph{localize} the update so that selection takes place in lower-dimensional subproblems. The other is to \emph{guide} particles toward regions that will receive high future weight. BPF uses localization to define the block approximation. TBPF then adds guidance through a change of measure on BPF resampling: it guide-tilts one offspring per block and applies a correction that preserves the finite-particle BPF normalizer expectation.

\subsection{High-dimensional and localized particle filtering}
\label{app:bg:block}

Localization exploits conditional or weak spatial dependence. \citet{rebeschini2015} gave a foundational analysis of local particle filtering on graphical models. Their BPF partitions the graph into blocks, performs local corrections, and obtains local error bounds whose constants can depend on block size and interaction neighborhoods rather than the total number of vertices. The benefit is accompanied by a blocking bias: replacing a joint law by a product of block marginals disrupts dependence across block boundaries. This bias--variance tradeoff is intrinsic to the approximation and is conceptually separate from Monte Carlo variance conditional on the chosen block recursion.

Several subsequent methods have pursued scalable SMC through localization or decomposition. Temporal block sampling was studied earlier by \citet{doucet2006}. \citet{johansen2015} combined blocking ideas with annealed importance sampling, and \citet{singh2017} used particle Gibbs within blocked time-series constructions. \citet{park2020} proposed the guided intermediate resampling filter (GIRF), which combines intermediate resampling with forecast-likelihood guides and was designed for moderately high-dimensional implicit dynamic models. \citet{ning2023} developed iterated BPF for high-dimensional parameter learning on graphical state-space models and established local error bounds, algorithmic convergence, and likelihood-maximization results. \citet{ionides2024} extended iterated BPF to estimate both shared and unit-specific parameters, demonstrating inference for a coupled measles model across twenty towns. In an applied setting, \citet{gu2026} used iterated BPF from \citet{ning2023}, alongside iterated bagged filtering, to estimate parameters in a spatially coupled, multi-strain SARS-CoV-2 transmission model and assess genomic-surveillance strategies for early variant detection. \citet{li2026hospital} developed  \DIFaddbegin \DIFadd{a blockwise agent-based iterated filter (BAIF) }\DIFaddend for agent-based transmission models on dynamic contact networks. Their hospital application partitions wards using the patient-transfer network to infer heterogeneous transmission and community-introduction parameters, illustrating how BPF variants can support inference beyond spatially regular models. Other developments include the divide-and-conquer SMC construction of \citet{crucinio2024}, which recursively merges low-dimensional filtering components and relaxes some factorization requirements of block filters.

TBPF inherits the spatial partition and local-update philosophy of this literature, but its new operation is not another blocking map. The deterministic block recursion is taken as the baseline approximation. TBPF then changes the \emph{finite-particle simulation law} of the block-resampling step and corrects for that change of measure. Consequently, the method does not remove the boundary bias created by blocking; instead it asks whether a given finite-particle block approximation can be simulated with more favorable normalizer variability.

\subsection{Look-ahead, auxiliary, and guided PFs}
\label{app:bg:guided}

Guidance uses information about future observations to reduce the chance that particles propagated from the current population will later receive negligible weight. The auxiliary PF of \citet{pitt1999} introduced predictive resampling weights for this purpose, and a large subsequent literature has studied look-ahead and auxiliary constructions. \citet{lin2013} reviewed look-ahead strategies, while block-sampling and intermediate-resampling approaches provide related mechanisms for incorporating future information \citep{doucet2006,park2020}.

GIRF is especially relevant to high-dimensional settings. It inserts intermediate propagation and resampling steps between observation times and uses a guide approximating forecast likelihoods of future observations \citep{park2020}. This can improve scaling without requiring an evaluable transition density, an important feature for simulator-defined partially observed Markov process (POMP) models. Conceptually, however, GIRF modifies the sequence of intermediate targets and resampling weights throughout propagation. TBPF instead leaves the propagation kernel and the baseline block proposal pool unchanged, and applies a finite-particle guide tilt only when offspring are selected within each block.

\citet{ionides2023bagged} developed bagged filters that combine replicated filtering trajectories using spatiotemporally localized weights. Their Figure~1 compares these methods with BPF and GIRF on a correlated Brownian-motion model. In that experiment, BPF maintains more stable log-likelihood estimates as the number of spatial units increases than the tested GIRF implementation. The comparison uses a general-purpose GIRF guide rather than the analytically available ideal Gaussian guide, so the finding concerns those implementations and settings. It provides an empirical motivation for retaining a localized BPF baseline while investigating additional guidance.

\subsection{TPFs and controlled SMC}
\label{app:bg:twisting}

Twisting can be understood as a change of measure designed to reduce the variance of SMC normalizing-constant estimates. In rare-event simulation and Feynman--Kac theory, positive eigenfunctions naturally generate Doob-type transforms or twisted kernels. \citet{whiteley2014} brought this principle to particle systems and characterized, within a broad class, an essentially unique transition attaining zero asymptotic-in-time normalized second-moment growth under their assumptions. The optimal transition is expressed through generalized eigenfunctions of the underlying nonnegative kernels. Their practical TPF modifies one distinguished particle in a way that can produce markedly different likelihood-estimation fluctuations even though most particles follow the standard mechanism.

Later work broadened the practical and conceptual scope of twisting. \citet{alaluhtala2016} developed accessible twisted-filter constructions with alternative resampling schemes and demonstrated their use in parameter estimation. \citet{guarniero2017} proposed the iterated auxiliary PF, learning approximations to optimal twisting functions through backward regression. \citet{heng2020} formulated controlled SMC as an approximate stochastic-control problem, iteratively improving proposal policies. \citet{bon2022} considered Monte Carlo twisting when exact sampling from twisted transitions or exact evaluation of twisted potentials is unavailable, using rejection sampling and unbiased random-weight approximations.

Recent work continues to connect SMC twisting with modern control and learning. \citet{luwang2024} developed a continuous-time perspective and a neural path-space twisting approach based on Kullback--Leibler (KL) optimization, while \citet{xue2025} proposed an online rolling controlled-SMC construction that learns twisting functions while filtering in real time. These developments emphasize increasingly flexible learned proposal or control mechanisms. TBPF addresses a different design point: it intentionally avoids changing the state propagation kernel and confines the change of measure to one offspring per block at resampling. This restriction makes the correction and the conditional offspring moments analytically explicit.

\subsection{Relationship of TBPF to the preceding literature}\label{app:bg:positioning}
TBPF combines the partitioned graphical architecture of BPFs \citep{rebeschini2015} with the particle-system change of measure used in TPFs \citep{whiteley2014}. Propagation remains unchanged: one child per block follows the guide-tilted law, and the correction uses an empirical average over all children. This differs from tilting every ancestor or learning a transition proposal, as in auxiliary and controlled SMC methods.

The dimension-local structure comes from the block architecture. TBPF adds an exactly corrected resampling twist whose conditional moments and correction stability can be analyzed directly. 

\DIFaddbegin \begin{table}[ht]\color{black}\centering\small
\caption{\color{black}Algorithmic design distinctions among BPF, TPF, GIRF, and TBPF.}\label{tab:method-design}
\begin{tabular}{lp{.14\linewidth}p{.30\linewidth}p{.35\linewidth}}\toprule
Method & Spatial blocks & Modified stage & Guide operations\\\midrule
BPF & Yes & Block weighting/resampling & None\\
TPF & No & Distinguished ancestor and propagation & Continuous integration and tilted sampling\\
GIRF & No & Guided intermediate resampling & Forecast guide evaluations and model simulation\\
TBPF & Yes & One resampled child per block & Finite empirical sums and categorical draws\\\bottomrule
\end{tabular}
\end{table}

\DIFaddend \appendixsection{A tractable twisted block resampling law}\label{supp:nav:a-tractable-twisted-block-resampling-law}
\paragraph{Proposal pool and offspring array.}
At time $t$, the propagation step draws $Z_t^i\sim f_t(X_{t-1}^i,\cdot)$ for $i=1,\ldots,N$. We write
\[
 Z_t=(Z_t^1,\ldots,Z_t^N)
\]
for this entire proposal pool, before resampling. Each $Z_t^i$ is a full state, and $Z_t^{i,K}$ is its restriction to block $K$.  \DIFaddbegin \DIFadd{Write }\DIFaddend $x=(x^1,\ldots,x^N)$  \DIFaddbegin \DIFadd{for the offspring array after resampling and recombination; the kernel acts on }\DIFaddend this whole array. Conditional on the realized pool, the ordinary block resampling law is
\[
 R_t(dx\mid\mathcal G_t)=\bigotimes_{K\in\cK}\prod_{j=1}^N\nu_t^K(dx^{j,K})\qquad\text{\normalfont and}\qquad \nu_t^K=\sum_iw_t^{i,K}\delta_{Z_t^{i,K}}.
\]
Thus each block child independently selects one proposal block with its normalized observation weight. Both kernels condition on $\mathcal G_t$, which contains the past, current proposal pool, and frozen guide-construction randomness; the observations are fixed throughout.

\paragraph{Guide averages and their meaning.}
Choose positive finite guides $\psi_t^K$ that are fixed before any current offspring are generated. A guide assigns larger values to block states judged useful for future observations. Define
\begin{equation}
 c_t^K=\nu_t^K(\psi_t^K)
       =\sum_iw_t^{i,K}\psi_t^K(Z_t^{i,K})\qquad\text{\normalfont and}\qquad H_t^K(x)=\frac{1}N\sum_{j=1}^N\psi_t^K(x^{j,K}).
\end{equation}
 \DIFaddbegin \DIFadd{Thus }\DIFaddend $c_t^K$ is  \DIFaddbegin \DIFadd{fixed before resampling, while }\DIFaddend $H_t^K(x)$  \DIFaddbegin \DIFadd{depends on the selected offspring}\DIFaddend. The twisted kernel is
\begin{equation}\label{supp:eq:twist}
 \widetilde R_t(dx\mid\mathcal G_t)
 =R_t(dx\mid\mathcal G_t)\prod_{K\in\cK}\frac{H_t^K(x)}{c_t^K}.
\end{equation}
For one block, arrays with $H_t^K(x)>c_t^K$ receive more probability than under ordinary resampling, and arrays with $H_t^K(x)<c_t^K$ receive less. The normalizing constant is exact:
\[
 \int H_t^K(x)R_t(dx\mid\mathcal G_t)
 =\frac{1}N\sum_j\int\psi_t^K(u)\nu_t^K(du)=c_t^K.
\]
Conditional independence across blocks therefore makes the product of ratios in \eqref{supp:eq:twist} integrate to one. 

\paragraph{Derivation from one distinguished child.}
Fix a time $t$ and block $K$. The tilted law for one child is
\[
 \nu_{t,\psi}^K(du)=\frac{\psi_t^K(u)}{c_t^K}\nu_t^K(du).
\]
Select a child label $J_t^K$ uniformly from $\{1,\ldots,N\}$. Conditional on $J_t^K=j$, draw that child from $\nu_{t,\psi}^K$ and all other children independently from $\nu_t^K$. Relative to the ordinary array law $(\nu_t^K)^{\otimes N}$, this has density $\psi_t^K(x^{j,K})/c_t^K$. Averaging over the uniform label gives
\[
 \frac{1}N\sum_{j=1}^N
 \left[\nu_{t,\psi}^K(dx^{j,K})\prod_{\ell\ne j}\nu_t^K(dx^{\ell,K})\right]
 =\left[\prod_{\ell=1}^N\nu_t^K(dx^{\ell,K})\right]
   \frac{N^{-1}\sum_j\psi_t^K(x^{j,K})}{c_t^K}.
\]
The last factor is precisely $H_t^K(x)/c_t^K$. Applying this construction independently in every block yields \eqref{supp:eq:twist}.  \DIFaddbegin \DIFadd{When }\DIFaddend $N=1$\DIFaddbegin \DIFadd{, the empirical proposal law is a point mass: $\nu_{t,\psi}^K=\nu_t^K$ and $H_t^K=c_t^K$. The twist is vacuous and TBPF coincides with BPF. Constant guides recover BPF for every $N$}\DIFaddend. Only one child per block is distinguished: tilting all children would instead produce a product over child-specific guide values and would define a different algorithm.

The ratio in \eqref{supp:eq:twist} changes the \emph{sampling law}. To preserve the baseline expectation, the normalizer estimator uses its reciprocal $c_t^K/H_t^K(X_t)$ after sampling, as proved in Theorem~\ref{supp:thm:rn}. Both guide averages are finite sums over proposal or offspring values; no continuous guide integral or transition-density evaluation is required.

\paragraph{Algorithm (resampling-stage TBPF).}
Algorithm~\ref{supp:alg:tbpf} gives the complete procedure. Every guide is constructed from the past, the full current proposal pool, and the prescribed observation look-ahead, and is frozen before any current offspring are drawn. The stored guide values are used in both the distinguished ancestor probabilities and the correction. All offspring draws are conditionally independent across blocks, given the pool and frozen guides.

\begin{algorithm}[!htb]
\caption{Resampling-stage TBPF}\label{supp:alg:tbpf}
\renewcommand{\baselinestretch}{1.18}\small
\begin{algorithmic}[1]
\setlength{\itemsep}{3pt}
\makeatletter
\let\tbpfOriginalAlgorithmGroup\ALC@g
\def\ALC@g{\tbpfOriginalAlgorithmGroup\setlength{\itemsep}{3pt}}
\makeatother
\REQUIRE Observations $y_{1:T}$, initial law $\pi_0$, transitions $(f_t)_{t=1}^T$, block likelihoods $g_t^K$, partition $\cK$, particle count $N$, and positive guide rules.
\ENSURE Corrected normalizer $\widehat Z_T^{TB}$, weighted block outputs $\widehat\pi_t^{TB}$, and recombined empirical outputs $\eta_t^N$.
\STATE Draw $X_0^1,\ldots,X_0^N$ independently with common law $\pi_0$; set $L_0\gets0$.
\FOR{$t=1,\ldots,T$}
  \STATE Draw $Z_t^i\sim f_t(X_{t-1}^i,\cdot)$ independently for $i=1,\ldots,N$.
  \FOR{each block $K\in\cK$}
    \STATE $G_i^K\gets g_t^K(Z_t^{i,K})$ for all $i$;
           $a_{t,N}^{K}\gets N^{-1}\sum_iG_i^K$;
           $w_t^{i,K}\gets G_i^K/\sum_jG_j^K$.
    \STATE $\nu_t^K\gets\sum_iw_t^{i,K}\delta_{Z_t^{i,K}}$.
    \STATE Construct and freeze $s_i^K\gets\psi_t^K(Z_t^{i,K})>0$ for all $i$.\label{supp:line:guide-values}
    \STATE If $t=T$, set $s_i^K\gets1$ for all $i$.
    \STATE $c_t^K\gets\sum_iw_t^{i,K}s_i^K$;
           $\widetilde w_t^{i,K}\gets w_t^{i,K}s_i^K/c_t^K$ for all $i$.\label{supp:line:tilted-weights}
  \ENDFOR
  \STATE $\widehat\pi_t^{TB}\gets\bigotimes_{K\in\cK}\nu_t^K$.
  \FOR{each block $K\in\cK$, using independent offspring randomness}\label{supp:line:block-resampling}
    \STATE Draw $J_t^K\sim\operatorname{Uniform}\{1,\ldots,N\}$.\label{supp:line:distinguished-label}
    \STATE Draw $A_t^{j,K}\sim\operatorname{Cat}(w_t^{1:N,K})$
           independently for $j\ne J_t^K$.\label{supp:line:ordinary-ancestors}
    \STATE Independently draw $A_t^{J_t^K,K}\sim\operatorname{Cat}(\widetilde w_t^{1:N,K})$.\label{supp:line:tilted-ancestor}
    \STATE Set $X_t^{j,K}\gets Z_t^{A_t^{j,K},K}$ for all $j$.\label{supp:line:copy-offspring}
    \STATE $H_t^K\gets N^{-1}\sum_{j=1}^N s_{A_t^{j,K}}^K$.\label{supp:line:empirical-guide}
  \ENDFOR
  \STATE Concatenate $X_t^j\gets(X_t^{j,K})_{K\in\cK}$ for $j=1,\ldots,N$.\label{supp:line:recombine}
  \STATE $\eta_t^N\gets N^{-1}\sum_j\delta_{X_t^j}$.
  \STATE $L_t\gets L_{t-1}+\sum_{K\in\cK}(\log a_{t,N}^{K}+\log c_t^K-\log H_t^K)$.
\ENDFOR
\STATE \textbf{return} $\widehat Z_T^{TB}=\exp(L_T)$ and $\{\widehat\pi_t^{TB},\eta_t^N\}_{t=1}^T$.
\end{algorithmic}
\end{algorithm}

The log-normalizer update is
\begin{equation}\label{supp:eq:logz}
 L_t=L_{t-1}+\sum_K\left[\log a_{t,N}^{K}+\log c_t^K-\log H_t^K(X_t)\right],
 \qquad L_0=0.
\end{equation}
\paragraph{Definition of the corrected block normalizer.}
The output of TBPF is the random estimator
\begin{equation}\label{supp:eq:tb-normalizer}
 \widehat Z_T^{TB}:=\exp(L_T)
 =\prod_{t=1}^T\prod_{K\in\cK}
       \left[a_{t,N}^{K}\frac{c_t^K}{H_t^K(X_t)}\right],
 \qquad \widehat Z_0^{TB}=1.
\end{equation}
The hat indicates a Monte Carlo estimator, and the superscript $TB$ identifies the twisted block algorithm. Each factor $a_{t,N}^{K}$ is the ordinary block observation-weight average; $c_t^K/H_t^K(X_t)$ corrects for the change in the resampling law. All factors are evaluated on the same history generated by TBPF. Thus \eqref{supp:eq:tb-normalizer} is the uncorrected product $\prod_{t,K}a_{t,N}^{K}$ along that history multiplied by its path correction $D_T$, defined in Theorem~\ref{supp:thm:rn}. 
Set $\psi_T^K=1$, since a twist of terminal resampling cannot improve an estimator that no longer depends on the resampled particles. Then $c_T^K=H_T^K=1$. Constant guides at every time recover BPF. The implementation evaluates weights and normalizer increments in log space. The weighted output is $\widehat\pi_t^{TB}=\bigotimes_K\nu_t^K$; for a block test $u$, its estimate is $\nu_t^K(u)=\sum_iw_t^{i,K}u(Z_t^{i,K})$. This pre-resampling output supplies the filtering means used in the experiments and is distinct from the recombined empirical output $\eta_t^N$. 

\paragraph{Empirical measure after resampling.}
For the recombined TBPF particles $X_t^1,\ldots,X_t^N$, define the random empirical probability measure
\[
 \eta_t^N:=\frac{1}N\sum_{i=1}^N\delta_{X_t^i}\qquad\text{\normalfont and}\qquad \eta_t^N(u):=\int u(x)\eta_t^N(dx)
                 =\frac{1}N\sum_{i=1}^N u(X_t^i),
\]
where $\delta_x$ is the unit point mass at $x$ and $u$ is a test function. Thus $\eta_t^N(u)$ is the equally weighted average of $u$ over the resampled full-state particles. It differs from the pre-resampling weighted block output $\widehat\pi_t^{TB}$. For particles generated by BPF, we analogously write $\eta_t^{B,N}$. For a block test $u:\cX^K\to\mathbb R$, we evaluate this same full-state empirical measure on the function $x\mapsto u(x^K)$, so $\eta_t^N(u)=N^{-1}\sum_i u(X_t^{i,K})$.

\begin{proposition}[Distinguished-offspring representation]\label{supp:prop:one}
 \DIFaddbegin \DIFadd{The distinguished-child sampling and recombination steps }\DIFaddend of Algorithm~\ref{supp:alg:tbpf} generate  \DIFaddbegin \DIFadd{exactly the twisted array law \eqref{supp:eq:twist}. The }\DIFaddend empirical guide average  \DIFaddbegin \DIFadd{supplies its }\DIFaddend normalizer correction.
\end{proposition}
\begin{proof}
Fix a time $t$ and block $K$, and condition on $\mathcal G_t$. Write $x^{1:N,K}=(x^{1,K},\ldots,x^{N,K})$ for the offspring array restricted to this block. We use $\widetilde R_t^K$ for its \emph{one-block} resampling kernel and reserve $\widetilde R_t$ (without a superscript $K$) for the \emph{full} resampling kernel over all blocks. Both kernels retain the conditioning on $\mathcal G_t$, as in \eqref{supp:eq:twist}.

Here $i$ indexes a proposal ancestor, whereas $j$ indexes an offspring position.  \DIFaddbegin \DIFadd{The guide-value step }\DIFaddend sets $s_i^K=\psi_t^K(Z_t^{i,K})$, and  \DIFaddbegin \DIFadd{the tilted-weight step }\DIFaddend computes
\[
 c_t^K=\sum_iw_t^{i,K}s_i^K=\nu_t^K(\psi_t^K)\qquad\text{\normalfont and}\qquad \widetilde w_t^{i,K}=\frac{w_t^{i,K}s_i^K}{c_t^K}.
\]
For a fixed distinguished position $J_t^K=j$,  \DIFaddbegin \DIFadd{the ordinary-ancestor step }\DIFaddend draws the other ancestors from $w_t^{1:N,K}$, while  \DIFaddbegin \DIFadd{the tilted-ancestor step }\DIFaddend draws ancestor $A_t^{j,K}$ from $\widetilde w_t^{1:N,K}$. For an ancestor-label vector $(a^1,\ldots,a^N)$, the conditional probability is therefore
\[
 \widetilde w_t^{a^j,K}\prod_{\ell\ne j}w_t^{a^\ell,K}
 =\left(\prod_{\ell=1}^N w_t^{a^\ell,K}\right)\frac{s_{a^j}^K}{c_t^K}.
\]
 \DIFaddbegin \DIFadd{The offspring-copying step }\DIFaddend sets $x^{\ell,K}=Z_t^{a^\ell,K}$, so $s_{a^j}^K=\psi_t^K(x^{j,K})$. Denote the conditional offspring law for this block given $J_t^K=j$ by $Q_{t,j}^K$. Then
\[
 Q_{t,j}^K(dx^{1:N,K}\mid\mathcal G_t)
 =\left[\prod_{\ell=1}^N\nu_t^K(dx^{\ell,K})\right]
   \frac{\psi_t^K(x^{j,K})}{c_t^K}.
\]
This statement also holds if different proposal labels have identical block states, because their guide values agree and their probabilities add.

 \DIFaddbegin \DIFadd{The uniform-label step }\DIFaddend assigns probability $1/N$ to each distinguished position in this block. Marginalizing $J_t^K$ by the law of total probability gives the one-block kernel
\begin{align*}
 \widetilde R_t^K(dx^{1:N,K}\mid\mathcal G_t)
 &=\frac{1}N\sum_{j=1}^N Q_{t,j}^K(dx^{1:N,K}\mid\mathcal G_t)=(\nu_t^K)^{\otimes N}(dx^{1:N,K})
   \underbrace{\frac{1}N\sum_{j=1}^N
     \frac{\psi_t^K(x^{j,K})}{\nu_t^K(\psi_t^K)}}_{H_t^K(x)/c_t^K}.
\end{align*}
The superscript $K$ is essential here: only the offspring in block $K$ have been generated. The sum over $j$ averages over the possible distinguished positions in  \DIFaddbegin \DIFadd{the uniform-label step}\DIFaddend; the algorithm draws only one such position per block. For the realized array,  \DIFaddbegin \DIFadd{the empirical guide average step }\DIFaddend computes
\[
 H_t^K(X_t)=N^{-1}\sum_j s_{A_t^{j,K}}^K
           =N^{-1}\sum_j\psi_t^K(X_t^{j,K}).
\]
That computation evaluates the correction after sampling; it is not a further sampling step.

Finally,  \DIFaddbegin \DIFadd{the independent block loop }\DIFaddend uses conditionally independent offspring randomness across blocks. After concatenation in  \DIFaddbegin \DIFadd{the recombination step}\DIFaddend, the full kernel on $x=(x^1,\ldots,x^N)$ is the product of the one-block kernels:
\begin{align*}
 \widetilde R_t(dx\mid\mathcal G_t)
 =\bigotimes_{K\in\cK}\widetilde R_t^K(dx^{1:N,K}\mid\mathcal G_t)
 &=\left[\bigotimes_{K\in\cK}(\nu_t^K)^{\otimes N}(dx^{1:N,K})\right]
       \prod_{K\in\cK}\frac{H_t^K(x)}{c_t^K}=R_t(dx\mid\mathcal G_t)\prod_{K\in\cK}\frac{H_t^K(x)}{c_t^K},
\end{align*}
which is precisely \eqref{supp:eq:twist}. Thus the one-block identity uses $\widetilde R_t^K$, whereas \eqref{supp:eq:twist} uses the full kernel $\widetilde R_t$.
\end{proof}

\paragraph{Auxiliary records and the reference law.}
Random guide construction requires an augmented history. Write $\Xi_t$ for the pre-offspring guide record, including the realized guide values and any retained guide-construction state. If future guides depend on resampling marks such as distinguished labels or ancestor labels, retain those marks in $A_t$. For the realized augmented history $h_{t-1}^+$ through time $t-1$, denote the guide-generation and conditional resampling-mark kernels, respectively, by
\begin{equation}\label{supp:eq:auxiliary-kernels}
 Q_t(d\xi_t\mid h_{t-1}^+,z_t),\qquad
 \widetilde K_t(da_t\mid h_{t-1}^+,z_t,\xi_t,x_t).
\end{equation}
Here $Q_t$ is the measurable guide-generation kernel used under both laws, and $\widetilde K_t$ is the conditional law of the retained marks under TBPF given the completed offspring array. The baseline extension first draws BPF offspring and then attaches marks using this same conditional kernel. Its particle-array marginal is still BPF because ordinary resampling does not depend on the guides or marks. If no later guide uses these marks, $A_t$ and $\widetilde K_t$ may be omitted. These auxiliary records are assumed to take values in standard Borel spaces, so the required conditional kernels exist; finite categorical marks satisfy this condition directly.

The following theorem restates Theorem~\ref{thm:rn} from the main text in the detailed notation used for its proof.

\begin{theorem}[Exact change of measure; restated]\label{supp:thm:rn}
Write $X_t=X_t^{1:N}$ and $Z_t=Z_t^{1:N}$ for the full particle and proposal arrays. With the preceding guide and mark kernels, define the augmented history
\[
 \mathcal H_T^+=(X_0,Z_1,\Xi_1,X_1,A_1,\ldots,Z_T,\Xi_T,X_T,A_T).
\]
Let $\mathbb P_B$ and $\mathbb P_{TB}$ denote the baseline extension and twisted laws of $\mathcal H_T^+$. Both use the same initialization, time-indexed propagation kernels $f_t$, and pre-offspring guide-generation kernels, and the common conditional mark kernel $\widetilde K_t$. Assume all guide values on each realized pool are finite and strictly positive. Then
\begin{equation}\label{supp:eq:rn}
 D_T:=\prod_{t,K}\frac{c_t^K}{H_t^K(X_t)}
 =\frac{d\mathbb P_B}{d\mathbb P_{TB}}\qquad\text{\normalfont and}\qquad \E_{TB}[D_TF]=\E_B[F]
\end{equation}
for every nonnegative measurable functional $F=\Phi(\mathcal H_T^+)$, and for every real-valued measurable functional satisfying $\E_B|F|<\infty$. The particle-array marginal of the baseline extension is the BPF law. In particular, the corrected estimator in \eqref{supp:eq:tb-normalizer} satisfies
 \DIFaddbegin \begin{equation}
 \DIFadd{\E_{TB}\widehat Z_T^{TB}=\E_B\widehat Z_T^B.
}\end{equation}\DIFaddend 
\end{theorem}
\begin{proof}
We factor the augmented laws, cancel their common auxiliary kernels, and then apply the resulting density identity to the block-normalizer functional. Let $h_T^+$ be an augmented history, with prefix $h_{t-1}^+$. Both algorithms use $\mu_0=\pi_0^{\otimes N}$ and
\[
 M_t(x_{t-1},dz_t)=\bigotimes_{i=1}^N f_t(x_{t-1}^i,dz_t^i).
\]
Recall the guide-generation kernel $Q_t$ and conditional resampling-mark kernel $\widetilde K_t$ in \eqref{supp:eq:auxiliary-kernels}. The augmented path laws factor as
\begin{align*}
 \mathbb P_B(dh_T^+)
 &=\mu_0(dx_0)\prod_{t=1}^T\Big[
 M_t(x_{t-1},dz_t)Q_t(d\xi_t\mid h_{t-1}^+,z_t)\\
 &\qquad\times R_t(dx_t\mid h_{t-1}^+,z_t,\xi_t)
 \widetilde K_t(da_t\mid h_{t-1}^+,z_t,\xi_t,x_t)\Big],\\
 \mathbb P_{TB}(dh_T^+)
 &=\mu_0(dx_0)\prod_{t=1}^T\Big[
 M_t(x_{t-1},dz_t)Q_t(d\xi_t\mid h_{t-1}^+,z_t)\\
 &\qquad\times\widetilde R_t(dx_t\mid h_{t-1}^+,z_t,\xi_t)
 \widetilde K_t(da_t\mid h_{t-1}^+,z_t,\xi_t,x_t)\Big].
\end{align*}
Here each product denotes sequential kernel composition: propagation first, then the frozen guide record, then the offspring array, then any retained marks.
The marked representation of the second law is a disintegration of the specified TBPF sampling procedure; it does not change that algorithm. The first law has the BPF particle marginal: at every step its propagation and offspring kernels are exactly the ordinary kernels, while integrating $Q_t$ and $\widetilde K_t$ gives one. This remains true when later guide records depend on the retained marks.

Conditional on the augmented prefix, current pool, and frozen guide record, \eqref{supp:eq:twist} gives
\[
 \frac{d\widetilde R_t(\cdot\mid h_{t-1}^+,z_t,\xi_t)}{dR_t(\cdot\mid h_{t-1}^+,z_t,\xi_t)}(x_t)
 =\prod_{K\in\cK}\frac{H_t^K(x_t)}{c_t^K}.
\]
Every ratio is finite and strictly positive on the common resampling support. The identical $M_t,Q_t,\widetilde K_t$ factors cancel, so multiplication of the conditional densities gives
\[
 \frac{d\mathbb P_{TB}}{d\mathbb P_B}(h_T^+)
 =\prod_{t=1}^T\prod_{K\in\cK}\frac{H_t^K(x_t)}{c_t^K}.
\]
The two augmented laws are therefore mutually absolutely continuous, with reciprocal density $D_T(h_T^+)$. For nonnegative measurable $\Phi$,
\begin{align*}
 \E_{TB}[D_T\Phi(\mathcal H_T^+)]
 =\int\Phi(h_T^+)\frac{d\mathbb P_B}{d\mathbb P_{TB}}(h_T^+)\,\mathbb P_{TB}(dh_T^+)=\int\Phi(h_T^+)\,\mathbb P_B(dh_T^+)
 =\E_B[\Phi(\mathcal H_T^+)].
\end{align*}
Applying this identity to $|\Phi|$ proves integrability under the corrected twisted law whenever $\E_B|\Phi|<\infty$; subtracting positive and negative parts gives the signed case.

Finally, choose the functional that ignores auxiliary records and multiplies the proposal-pool likelihood averages:
\[
 \mathcal Z_T(h_T^+)=\prod_{t=1}^T\prod_{K\in\cK}a_{t,N}^{K}(z_t)\qquad\text{\normalfont and}\qquad a_{t,N}^{K}(z_t)=\frac{1}N\sum_{i=1}^N g_t^K(z_t^{i,K}).
\]
Under the baseline extension its distribution is that of $\widehat Z_T^B$, because the particle marginal is BPF. Under TBPF, \eqref{supp:eq:tb-normalizer} gives $D_T\mathcal Z_T(\mathcal H_T^+)=\widehat Z_T^{TB}$. Applying the expectation identity proves the assertion.
\end{proof}

\appendixsection{Local sampling and global correction behavior}\label{supp:nav:local-sampling-and-global-correction-behavior}
\begin{lemma}[Deterministic one-stage replacement bound]\label{supp:lem:replacement}
Condition on $\mathcal G_t$, before generating any current offspring. First draw an ordinary reference array with independent block columns of law $(\nu_t^K)^{\otimes N}$. Independently choose each $J_t^K$ uniformly, draw $V_t^K\sim\nu_{t,\psi}^K$ independently of all reference children, and replace entry $J_t^K$ in block $K$ by $V_t^K$. The reference array has law $R_t(\cdot\mid\mathcal G_t)$, and the replaced array has law $\widetilde R_t(\cdot\mid\mathcal G_t)$. Denote their recombined full-state rows by $X^{i,\mathrm{ref}}$ and $X^{i,\mathrm{rep}}$, respectively, and define
\[
 \eta_N^{\mathrm{ref}}
 :=\frac{1}N\sum_{i=1}^N\delta_{X^{i,\mathrm{ref}}}\qquad\text{\normalfont and}\qquad \eta_N^{\mathrm{rep}}
 :=\frac{1}N\sum_{i=1}^N\delta_{X^{i,\mathrm{rep}}}.
\]
 In particular, $\eta_N^{\mathrm{ref}}(u)=N^{-1}\sum_i u(X^{i,\mathrm{ref}})$, and likewise for the replaced array. If $u$ depends on a set $J$ meeting $q(J)$ blocks, then their empirical averages obey, pathwise,
\[
 |\eta_N^{\mathrm{rep}}(u)-\eta_N^{\mathrm{ref}}(u)|
 \le\frac{2q(J)\|u\|_\infty}{N}.
\]
For a single block the constant is $2\|u\|_\infty/N$. This compares one stage with the same pool, not the two recursive algorithms with different histories.
\end{lemma}
\begin{proof}
For a fixed label, all non-replaced entries remain independent $\nu_t^K$ draws and the replacement has law $\nu_{t,\psi}^K$. Averaging over the fresh uniform label gives Proposition~\ref{supp:prop:one}; block independence gives the product law. Only the distinguished row labels of blocks meeting $J$ can change the test value. There are at most $q(J)$ such labels, each changing the sum by at most $2\|u\|_\infty$.
\end{proof}

\paragraph{Integration and variance notation.}
For a probability measure $\rho$ on $\cX^K$ and a bounded measurable function $u:\cX^K\to\mathbb R$, write
\[
 \rho(u):=\int_{\cX^K}u(x)\rho(dx)\qquad\text{\normalfont and}\qquad \operatorname{Var}_{\rho}(u)
 :=\int_{\cX^K}[u(x)-\rho(u)]^2\rho(dx)
 =\rho(u^2)-[\rho(u)]^2,
\]
where  $\operatorname{Var}_{\rho}(u)$ is the variance of $u(Y)$ when $Y$ has distribution $\rho$. In the conditional results below, the pool and frozen guide information are held fixed. If $\rho$ and $u$ depend on that information, the displayed quantity is the corresponding conditional variance and may be random before conditioning.

The following theorem restates Theorem~\ref{thm:moments} from the main text in the detailed notation used for its proof.

\begin{theorem}[Exact conditional local moments; restated]\label{supp:thm:moments}
Fix a time $t$ and block $K$ with state space $\cX^K$. Recall the tilted block law
$\nu_{t,\psi}^K(dx^K)=\psi_t^K(x^K)\nu_t^K(dx^K)/\nu_t^K(\psi_t^K)$. Use the empirical measure $\eta_t^N$ defined above, evaluated on block tests by $\eta_t^N(u)=N^{-1}\sum_i u(X_t^{i,K})$. \DIFaddbegin \DIFadd{Let $u(x^K)$ be a bounded random block function whose evaluation is jointly measurable in $\mathcal G_t$ and the block state. The realized function is fixed before the current offspring are drawn and is held fixed when conditioning on $\mathcal G_t$.} \DIFaddend Conditional on $\mathcal G_t$,
 \DIFaddbegin \begin{gather}
 \DIFadd{\E_{TB}[\eta_t^N(u)\mid\mathcal G_t]=\nu_t^K(u)+\frac{\nu_{t,\psi}^K(u)-\nu_t^K(u)}N,\label{supp:eq:mean}}\\
 \DIFadd{\Var_{TB}(\eta_t^N(u)\mid\mathcal G_t)=\frac{(N-1)\Var_{\nu_t^K}(u)+\Var_{\nu_{t,\psi}^K}(u)}{N^2}.\label{supp:eq:var}
}\end{gather}\DIFaddend 
If $|u|\le1$, then
 \DIFaddbegin \begin{equation}\DIFadd{\label{supp:eq:mse}
 \E_{TB}[|\eta_t^N(u)-\nu_t^K(u)|^2\mid\mathcal G_t]\le\frac{1}N+\frac{4}{N^2}.
}\end{equation}\DIFaddend 
For a sequence of proposal pools at this fixed $t,K$, write $\nu_{t,N}^K$ for $\nu_t^K$ with its particle-count dependence made explicit, and $\nu_{t,N,\psi}^K$ for the corresponding tilted law. More precisely, let $(u_N)_{N\ge1}$ be a sequence of measurable real-valued random block functions $u_N:\cX^K\to\mathbb R$. Each function has a  \DIFaddbegin \DIFadd{jointly measurable evaluation in the pool/guide information and the state; its realization is fixed before drawing }\DIFaddend current offspring. Assume the functions are uniformly bounded: $\sup_N\|u_N\|_\infty\le M<\infty$ almost surely for a deterministic $M$, where $\|u_N\|_\infty=\sup_{x\in\cX^K}|u_N(x)|$. In the notation just defined, set
\[
 v_N:=\operatorname{Var}_{\nu_{t,N}^K}(u_N)
     =\nu_{t,N}^K(u_N^2)-[\nu_{t,N}^K(u_N)]^2.
\]
If $v_N\to\sigma^2$ in probability for a deterministic $\sigma^2\ge0$, then the conditional characteristic functions of $\sqrt N\{\eta_t^N(u_N)-\nu_{t,N}^K(u_N)\}$ converge in probability to those of  \DIFaddbegin \DIFadd{$\mathcal N(0,\sigma^2)$. This one-stage limit is conditional on a common frozen pool; it is neither a corrected-normalizer limit theorem nor a recursive BPF--TBPF comparison}\DIFaddend.
\end{theorem}
\begin{proof}
 \DIFaddbegin \DIFadd{Condition on $\mathcal G_t$, so that the test $u$ is fixed, and on the distinguished label $J_t^K=j$ from Algorithm~\ref{alg:tbpf}. The $N-1$ ordinary children have law $\nu_t^K$, the remaining child has law }\DIFaddend $\nu_{t,\psi}^K$\DIFaddbegin \DIFadd{, and these draws are independent. Their average has conditional mean $\nu_t^K(u)+[\nu_{t,\psi}^K(u)-\nu_t^K(u)]/N$ and conditional variance given in \eqref{supp:eq:var}. Both expressions are }\DIFaddend independent of $j$.  \DIFaddbegin \DIFadd{Total expectation and total variance therefore give \eqref{supp:eq:mean}--\eqref{supp:eq:var}; the between-label variance is zero}\DIFaddend.  \DIFaddbegin \DIFadd{For }\DIFaddend $|u|\le1$,  \DIFaddbegin \DIFadd{the variance is at most $1/N$ and squared bias at most $4/N^2$, }\DIFaddend proving \eqref{supp:eq:mse}. \DIFaddbegin \DIFadd{Marginalizing the label does not preserve independence of the children.
}\DIFaddend 

\DIFaddbegin \DIFadd{For the one-stage limit, condition on $\mathcal G_{t,N}$, the sigma-field $\mathcal G_t$ for the system with $N$ particles. The realized function $u_N$ is then fixed. Draw $Y_{t,N}^{1,K},\ldots,Y_{t,N}^{N,K}$ independently from $\nu_{t,N}^K$ and define the reference empirical block measure and its test average by
}\[
\DIFadd{\bar\eta_t^{K,N}:=\frac1N\sum_{i=1}^N\delta_{Y_{t,N}^{i,K}},\qquad
\bar\eta_t^{K,N}(u_N)=\frac1N\sum_{i=1}^N u_N(Y_{t,N}^{i,K}).}
\]
\DIFadd{Replace a uniformly chosen entry by an independent draw from $\nu_{t,N,\psi}^K$. This couples the reference measure to the TBPF block offspring and gives
}\[
\DIFadd{\sqrt N\,|\eta_t^N(u_N)-\bar\eta_t^{K,N}(u_N)|\le2M/\sqrt N.}
\]\DIFaddend
 \DIFaddbegin \DIFadd{For $\xi_{t,N}^{i,K}=u_N(Y_{t,N}^{i,K})-\nu_{t,N}^K(u_N)$, conditional centering, boundedness $|\xi_{t,N}^{i,K}|\le2M$, and variance $v_N$ give, for each fixed real $z$,
}\[
\DIFadd{\E[e^{\mathrm i z\xi_{t,N}^{1,K}/\sqrt N}\mid\mathcal G_{t,N}]
=1-z^2v_N/(2N)+O(N^{-3/2}).
}\]\DIFaddend 
 The remainder is  \DIFaddbegin \DIFadd{uniformly bounded over pools. Conditional independence raises this expression to the $N$th power, converging in probability }\DIFaddend to $e^{-z^2\sigma^2/2}$.  \DIFaddbegin \DIFadd{The replacement bound and $|e^{ia}-e^{ib}|\le|a-b|$ transfer the characteristic-function }\DIFaddend limit to the twisted  \DIFaddbegin \DIFadd{average. This proves the conditional one-stage claim.
}\DIFaddend \end{proof}

\paragraph{Notation for the resampling correction.}
Fix a time $t$ and let $\mathcal G_t$ contain the past, the current proposal pool $Z_t^{1:N}$, and the frozen guides, before drawing current offspring. For each block $K\in\cK$, recall
\begin{gather*}
 \nu_t^K=\sum_{i=1}^Nw_t^{i,K}\delta_{Z_t^{i,K}},\qquad
 c_t^K=\nu_t^K(\psi_t^K)=\sum_{i=1}^Nw_t^{i,K}\psi_t^K(Z_t^{i,K}),\\
 H_t^K(X_t)=\frac{1}N\sum_{j=1}^N\psi_t^K(X_t^{j,K}),
 \qquad\text{\normalfont and}\qquad C_t^K=\frac{c_t^K}{H_t^K(X_t)}.
\end{gather*}
Here $Z_t^{i,K}$ is proposal $i$ restricted to block $K$, whereas $X_t^{j,K}$ is resampled offspring $j$ in that block. The full offspring array is $X_t=(X_t^1,\ldots,X_t^N)$. The weighted proposal law $\nu_t^K$ and guide average $c_t^K$ are fixed given $\mathcal G_t$; the empirical average $H_t^K(X_t)$ and correction $C_t^K$ depend on the offspring drawn at time $t$.

\DIFadd{Let $\alpha_t^K$ and $\beta_t^K$ be the minimum and maximum of $\psi_t^K$ on the finite proposal support. These are positive finite $\mathcal G_t$-measurable quantities, and for every $z^K\in\operatorname{supp}(\nu_t^K)$,}
\[
 0<\alpha_t^K\le\psi_t^K(z^K)\le\beta_t^K<\infty
.
\]
Define the range ratio $r_t^K=\beta_t^K/\alpha_t^K$\DIFadd{, as in Proposition~\ref{prop:correction}}. The symbols $\alpha_t^K,\beta_t^K$ denote \DIFadd{the guide extrema}, distinct from the observation-weight average $a_{t,N}^{K}$. The scalar $r_t^K$ is a guide-range ratio, distinct from the resampling kernel $R_t$.

The one-stage conditional bound below allows $r_t^K$ to depend on the realized
history and proposal pool. For the deterministic cumulative bound
\eqref{supp:eq:global}, we impose the additional assumption that there are finite
constants  \DIFaddbegin \DIFadd{$\rho_t^K$ with $r_t^K\le \rho_t^K$ }\DIFaddend almost surely under the algorithm's
history and pool law. Thus the same constant must cover all possible histories
and pools, apart from a set of probability zero; it may vary with $t$ and $K$
but cannot be chosen retrospectively from a particular run. Observations are
held fixed in this statement. Finiteness of the ratio on each finite pool does
not itself imply a uniform bound over random pools. For example, a positive
Gaussian-shaped guide can take arbitrarily small values at sufficiently distant
proposals. Clipping guide values to a fixed interval $[c_{\min},c_{\max}]$, with $0<c_{\min}\le c_{\max}<\infty$,
would ensure $r_t^K\le c_{\max}/c_{\min}$ for every pool. This extra assumption is sufficient
for the stated cumulative bound, rather than necessary for the algorithm or
the conditional moment identities to be valid.

For quantities involving this block, $\E_B[\cdot\mid\mathcal G_t]$ means integration against the ordinary law $(\nu_t^K)^{\otimes N}$, and $\E_{TB}[\cdot\mid\mathcal G_t]$ means integration against the twisted block law $\widetilde R_t^K$, at the same realized pool.
Define the time-$t$ correction  \DIFaddbegin \DIFadd{$\mathcal C_t=\prod_{K\in\cK}C_t^K$}\DIFaddend.
The cumulative correction from \eqref{supp:eq:rn} can then be recalled as
 \DIFaddbegin \[
 \DIFadd{D_t=\prod_{s=1}^t \mathcal C_s\qquad\text{\normalfont and}\qquad D_0=1.
}\]\DIFaddend 

The following proposition restates Proposition~\ref{prop:correction} from the main text in the detailed notation used for its proof.

\begin{proposition}[Correction second moment; restated]\label{supp:prop:second}
Under the guide bounds and notation above, for every time $t$ and block $K$,
 \DIFaddbegin \[
 \DIFadd{\E_{TB}[C_t^K\mid\mathcal G_t]=1,\qquad \E_{TB}[(C_t^K)^2\mid\mathcal G_t]
 =1+\E_B\!\left[\frac{(H_t^K(X_t)-c_t^K)^2}{c_t^K H_t^K(X_t)}
                    \,\middle|\,\mathcal G_t\right]\le1+\frac{(\sqrt{r_t^K}-1)^2}{N}.
}\]\DIFaddend 
If deterministic finite constants  \DIFaddbegin \DIFadd{$\rho_t^K\ge1$ satisfy $r_t^K\le \rho_t^K$ }\DIFaddend for every realized history and pool, then
 \DIFaddbegin \begin{equation}\DIFadd{\label{supp:eq:global}
 \E_{TB}D_T^2
 \le\prod_{t=1}^T\prod_{K\in\cK}
       \left[1+\frac{(\sqrt{\rho_t^K}-1)^2}{N}\right]
 \le\exp\left\{\frac{1}N\sum_{t=1}^T\sum_{K\in\cK}(\sqrt{\rho_t^K}-1)^2\right\}.
}\end{equation}\DIFaddend 
\end{proposition}
\begin{proof}
We retain the time and block indices throughout. First fix $t,K$ and condition on $\mathcal G_t$; we subsequently combine the conditional bounds over blocks and times.

\paragraph{One-block change of measure.}
The distinguished-offspring representation gives the density $H_t^K(X_t)/c_t^K$ of the twisted block law relative to $(\nu_t^K)^{\otimes N}$. Thus for a nonnegative measurable function $F_t^K$ of the block offspring array,
\[
 \E_{TB}[F_t^K(X_t^{1:N,K})\mid\mathcal G_t]
 =\E_B\!\left[\frac{H_t^K(X_t)}{c_t^K}F_t^K(X_t^{1:N,K})
                       \,\middle|\,\mathcal G_t\right].
\]
This identity compares two resampling laws at the same pool. Since the offspring belong to the proposal support,
\[
 \alpha_t^K\le H_t^K(X_t)\le\beta_t^K\qquad\text{\normalfont and}\qquad \alpha_t^K\le c_t^K\le\beta_t^K.
\]
Apply the identity to the correction $C_t^K$ and its square:
\begin{gather*}
 \E_{TB}[C_t^K\mid\mathcal G_t]
 =\E_B\!\left[\frac{H_t^K(X_t)}{c_t^K}
                  \frac{c_t^K}{H_t^K(X_t)}\,\middle|\,\mathcal G_t\right]=1,\\
 \E_{TB}[(C_t^K)^2\mid\mathcal G_t]
 =\E_B\!\left[\frac{H_t^K(X_t)}{c_t^K}
                  \frac{(c_t^K)^2}{[H_t^K(X_t)]^2}\,\middle|\,\mathcal G_t\right]=\E_B\!\left[\frac{c_t^K}{H_t^K(X_t)}\,\middle|\,\mathcal G_t\right].
\end{gather*}
Under the ordinary conditional law, $X_t^{1,K},\ldots,X_t^{N,K}$ are i.i.d. from $\nu_t^K$, so
\[
 \E_B[H_t^K(X_t)\mid\mathcal G_t]
 =\frac{1}N\sum_{j=1}^N\nu_t^K(\psi_t^K)=c_t^K.
\]
The algebraic identity
\[
 \frac{(H_t^K(X_t)-c_t^K)^2}{c_t^K H_t^K(X_t)}
 =\frac{H_t^K(X_t)}{c_t^K}-2+\frac{c_t^K}{H_t^K(X_t)}
\]
therefore yields
\begin{align*}
 \E_B\!\left[\frac{(H_t^K(X_t)-c_t^K)^2}{c_t^K H_t^K(X_t)}
                 \,\middle|\,\mathcal G_t\right]
 &=\frac{\E_B[H_t^K(X_t)\mid\mathcal G_t]}{c_t^K}-2
       +\E_B\!\left[\frac{c_t^K}{H_t^K(X_t)}\,\middle|\,\mathcal G_t\right]\\
 &=\E_{TB}[(C_t^K)^2\mid\mathcal G_t]-1.
\end{align*}
This proves the exact conditional second-moment identity.

\paragraph{Bounding the one-block second moment.}
 \DIFaddbegin \DIFadd{Condition throughout on }\DIFaddend $\mathcal G_t$\DIFaddbegin \DIFadd{. Since $(\psi_t^K-\alpha_t^K)(\beta_t^K-\psi_t^K)\ge0$ on the pool, taking its $\nu_t^K$ expectation yields
}\[
\DIFadd{\Var_{\nu_t^K}(\psi_t^K)\le(c_t^K-\alpha_t^K)(\beta_t^K-c_t^K).
}\]\DIFaddend 
 \DIFaddbegin \DIFadd{Under the baseline law, $\Var_B(H_t^K\mid\mathcal G_t)=\Var_{\nu_t^K}(\psi_t^K)/N$. Consequently,
}\begin{align*}
\DIFadd{\E_{TB}[(C_t^K)^2\mid\mathcal G_t]-1
}&\DIFadd{\le\frac{\Var_{\nu_t^K}(\psi_t^K)}{Nc_t^K\alpha_t^K}
\le\frac{(c_t^K-\alpha_t^K)(\beta_t^K-c_t^K)}{Nc_t^K\alpha_t^K}
\le\frac{(\sqrt{r_t^K}-1)^2}{N}.
}\end{align*}\DIFaddend 
 \DIFaddbegin \DIFadd{For the last inequality set $x=c_t^K/\alpha_t^K$ and use
$(x-1)(r_t^K-x)/x=r_t^K+1-x-r_t^K/x\le(\sqrt{r_t^K}-1)^2$}\DIFaddend.

\paragraph{Combining blocks at time $t$.}
Given $\mathcal G_t$, all guides and proposal weights are fixed, and the block offspring arrays are independent under the twisted kernel. Each $C_t^K$ depends only on the time-$t$ offspring in block $K$ and the fixed conditioning information. Therefore
 \DIFaddbegin \begin{align*}
 \DIFadd{\E_{TB}[\mathcal C_t^2\mid\mathcal G_t]
 }&\DIFadd{=\E_{TB}\!\left[\prod_{K\in\cK}(C_t^K)^2\,\middle|\,\mathcal G_t\right]=\prod_{K\in\cK}\E_{TB}[(C_t^K)^2\mid\mathcal G_t]}\\
 &\DIFadd{\le\prod_{K\in\cK}\left[1+\frac{(\sqrt{r_t^K}-1)^2}{N}\right]
 \le\prod_{K\in\cK}\left[1+\frac{(\sqrt{\rho_t^K}-1)^2}{N}\right]=:B_t.
}\end{align*}\DIFaddend 
The assumed uniform deterministic bounds  \DIFaddbegin \DIFadd{$\rho_t^K$ }\DIFaddend make $B_t$ deterministic. This factorization does not assert independence before conditioning on the pool or independence across times.

\paragraph{Accumulation over time.}\label{supp:correction-filtration}
Let $\mathcal F_{t-1}$ contain all variables generated through resampling at time $t-1$. Then $\mathcal F_{t-1}\subseteq\mathcal G_t$, and $D_{t-1}$ is $\mathcal F_{t-1}$-measurable. Since  \DIFaddbegin \DIFadd{$D_t=D_{t-1}\mathcal C_t$}\DIFaddend, the tower property gives
 \DIFaddbegin \begin{align*}
 \DIFadd{\E_{TB}[D_t^2\mid\mathcal F_{t-1}]
 }&\DIFadd{=D_{t-1}^2\,
   \E_{TB}\!\left[\E_{TB}[\mathcal C_t^2\mid\mathcal G_t]
                   \,\middle|\,\mathcal F_{t-1}\right]\le B_tD_{t-1}^2.
}\end{align*}\DIFaddend 
Taking expectations and iterating from $D_0=1$ yields
\[
 \E_{TB}D_T^2\le B_T\E_{TB}D_{T-1}^2
 \le\cdots\le\prod_{t=1}^T B_t.
\]
Finally, $1+x\le e^x$ for $x\ge0$ applied to every factor in this product proves \eqref{supp:eq:global}. The time bound thus uses successive conditional expectations rather than temporal independence.
\end{proof}

\DIFaddbegin \subsection{\DIFadd{Variance comparison and conditional guide quality}}\label{supp:variance-guides}
\paragraph{\DIFadd{Exact variance difference (recalled).}}
\DIFadd{Recall $\mathcal Z_T=\prod_{t,K}a_{t,N}^{K}$ and $D_T=d\mathbb P_B/d\mathbb P_{TB}$. If $\mathcal Z_T$ under BPF and $D_T\mathcal Z_T$ under TBPF have finite second moments, Equation~\eqref{eq:variance-identity} states
}
\begin{equation*}
\DIFadd{\Var_{TB}(D_T\mathcal Z_T)-\Var_B(\mathcal Z_T)
=\E_B[(D_T-1)\mathcal Z_T^2].}
\tag{\ref{eq:variance-identity}}
\end{equation*}
\begin{proof}
\DIFadd{The change-of-measure theorem gives $\E_{TB}(D_T\mathcal Z_T)=\E_B\mathcal Z_T$ and $\E_{TB}(D_T^2\mathcal Z_T^2)=\E_B(D_T\mathcal Z_T^2)$. Subtracting the common squared mean proves \eqref{eq:variance-identity}. For any deterministic $z$, the two squared-error risks equal their respective variances plus the same $(\E_B\mathcal Z_T-z)^2$. 
}

\end{proof}

\DIFadd{The following proposition restates Proposition~\ref{prop:guide-improvement} from the main text and adds an exact finite-perturbation identity.}

\begin{proposition}[Conditional guide improvement; restated with a finite-perturbation identity]\label{supp:prop:guide-improvement}
Condition on a fixed block pool with weighted law $\nu$ and offspring law $R=\nu^{\otimes N}$. Let $X=(X^1,\ldots,X^N)$ and $U$ have baseline joint law $P_0(dx,du)=R(dx)\mathsf M(x,du)$, where $\mathsf M$ is the continuation kernel. Fix a nonnegative measurable payoff $W$ with $\mathcal A(x):=\int W(x,u)^2\mathsf M(x,du)$ integrable under $R$, and a bounded measurable block function $h$. For a real parameter $\theta$ satisfying $|\theta|\|h\|_\infty<1$, use $\psi_\theta=1+\theta h$, holding the pool, $\mathsf M$, $W$, and $h$ fixed. Set $m_h=\nu(h)$ and $\bar h(x)=N^{-1}\sum_jh(x^j)$, and define
\[
\begin{gathered}
 \Lambda_\theta(x)=\frac{1+\theta\bar h(x)}{1+\theta m_h},\quad
 Q_\theta(dx,du)=\Lambda_\theta(x)P_0(dx,du),\\
 \mathcal V(\theta)=\Var_{Q_\theta}\!\left(\frac{W(X,U)}{\Lambda_\theta(X)}\right).
\end{gathered}
\]
Then $\mathcal V'(0)=-\operatorname{Cov}_R(\mathcal A,\bar h)$. Some positive admissible $\theta$ reduces variance if and only if this covariance is positive; in that case, all sufficiently small positive $\theta$ do so.

Moreover, write $s(x)=\bar h(x)-m_h$, $\alpha_\theta=\theta/(1+\theta m_h)$, and $c_{\mathrm{step}}=\E_R[\mathcal A s]=\operatorname{Cov}_R(\mathcal A,\bar h)$. For every admissible $\theta$,
\begin{equation}\label{supp:eq:finite-guide-variance}
 \mathcal V(\theta)-\mathcal V(0)
 =-\alpha_\theta c_{\mathrm{step}}+\alpha_\theta^2
 \E_R\!\left[\frac{\mathcal A s^2}{1+\alpha_\theta s}\right].
\end{equation}
\end{proposition}
\begin{proof}
All laws in this proof are conditional on the fixed pool. For an integrable function $F$, define $$\E_{P_0}F=\int F(x,u)P_0(dx,du)\qquad \text{and}\qquad \E_{Q_\theta}F=\int F(x,u)Q_\theta(dx,du);$$ for a function $a$ of the array alone, $\E_Ra=\int a(x)R(dx)$. Thus an expectation subscript specifies a probability law, not conditioning on $\Lambda_\theta$. The function $\Lambda_\theta$ is fixed, but its evaluation $\Lambda_\theta(X)$ remains random as $X$ varies.

Since $\E_R\bar h=m_h$, the density satisfies $\E_R\Lambda_\theta=1$. The restriction on $\theta$ makes $\Lambda_\theta$ strictly positive and bounded away from zero, so $Q_\theta$ is a probability law and the corrected payoff has a finite second moment. Direct integration gives
\begin{align*}
 \E_{Q_\theta}\!\left[\frac{W(X,U)}{\Lambda_\theta(X)}\right]
 &=\iint\frac{W(x,u)}{\Lambda_\theta(x)}\Lambda_\theta(x)R(dx)\mathsf M(x,du)
 =\E_{P_0}W,\\
 \E_{Q_\theta}\!\left[\frac{W(X,U)^2}{\Lambda_\theta(X)^2}\right]
 &=\iint\frac{W(x,u)^2}{\Lambda_\theta(x)}R(dx)\mathsf M(x,du)
 =\E_R\!\left[\frac{\mathcal A}{\Lambda_\theta}\right].
\end{align*}
The cancellation uses the density in $Q_\theta$, not constancy of $\Lambda_\theta(X)$. Subtracting the common squared mean therefore yields
\[
 \mathcal V(\theta)-\mathcal V(0)
 =\E_R\!\left[\mathcal A\left(\frac1{\Lambda_\theta}-1\right)\right].
\]
Now $\Lambda_\theta=1+\alpha_\theta s$ and $\E_Rs=0$. Substituting
\[
 \frac1{1+\alpha_\theta s}-1
 =-\alpha_\theta s+\frac{\alpha_\theta^2s^2}{1+\alpha_\theta s}
\]
proves \eqref{supp:eq:finite-guide-variance}. Because $s$ is bounded and $\mathcal A$ is integrable, the expectation in the remainder is bounded in a neighborhood of zero. Also $\alpha_\theta=\theta+O(\theta^2)$, so
$$\mathcal V(\theta)-\mathcal V(0)=-\theta c_{\mathrm{step}}+O(\theta^2),$$ proving the derivative formula.

If $c_{\mathrm{step}}>0$, this expansion gives strict reduction for all sufficiently small positive $\theta$. Conversely, every positive admissible $\theta$ has $\alpha_\theta>0$, and the remainder in \eqref{supp:eq:finite-guide-variance} is nonnegative. If $c_{\mathrm{step}}\le0$, both terms on its right-hand side are nonnegative, ruling out strict reduction for any such $\theta$. This proves the equivalence, including the constant-guide case $s=0$.
\end{proof}

\paragraph{\DIFadd{A conditional log-variance criterion.}}
\DIFadd{The following proposition restates the baseline log-variance identity \eqref{eq:log-guide-improvement} from the main text and extends it to every admissible $\theta$, with the pool and continuation fixed as in Proposition~\ref{prop:guide-improvement}.
}\begin{proposition}[Derivative of log variance]\label{supp:prop:log-derivative}
Use the fixed-pool laws $P_0$ and $Q_\theta$ and the bounded perturbation $h$ defined in Proposition~\ref{supp:prop:guide-improvement}. Assume $W(X,U)>0$ almost surely under $P_0$ and $\E_{P_0}[(\log W)^2]<\infty$. The pool, continuation kernel, payoff, and $h$ remain fixed as $\theta$ varies. Recall the centered array statistic
\[
 s(x)=\bar h(x)-m_h=\frac1N\sum_{j=1}^N[h(x^j)-\nu(h)].
\]
For $|\theta|\|h\|_\infty<1$, define the corrected log payoff, its mean, and its variance by
\[
 Y_{\log,\theta}=\log W(X,U)-\log\Lambda_\theta(X),\qquad
 \mu_{\log,\theta}=\E_{Q_\theta}Y_{\log,\theta},\qquad
 \mathcal V_{\log}(\theta)=\Var_{Q_\theta}(Y_{\log,\theta}).
\]
Also define the log-density derivative
\[
 \zeta_\theta(x):=\frac{\partial}{\partial\theta}\log\Lambda_\theta(x)
 =\frac{s(x)}{[1+\theta\bar h(x)](1+\theta m_h)}.
\]
Then $\mathcal V_{\log}$ is continuously differentiable throughout this interval, with
\begin{equation}\label{supp:eq:log-derivative-general}
 \mathcal V_{\log}'(\theta)
 =\E_{Q_\theta}\!\left[\zeta_\theta(X)
 \left\{(Y_{\log,\theta}-\mu_{\log,\theta})^2
       -2(Y_{\log,\theta}-\mu_{\log,\theta})\right\}\right].
\end{equation}
In particular, set $Y_{\log}=Y_{\log,0}=\log W(X,U)$ and $\mu_{\log}=\mu_{\log,0}$. At the baseline,
\begin{equation}\label{supp:eq:log-derivative}
 \mathcal V_{\log}'(0)
 =\E_{P_0}\!\left[s(X)\{(Y_{\log}-\mu_{\log})^2-2(Y_{\log}-\mu_{\log})\}\right]
 =\operatorname{Cov}_R(\bar h,B_{\log}),
\end{equation}
where
\[
 B_{\log}(x)=\int\left\{[\log W(x,u)-\mu_{\log}]^2
             -2[\log W(x,u)-\mu_{\log}]\right\}\mathsf M(x,du).
\]
A negative baseline covariance gives strict log-variance reduction for all sufficiently small positive $\theta$. More generally, a negative derivative in \eqref{supp:eq:log-derivative-general} gives a local reduction as $\theta$ increases from that interior point.
\end{proposition}
\begin{proof}
First, differentiating the density explicitly gives
\[
 \Lambda'_\theta(x)
 =\frac{\bar h(x)(1+\theta m_h)-m_h[1+\theta\bar h(x)]}{(1+\theta m_h)^2}
 =\frac{s(x)}{(1+\theta m_h)^2}.
\]
Thus $\Lambda'_0(x)=s(x)$ is a consequence of the formula for $\Lambda_\theta$, not a definition of $s$. Moreover,
$\zeta_\theta=\Lambda'_\theta/\Lambda_\theta$, so $\zeta_0=s$ as well. The fixed-pool identity $\E_Rs=0$ implies
\[
 \E_{Q_\theta}\zeta_\theta(X)
 =\E_R[\Lambda_\theta\zeta_\theta]
 =\E_R\Lambda'_\theta=0.
\]
The subscripts $P_0$ and $Q_\theta$ specify integration laws. 

On every compact subinterval of $|\theta|\|h\|_\infty<1$, boundedness of $h$ bounds $\Lambda_\theta$ above and away from zero, and uniformly bounds $\Lambda'_\theta$, $\log\Lambda_\theta$, and $\zeta_\theta$. Consequently the first two log-moment integrands, written under $P_0$, and their derivatives are dominated by a constant times $1+(\log W)^2$. This is integrable by assumption, so differentiation under the integral is valid and the resulting derivatives are continuous.

Since $\partial_\theta Y_{\log,\theta}=-\zeta_\theta(X)$, differentiating both the density and the log payoff gives
\begin{align*}
 \mu'_{\log,\theta}
 &=\frac{d}{d\theta}\E_{P_0}[\Lambda_\theta(X)Y_{\log,\theta}]
 =\E_{Q_\theta}[\zeta_\theta(X)(Y_{\log,\theta}-1)],\\
 \frac{d}{d\theta}\E_{Q_\theta}[Y_{\log,\theta}^2]
 &=\frac{d}{d\theta}\E_{P_0}[\Lambda_\theta(X)Y_{\log,\theta}^2]
 =\E_{Q_\theta}[\zeta_\theta(X)(Y_{\log,\theta}^2-2Y_{\log,\theta})].
\end{align*}
Unlike the corrected ordinary mean in Proposition~\ref{supp:prop:guide-improvement}, the corrected log mean $\mu_{\log,\theta}$ generally varies with $\theta$. Subtracting $2\mu_{\log,\theta}\mu'_{\log,\theta}$ from the second-moment derivative and using $\E_{Q_\theta}\zeta_\theta(X)=0$ gives \eqref{supp:eq:log-derivative-general}.

At $\theta=0$, $Q_0=P_0$, $\Lambda_0=1$, and $\zeta_0=s$, yielding the first equality in \eqref{supp:eq:log-derivative}. Integrating over $U$ conditional on $X=x$ gives $\E_R[sB_{\log}]$. Since $s=\bar h-\E_R\bar h$, this is $\operatorname{Cov}_R(\bar h,B_{\log})$. The local reduction statements follow from a negative derivative at an interior point.
\end{proof}
\begin{corollary}[A quantitative finite-step guarantee]\label{supp:cor:centered-step}
Under Proposition~\ref{supp:prop:guide-improvement}, recall $m_h=\nu(h)$ and set
\[
 h_0=h-m_h,\qquad s(x)=\frac1N\sum_{j=1}^N h_0(x^j),\qquad
 H_0=\|h_0\|_\infty,\qquad
 c_{\mathrm{step}}=\E_R[\mathcal A s]>0,\qquad
 b_{\mathrm{step}}=\E_R[\mathcal A s^2].
\]
For the centered guide $\psi_\alpha^c=1+\alpha h_0$ with $|\alpha|H_0<1$, define
\[
 \Lambda_\alpha^c(x)=1+\alpha s(x),\qquad
 Q_\alpha^c(dx,du)=\Lambda_\alpha^c(x)P_0(dx,du),\qquad
 \mathcal V_c(\alpha)=\Var_{Q_\alpha^c}\!\left(\frac{W(X,U)}{\Lambda_\alpha^c(X)}\right).
\]
Then $H_0>0$ and $b_{\mathrm{step}}>0$, and for $0<\alpha<1/H_0$,
\begin{equation}\label{supp:eq:step-bound}
 \mathcal V_c(\alpha)-\mathcal V_c(0)
 \le-\alpha c_{\mathrm{step}}+
 \frac{\alpha^2b_{\mathrm{step}}}{1-\alpha H_0}.
\end{equation}
Thus every $0<\alpha<c_{\mathrm{step}}/(b_{\mathrm{step}}+H_0c_{\mathrm{step}})$ gives strict variance reduction. In particular, the explicit choice
\begin{equation}\label{supp:eq:explicit-guide-step}
 \alpha_{\mathrm{safe}}=
 \frac{c_{\mathrm{step}}}{2(b_{\mathrm{step}}+H_0c_{\mathrm{step}})}
 \quad\text{gives}\quad
 \mathcal V_c(\alpha_{\mathrm{safe}})-\mathcal V_c(0)
 \le-\frac{c_{\mathrm{step}}^2}{2(2b_{\mathrm{step}}+H_0c_{\mathrm{step}})}<0.
\end{equation}
Moreover, $\mathcal V_c$ is strictly convex on $|\alpha|H_0<1$, with
\[
 \mathcal V_c''(\alpha)=2\E_R\!\left[\frac{\mathcal A s^2}{(1+\alpha s)^3}\right]>0.
\]
For an admissible parameter $\theta$ in the original guide family, $\mathcal V(\theta)=\mathcal V_c(\alpha_\theta)$ whenever $|\alpha_\theta|H_0<1$. Consequently the finite-step bound applies to that family when $\alpha_\theta$ belongs to the sufficient interval above.
\end{corollary}
\begin{proof}
By construction, $\nu(h_0)=0$, $\E_Rs=0$, and $|s|\le H_0$. The integrability of $\mathcal A$ and boundedness of $s$ make both $c_{\mathrm{step}}$ and $b_{\mathrm{step}}$ finite. If $H_0=0$, then $s=0$ almost surely. If $b_{\mathrm{step}}=0$, nonnegativity gives $\mathcal A s^2=0$ almost surely and hence $\mathcal A s=0$ almost surely. Either case contradicts $c_{\mathrm{step}}>0$, so $H_0,b_{\mathrm{step}}>0$.

For $|\alpha|H_0<1$, the centered guide and array density are strictly positive, $\E_R\Lambda_\alpha^c=1$, and $\Lambda_\alpha^c\ge1-|\alpha|H_0$. Thus $Q_\alpha^c$ is a probability law. The change-of-measure calculation in Proposition~\ref{supp:prop:guide-improvement} gives
\[
 \E_{Q_\alpha^c}\!\left[\frac{W}{\Lambda_\alpha^c(X)}\right]=\E_{P_0}W\qquad \text{and}\qquad
 \mathcal V_c(\alpha)=\E_R\!\left[\frac{\mathcal A}{1+\alpha s}\right]-(\E_{P_0}W)^2.
\]
Subtracting the value at zero and using
$(1+\alpha s)^{-1}-1=-\alpha s+\alpha^2s^2/(1+\alpha s)$ yields the exact identity
\[
 \mathcal V_c(\alpha)-\mathcal V_c(0)
 =-\alpha c_{\mathrm{step}}+
 \alpha^2\E_R\!\left[\frac{\mathcal A s^2}{1+\alpha s}\right].
\]
For $0<\alpha<1/H_0$, bound the denominator below by $1-\alpha H_0$ to obtain \eqref{supp:eq:step-bound}. Its upper bound factors as
\[
 -\alpha c_{\mathrm{step}}+
 \frac{\alpha^2b_{\mathrm{step}}}{1-\alpha H_0}
 =\frac{\alpha\{\alpha(b_{\mathrm{step}}+H_0c_{\mathrm{step}})-c_{\mathrm{step}}\}}
        {1-\alpha H_0}.
\]
The denominator is positive, and the numerator is negative on the stated sufficient interval. Since $b_{\mathrm{step}}>0$,
\[
 0<\frac{c_{\mathrm{step}}}{b_{\mathrm{step}}+H_0c_{\mathrm{step}}}<\frac1{H_0},
\]
so that entire interval is admissible. The choice $\alpha_{\mathrm{safe}}$ is its midpoint and satisfies $\alpha_{\mathrm{safe}}H_0<1/2$. Substituting it into the factored upper bound gives exactly the negative bound in \eqref{supp:eq:explicit-guide-step}.

For convexity, on any compact subinterval of $|\alpha|H_0<1$ there is a constant $\delta>0$ such that $1+\alpha s\ge\delta$. The absolute first and second derivatives of $\mathcal A/(1+\alpha s)$ are bounded by $\mathcal A H_0/\delta^2$ and $2\mathcal A H_0^2/\delta^3$, respectively. These are integrable, so differentiation under the expectation gives
\[
 \mathcal V_c'(\alpha)=-\E_R\!\left[\frac{\mathcal A s}{(1+\alpha s)^2}\right]\qquad \text{and}\qquad
 \mathcal V_c''(\alpha)=2\E_R\!\left[\frac{\mathcal A s^2}{(1+\alpha s)^3}\right].
\]
Because $b_{\mathrm{step}}=\E_R[\mathcal A s^2]>0$, the second derivative is strictly positive, proving strict convexity.

Finally, for $\alpha_\theta=\theta/(1+\theta m_h)$,
\[
 \Lambda_\theta(x)=\frac{1+\theta\bar h(x)}{1+\theta m_h}
 =1+\alpha_\theta[\bar h(x)-m_h]=\Lambda_{\alpha_\theta}^c(x).
\]
Whenever both parameters are in their stated domains, the joint tilted laws and corrected payoffs therefore coincide, proving $\mathcal V(\theta)=\mathcal V_c(\alpha_\theta)$.
\end{proof}

\DIFaddend \subsection{ \DIFaddbegin \DIFadd{An exact finite-pool guide-strength example}\DIFaddend }\DIFaddbegin \label{supp:pool-example}
\DIFadd{We detail the example following Proposition~\ref{prop:guide-improvement}. Consider a single block with $N=2$ and $T\ge2$, and condition on the past and an equal-weight proposal pool $\{0,1\}$ at time $T-1$, before drawing the offspring. Its weighted law is $\nu=(\delta_0+\delta_1)/2$, and the baseline offspring array $X=(X^1,X^2)$ has law $R=\nu^{\otimes2}$. At time $T$, use identity propagation, likelihood values $g_{\mathrm{ex}}(0)=1$ and $g_{\mathrm{ex}}(1)=3$, and the prescribed terminal guide $1$. The next normalizer increment is therefore
}\[
 W_{\mathrm{ex}}(X)=\frac{g_{\mathrm{ex}}(X^1)+g_{\mathrm{ex}}(X^2)}2.
\]
\DIFadd{Factors accumulated before this resampling step are fixed under the conditioning and are omitted from the calculation. For any positive guide $\psi$ at time $T-1$, define
}\[
 \begin{aligned}
 c_\psi=\nu(\psi)&=\frac{\psi(0)+\psi(1)}2,\qquad
 & H_\psi(x)&=\frac{\psi(x^1)+\psi(x^2)}2,\\
 \Lambda_\psi(x)&=\frac{H_\psi(x)}{c_\psi},
 & \widetilde R_\psi(dx)&=\Lambda_\psi(x)R(dx).
 \end{aligned}
\]
\DIFadd{The time-$(T-1)$ correction is $1/\Lambda_\psi(X)$, so the remaining normalizer contribution is $W_{\mathrm{ex}}(X)/\Lambda_\psi(X)$ under $\widetilde R_\psi$, with mean $\E_R W_{\mathrm{ex}}=2$. Each probability tuple and payoff tuple below lists the groups $(0,0)$, $\{(0,1),(1,0)\}$, and $(1,1)$, in that order. The middle probability combines the two mixed arrays. Means and variances use the law in the corresponding column; lower variance is better.
}\begin{center}
\begin{tabular}{lccc}\toprule
 & \DIFadd{Baseline $\psi=1$ }& \DIFadd{Aligned $\psi=g_{\mathrm{ex}}$ }& \DIFadd{Opposing $\psi=4-g_{\mathrm{ex}}$}\\\midrule
\DIFadd{Group probabilities }& \DIFadd{$(1/4,1/2,1/4)$ }& \DIFadd{$(1/8,1/2,3/8)$ }& \DIFadd{$(3/8,1/2,1/8)$}\\
\DIFadd{Corrected payoff $W_{\mathrm{ex}}/\Lambda_\psi$ }& \DIFadd{$(1,2,3)$ }& \DIFadd{$(2,2,2)$ }& \DIFadd{$(2/3,2,6)$}\\
\DIFadd{Mean }& \DIFadd{$2$ }& \DIFadd{$2$ }& \DIFadd{$2$}\\
\DIFadd{Variance }& \DIFadd{$1/2$ }& \DIFadd{$0$ }& \DIFadd{$8/3$}\\\bottomrule
\end{tabular}
\end{center}
\DIFadd{For the aligned guide, $c_\psi=2$ and $H_\psi=W_{\mathrm{ex}}$, so the corrected payoff equals $2$ for every array. For the opposing guide, $c_\psi=2$ and $H_\psi=4-W_{\mathrm{ex}}$; its variance relative to the baseline is $(8/3)/(1/2)=16/3$.
}

\DIFadd{More generally, let $\psi_\beta=g_{\mathrm{ex}}^\beta$ for $\beta\in\mathbb R$ and put $q_\beta=3^\beta$. The guide values are $(1,q_\beta)$, so $c_{\psi_\beta}=(1+q_\beta)/2$ and $H_{\psi_\beta}$ takes values $1,(1+q_\beta)/2,q_\beta$ on the three groups. Using the baseline group probabilities $(1/4,1/2,1/4)$ gives
}\begin{align*}
 \mathcal V_{\mathrm{ex}}(\beta)
 :=\Var_{\widetilde R_{\psi_\beta}}\!\left(\frac{W_{\mathrm{ex}}}{\Lambda_{\psi_\beta}}\right)
 =\E_R\!\left[\frac{W_{\mathrm{ex}}^2}{\Lambda_{\psi_\beta}}\right]-4=\frac{1+q_\beta}{8}+2+\frac{9(1+q_\beta)}{8q_\beta}-4
 =\frac{(q_\beta-3)^2}{8q_\beta}.
\end{align*}
\DIFadd{The constant guide $\beta=0$ gives variance $1/2$, and the unique minimum is zero at $\beta=1$. The variance is below the baseline exactly when $0<\beta<2$, and returns to $1/2$ at $\beta=2$. The opposing guide is equivalent to $\beta=-1$, since multiplying a guide by a positive constant leaves $\Lambda_\psi$ unchanged. These identities describe how guide strength changes the corrected variance for this fixed pool and deterministic continuation.
}

\subsection{\DIFadd{A full-history total-variation consequence}}\label{supp:tv}
\DIFadd{Let $S=\sum_{t,K}(\sqrt{\rho_t^K}-1)^2$ and define $\|P-Q\|_{\mathrm{TV}}=\sup_A|P(A)-Q(A)|$. Since $\E_{TB}D_T=1$, Cauchy--Schwarz and the correction bound give
}\[
\DIFadd{\|\mathbb P_B-\mathbb P_{TB}\|_{\mathrm{TV}}
=\tfrac12\E_{TB}|D_T-1|
\le\tfrac12\sqrt{\E_{TB}D_T^2-1}
\le\tfrac12\sqrt{e^{S/N}-1}.
}\]
\DIFadd{For a sequence of settings, write $S_N$ for the corresponding value of $S$. If $S_N/N\to0$, the laws of any identical measurable history statistic under the two measures become arbitrarily close in total variation. Consequently any baseline distributional limit transfers for that same statistic, even when it depends on $N$. 
}

\appendixsection{Fixed-horizon consistency}\label{supp:sec:theory}

\DIFaddend \label{supp:subsec:consistency}
\paragraph{Assumptions and notation for consistency.}
Assume each component of $\cX=\prod_{v\in V}\cX^v$ is Polish, and fix
$d=|V|$, $T$, and the partition $\cK$. Write $C_b(\cX)$ for the bounded
continuous real-valued functions on $\cX$. Assume
\[
 f_tu(x):=\int u(z)f_t(x,dz)\in C_b(\cX)\quad(u\in C_b(\cX))\qquad\text{\normalfont and}\qquad g_t^v\in C_b(\cX^v),\quad g_t^v>0.
\]
Thus each $f_t$ is Feller; the displayed conditions hold for every $t=1,\ldots,T$ and $v\in V$. Initialize $X_0^1,\ldots,X_0^N$ independently from the
 \DIFaddbegin \DIFadd{block-product law $\pi_0=\bigotimes_{K\in\cK}\pi_0^K=\widetilde\pi_0$}\DIFaddend, and use Algorithm~\ref{supp:alg:tbpf}.
Guides may depend on $N$ and on $\mathcal G_t$ (the past, current pool, and
recorded pre-offspring guide randomness), but satisfy
\[
 0<\psi_t^K(Z_t^{i,K})<\infty\quad\text{\normalfont almost surely for every }t,K,i.
\]
Recall $\eta_t^N=N^{-1}\sum_i\delta_{X_t^i}$ and denote
convergence in probability in the weak topology by
 \DIFaddbegin \DIFadd{$\xrightarrow{\mathrm{weak},\mathbb P}$}\DIFaddend. Expectations and probabilities below
are under the TBPF law for the indicated $N$. All limits in the following
theorem and proof are taken as the particle count $N\to\infty$, with
$d$, $T$, $\cK$, and the observation sequence $y_{1:T}$ fixed.

The following theorem restates Theorem~\ref{thm:consistency} from the main text in the detailed notation used for its proof.

\begin{theorem}[Fixed-horizon consistency; restated]\label{supp:thm:consistent}
Under the preceding assumptions, as $N\to\infty$, for every fixed  \DIFaddbegin \DIFadd{$1\le t\le T$,
}\[
 \DIFadd{\eta_t^N,\ \widehat\pi_t^{TB}\xrightarrow{\mathrm{weak},\mathbb P}\widetilde\pi_t\quad(t\ge1),\qquad\text{\normalfont and}\qquad \eta_t^N(u)\xrightarrow{\mathbb P}\widetilde\pi_t(u)
 \quad\text{\normalfont for every }u\in C_b(\cX),
}\]\DIFaddend 
where $\widetilde\pi_t$ is defined by \eqref{eq:block}.
If, in addition, deterministic finite constants  \DIFaddbegin \DIFadd{$\rho_t^K$}\DIFaddend, independent of $N$,
satisfy
 \DIFaddbegin \[
 \DIFadd{\frac{\max_{1\le i\le N}\psi_t^K(Z_t^{i,K})}{\min_{1\le i\le N}\psi_t^K(Z_t^{i,K})}
 \le \rho_t^K\quad\text{\normalfont almost surely for every }N,t,K,
}\]\DIFaddend 
then, as $N\to\infty$, for $\widehat Z_T^{TB}$ defined in \eqref{supp:eq:tb-normalizer},
\begin{equation}
 D_T\longrightarrow1\qquad\text{\normalfont and}\qquad \widehat Z_T^{TB}\longrightarrow
 Z_T^{\mathrm{block}}:=\prod_{t=1}^T\prod_{K\in\cK}
 (P_t\widetilde\pi_{t-1})^K(g_t^K)
 \quad\text{\normalfont in }L^2(\mathbb P_{TB}).
\end{equation}
\end{theorem}
\begin{proof}\leavevmode
The proof proceeds in five steps. Step 1 establishes consistency at initialization
and relates scalar convergence to weak convergence in probability. Steps 2 and 3
propagate consistency through prediction and block weighting, respectively.
Step 4 controls twisted resampling and recombination, completing the induction
for filtering consistency. Under the additional uniform guide-ratio bound,
Step 5 proves $L^2$ convergence of the cumulative correction and the corrected
normalizer.

\paragraph{Step 1. Initialization and the weak topology.}\label{supp:weak-metric}
For $u\in C_b(\cX)$, independence at initialization gives
\[
 \E_{TB}\eta_0^N(u)=\pi_0(u)\qquad\text{\normalfont and}\qquad \E_{TB}|\eta_0^N(u)-\pi_0(u)|^2
 =\frac{\pi_0(u^2)-\pi_0(u)^2}{N}
 \le\frac{\|u\|_\infty^2}{N}.
\]
 \DIFaddbegin \DIFadd{For a Polish state space, use the standard weak-topology construction \citep{billingsley1999} and }\DIFaddend choose a countable convergence-determining family  \DIFaddbegin \DIFadd{$(u_m)$ in the unit ball of $C_b(\cX)$ and }\DIFaddend the weak metric
 \DIFaddbegin \[
 \DIFadd{d_*(\mu,\nu)=\sum_{m\ge1}2^{-m}|\mu(u_m)-\nu(u_m)|.
}\]\DIFaddend 
 \DIFaddbegin \DIFadd{For any finite initial segment, scalar convergence in probability follows from the preceding variance bound and a union bound; the remaining tail is deterministically at most $2^{1-L}$. Letting the segment length }\DIFaddend $L$  \DIFaddbegin \DIFadd{grow establishes weak convergence in probability at initialization. Conversely, integration of any fixed bounded continuous function }\DIFaddend is continuous in the weak topology.  \DIFaddbegin \DIFadd{These facts justify the induction below.}\DIFaddend.

\paragraph{Step 2. Prediction.}\label{supp:prediction-notation}
Suppose  \DIFaddbegin \DIFadd{$\eta_{t-1}^N\xrightarrow{\mathrm{weak},\mathbb P}\widetilde\pi_{t-1}$}\DIFaddend.
Recall that $\mathcal F_{t-1}$ contains all variables generated through
time $t-1$, and define
\[
 q_t^N:=\frac{1}N\sum_{i=1}^N\delta_{Z_t^i}\qquad\text{\normalfont and}\qquad q_t:=P_t\widetilde\pi_{t-1}=\widetilde\pi_{t-1}f_t.
\]
Conditionally on $\mathcal F_{t-1}$, the proposals are independent with
$Z_t^i\sim f_t(X_{t-1}^i,\cdot)$. Consequently, for $u\in C_b(\cX)$,
\begin{align*}
 \E_{TB}[q_t^N(u)\mid\mathcal F_{t-1}]
 &=\frac{1}N\sum_i f_tu(X_{t-1}^i)=\eta_{t-1}^N(f_tu),\\
 \Var_{TB}(q_t^N(u)\mid\mathcal F_{t-1})
 &=\frac{1}{N^2}\sum_i\left\{f_t(u^2)(X_{t-1}^i)
                         -[f_tu(X_{t-1}^i)]^2\right\}
 \le\frac{\|u\|_\infty^2}{N}.
\end{align*}
For any $\varepsilon>0$, conditional Chebyshev and the tower property give
\begin{align*}
 \mathbb P_{TB}(|q_t^N(u)-q_t(u)|>\varepsilon)
 &\le \mathbb P_{TB}\left(|q_t^N(u)-\eta_{t-1}^N(f_tu)|>\varepsilon/2\right)
+\mathbb P_{TB}\left(|\eta_{t-1}^N(f_tu)-\widetilde\pi_{t-1}(f_tu)|>\varepsilon/2\right)\\
 &\le\frac{4\|u\|_\infty^2}{N\varepsilon^2}+\mathbb P_{TB}\left(
 |\eta_{t-1}^N(f_tu)-\widetilde\pi_{t-1}(f_tu)|>\varepsilon/2\right)
 \longrightarrow0.
\end{align*}
The last term tends to zero since $f_tu\in C_b(\cX)$ by the Feller assumption for this fixed $t$. 
The countable-family argument above therefore yields
 \DIFaddbegin \DIFadd{$q_t^N\xrightarrow{\mathrm{weak},\mathbb P}q_t$}\DIFaddend.

\paragraph{Step 3. Block weighting.}\label{supp:block-weighting-notation}
Let $p_K:\cX\to\cX^K$ be the coordinate projection. For
$\varphi\in C_b(\cX^K)$, the definitions in \eqref{eq:weights} give
\[
 a_{t,N}^{K}=(q_t^N)^K(g_t^K)\qquad\text{\normalfont and}\qquad \nu_t^K(\varphi)
 =\frac{q_t^N((g_t^K\varphi)\circ p_K)}{q_t^N(g_t^K\circ p_K)}.
\]
Since $g_t^v\in C_b(\cX^v)$ for every $v\in K$ and $K$ is finite,
\[
 g_t^K=\prod_{v\in K}g_t^v\in C_b(\cX^K)\qquad\text{\normalfont and}\qquad \|g_t^K\|_\infty\le\prod_{v\in K}\|g_t^v\|_\infty<\infty.
\]
Thus $(g_t^K\varphi)\circ p_K$ and $g_t^K\circ p_K$ both belong to
$C_b(\cX)$. Define the deterministic likelihood increment
$a_{t,*}^K:=q_t^K(g_t^K)>0$. The block recursion \eqref{eq:block} gives
\[
 \widetilde\pi_t^K(\varphi)=\frac{q_t^K(g_t^K\varphi)}{a_{t,*}^K}.
\]
The prediction step established
$q_t^N(u)\xrightarrow{\mathbb P}q_t(u)$ for every fixed $u\in C_b(\cX)$.
Applying this statement with $u=(g_t^K\varphi)\circ p_K$ and
$u=g_t^K\circ p_K$, respectively, gives
\begin{align*}
 (q_t^N)^K(g_t^K\varphi)
 =q_t^N((g_t^K\varphi)\circ p_K)
 \xrightarrow{\mathbb P}q_t((g_t^K\varphi)\circ p_K)
 =q_t^K(g_t^K\varphi),\\
 a_{t,N}^{K}=(q_t^N)^K(g_t^K)
 =q_t^N(g_t^K\circ p_K)
 \xrightarrow{\mathbb P}q_t(g_t^K\circ p_K)
 =q_t^K(g_t^K)=a_{t,*}^K.
\end{align*}
Here the equalities use the definition of the $K$-marginal through the
coordinate projection $p_K$.
Since $a_{t,*}^K>0$ and $a_{t,N}^{K}\xrightarrow{\mathbb P}a_{t,*}^K$,
\[
 \mathbb P_{TB}\left(a_{t,N}^{K}<a_{t,*}^K/2\right)
 \le\mathbb P_{TB}\left(|a_{t,N}^{K}-a_{t,*}^K|>a_{t,*}^K/2\right)
 \longrightarrow0.
\]
On the event $a_{t,N}^{K}\ge a_{t,*}^K/2$, subtracting the two ratios gives
\begin{align*}
 |\nu_t^K(\varphi)-\widetilde\pi_t^K(\varphi)|
 &=\left|\frac{(q_t^N)^K(g_t^K\varphi)}{a_{t,N}^{K}}
       -\frac{q_t^K(g_t^K\varphi)}{a_{t,*}^K}\right|\\
 &\le \frac{|(q_t^N)^K(g_t^K\varphi)-q_t^K(g_t^K\varphi)|}{a_{t,N}^{K}}
 +\frac{|q_t^K(g_t^K\varphi)|}{a_{t,N}^{K} a_{t,*}^K}|a_{t,N}^{K}-a_{t,*}^K|\\
 &\le\frac{2}{a_{t,*}^K}
       |(q_t^N)^K(g_t^K\varphi)-q_t^K(g_t^K\varphi)|+\frac{2\|\varphi\|_\infty}{a_{t,*}^K}
       |a_{t,N}^{K}-a_{t,*}^K|,
\end{align*}
where we used $|q_t^K(g_t^K\varphi)|\le a_{t,*}^K\|\varphi\|_\infty$.
Both terms on the right converge to zero in probability, and the event
on which this bound holds has probability tending to one. Hence
\[
 \nu_t^K(\varphi)\xrightarrow{\mathbb P}\widetilde\pi_t^K(\varphi),
 \qquad\text{\normalfont for every }\varphi\in C_b(\cX^K).
\]
Applying the countable-family argument on the Polish space $\cX^K$ yields
 \DIFaddbegin \DIFadd{$\nu_t^K\xrightarrow{\mathrm{weak},\mathbb P}\widetilde\pi_t^K$}\DIFaddend.
There are finitely many blocks, and taking finite products of probability
measures is continuous for weak convergence. Recalling the weighted output,
it follows that
 \DIFaddbegin \[
 \DIFadd{\widehat\pi_t^{TB}=\bigotimes_{K\in\cK}\nu_t^K
 \xrightarrow{\mathrm{weak},\mathbb P}
 \bigotimes_{K\in\cK}\widetilde\pi_t^K=\widetilde\pi_t.
}\]\DIFaddend 

\paragraph{Step 4. Twisted resampling and recombination.}
Condition on $\mathcal G_t$. For each block, draw independent ordinary children
$\bar X_t^{1,K},\ldots,\bar X_t^{N,K}\sim\nu_t^K$, independently across blocks.
Also draw independent uniform labels $J_t^K$ and independent replacements
$V_t^K$ with the tilted law introduced in the distinguished-child construction,
\[
 \nu_{t,\psi}^K(dz^K)
 =\frac{\psi_t^K(z^K)}{\nu_t^K(\psi_t^K)}\nu_t^K(dz^K).
\]
The assumed positive finite guide values make this a probability law. Set
\[
 X_t^{i,K}=\begin{cases}
 V_t^K,&i=J_t^K,\\
 \bar X_t^{i,K},&i\ne J_t^K.
 \end{cases}
\]
By Proposition~\ref{supp:prop:one}, this construction has exactly the TBPF
resampling law. The ordinary recombined rows
$\bar X_t^i=(\bar X_t^{i,K})_{K\in\cK}$ are conditionally i.i.d. with law
$\widehat\pi_t^{TB}$. For $\bar\eta_t^N=N^{-1}\sum_i\delta_{\bar X_t^i}$,
\[
 \E_{TB}[\bar\eta_t^N(u)\mid\mathcal G_t]=\widehat\pi_t^{TB}(u)\qquad\text{\normalfont and}\qquad \E_{TB}\left[|\bar\eta_t^N(u)-\widehat\pi_t^{TB}(u)|^2
                 \mid\mathcal G_t\right]\le\frac{\|u\|_\infty^2}{N}.
\]
Only rows with indices in $\{J_t^K:K\in\cK\}$ can change, so, pathwise,
\[
 |\eta_t^N(u)-\bar\eta_t^N(u)|
 \le\frac{1}N\sum_{i\in\{J_t^K:K\in\cK\}}
             |u(X_t^i)-u(\bar X_t^i)|
 \le\frac{2|\cK|\|u\|_\infty}{N}.
\]
Combining this with the preceding conditional variance bound yields
\begin{align*}
 \E_{TB}|\eta_t^N(u)-\widehat\pi_t^{TB}(u)|^2
 &\le 2\E_{TB}|\eta_t^N(u)-\bar\eta_t^N(u)|^2
       +2\E_{TB}|\bar\eta_t^N(u)-\widehat\pi_t^{TB}(u)|^2\\
 &\le\frac{8|\cK|^2\|u\|_\infty^2}{N^2}
       +\frac{2\|u\|_\infty^2}{N}\longrightarrow0.
\end{align*}
Here the first inequality uses $|a+b|^2\le2|a|^2+2|b|^2$, and the second
uses the pathwise bound and the tower property for the conditional variance bound.
Since
\[
 \eta_t^N(u)-\widetilde\pi_t(u)
 =[\eta_t^N(u)-\widehat\pi_t^{TB}(u)]
  +[\widehat\pi_t^{TB}(u)-\widetilde\pi_t(u)],
\]
we obtain convergence in probability for each $u\in C_b(\cX)$ and hence
 \DIFaddbegin \DIFadd{$\eta_t^N\xrightarrow{\mathrm{weak},\mathbb P}\widetilde\pi_t$}\DIFaddend.
This completes the induction. 

\paragraph{Step 5. Cumulative correction and corrected normalizer.}
Now assume the uniform constants  \DIFaddbegin \DIFadd{$\rho_t^K$ }\DIFaddend exist and set
 \DIFaddbegin \[
 \DIFadd{S:=\sum_{t=1}^T\sum_{K\in\cK}(\sqrt{\rho_t^K}-1)^2<\infty.
}\]\DIFaddend 
Theorem~\ref{supp:thm:rn} gives $\E_{TB}D_T=1$, and
Proposition~\ref{supp:prop:second} gives $\E_{TB}D_T^2\le e^{S/N}$. Therefore
\[
 \|D_T-1\|_2^2
 =\E_{TB}D_T^2-2\E_{TB}D_T+1
 =\E_{TB}D_T^2-1
 \le e^{S/N}-1\longrightarrow0.
\]

To establish convergence of the corrected normalizer, first consider the product of block weights. The block-weighting step established $a_{t,N}^{K}\to a_{t,*}^K$ in
probability. Since the product contains finitely many factors,
\[
 F_N:=\prod_{t=1}^T\prod_{K\in\cK}a_{t,N}^{K}
 \xrightarrow{\mathbb P}
 \prod_{t=1}^T\prod_{K\in\cK}a_{t,*}^K=Z_T^{\mathrm{block}}.
\]
Bounded likelihoods imply
\[
 0<F_N\le M:=\prod_{t=1}^T\prod_{K\in\cK}\|g_t^K\|_\infty<\infty\qquad\text{\normalfont and}\qquad 0<Z_T^{\mathrm{block}}\le M.
\]
For every $\varepsilon>0$ this deterministic bound gives
\begin{align*}
 \E_{TB}|F_N-Z_T^{\mathrm{block}}|^2
 &=\E_{TB}\left[|F_N-Z_T^{\mathrm{block}}|^2
       \mathbf 1_{\{|F_N-Z_T^{\mathrm{block}}|\le\varepsilon\}}\right]
 +\E_{TB}\left[|F_N-Z_T^{\mathrm{block}}|^2
       \mathbf 1_{\{|F_N-Z_T^{\mathrm{block}}|>\varepsilon\}}\right]\\
 &\le\varepsilon^2+4M^2\mathbb P_{TB}
          (|F_N-Z_T^{\mathrm{block}}|>\varepsilon).
\end{align*}
Here we used $|F_N-Z_T^{\mathrm{block}}|\le2M$ on the second event.
Taking $N\to\infty$ and then $\varepsilon\downarrow0$ proves
$\|F_N-Z_T^{\mathrm{block}}\|_2\to0$.
 Recalling 
\[
\widehat Z_T^{TB}
=\prod_{t=1}^T\prod_{K\in\cK}
\left[a_{t,N}^{K}\frac{c_t^K}{H_t^K(X_t)}\right]
=D_T\prod_{t=1}^T\prod_{K\in\cK}a_{t,N}^{K},
\]
we have $\widehat Z_T^{TB}=D_TF_N$.
Finally, 
\[
 \|\widehat Z_T^{TB}-Z_T^{\mathrm{block}}\|_2
 \le\|(D_T-1)F_N\|_2+\|F_N-Z_T^{\mathrm{block}}\|_2 \le M\sqrt{e^{S/N}-1}+\|F_N-Z_T^{\mathrm{block}}\|_2
 \longrightarrow0.
\]

\end{proof}
\begin{remark}[Different assumptions for filtering and normalizers]\label{supp:rem:consistency-scope}
Filtering consistency here is weak convergence in probability even for $N$-dependent positive guides whose tilted law concentrates on a single proposal particle: only one child per block is replaced. The $L^2$ conclusions for $D_T$ and $\widehat Z_T^{TB}$ additionally require deterministic range bounds uniform in $N$, histories, and pools. 
\end{remark} 

\appendixsection{Detailed experimental models and comparisons}\label{supp:sec:experiments}
Section~\ref{supp:subsec:models} specifies the coupled models and campaign design, Section~\ref{supp:subsec:guide-design} details the TBPF guide construction, and Section~\ref{supp:subsec:tpf} defines the TPF comparator. Section~\ref{supp:subsec:validation} summarizes simulation results and method comparisons, and Section~\ref{supp:sec:real_data} describes the real panel.

\subsection{Coupled models and design}\label{supp:subsec:models}
\citet[Section 4.4]{whiteley2014} studied an LG model with independent standard Gaussian noise and a scalar SV model for pound/dollar returns. Their variance-growth figures use 10,000 independent filter runs for each fixed observation sequence. Motivated by those examples, our simulation study constructs spatially coupled, high-dimensional LG and SV models, specified below. These are new simulation settings rather than reproductions of the original exchange-rate dataset, fitted parameters, or Monte Carlo precision. 
We compare the bootstrap PF, BPF, the TPF (Algorithm~2 of \citet{whiteley2014}), and the resampling-stage TBPF. The blocked methods use block sizes 1, 2, 3, and 4; PF and TPF operate on the full state.

Let $v\in\{1,\ldots,d\}$ index sites on a periodic ring, with
$X_t^0=X_t^d$ and $X_t^{d+1}=X_t^1$. Both simulation models use the latent transition
 \DIFaddbegin \begin{equation}\DIFadd{\label{supp:eq:model}
 X_t^v=(0.9-2\kappa_{\mathrm{sp}})X_{t-1}^v+\kappa_{\mathrm{sp}}(X_{t-1}^{v-1}+X_{t-1}^{v+1})+\sigma\epsilon_t^v,
 \quad X_0\sim\mathcal N(0,I_d),\quad\kappa_{\mathrm{sp}}=0.1.
}\end{equation}\DIFaddend 
The parameter $\kappa_{\mathrm{sp}}$ controls spatial interaction strength. The transition mean is equivalently
$0.9X_{t-1}^v+\kappa_{\mathrm{sp}}(X_{t-1}^{v-1}-2X_{t-1}^v+X_{t-1}^{v+1})$:
the second term adds nearest-neighbor spatial coupling to the scalar autoregression.
If $\kappa_{\mathrm{sp}}=0$ and $\sigma=1$, the coordinates become independent copies of
the scalar latent model of \citet{whiteley2014}, $X_t=0.9X_{t-1}+V_t$, $Y_t=X_t+W_t$, where $V_t$ and $W_t$ are mutually independent standard Gaussian noises, independent across times and of the initial state. Our experiments instead fix $\kappa_{\mathrm{sp}}=0.1$,
so the coefficients of the site's own state and its two neighbors are $0.7$, $0.1$,
and $0.1$, respectively. Thus these experiments extend the scalar model;
they do not use its uncoupled transition.

\paragraph{LG model.}
We use \eqref{supp:eq:model} with $\sigma=1$ and the observation equation
 \DIFaddbegin \begin{equation}\DIFadd{\label{supp:eq:lg-observation}
 Y_t^v=X_t^v+e_t^v\qquad\text{\normalfont and}\qquad Y_t^v\mid X_t\sim\mathcal N(X_t^v,1).
}\end{equation}\DIFaddend 
Together, \eqref{supp:eq:model} and \eqref{supp:eq:lg-observation} form an LG
state-space model. Setting $\kappa_{\mathrm{sp}}=0$ would recover independent copies of both
equations in the original scalar Gaussian example.

\paragraph{SV model.}
The full simulation model is
 \DIFaddbegin \begin{equation}\DIFadd{\label{supp:eq:sv-model}
 \begin{aligned}
 X_t^v&=0.7X_{t-1}^v+0.1(X_{t-1}^{v-1}+X_{t-1}^{v+1})+0.3\epsilon_t^v,\\
 Y_t^v&=\exp(X_t^v/2)e_t^v,
 \qquad Y_t^v\mid X_t\sim\mathcal N(0,\exp(X_t^v)).
 \end{aligned}
}\end{equation}\DIFaddend 
Here $X_t^v$ is the latent log variance, and $\exp(X_t^v/2)$ is the conditional
standard deviation of the observation. The latent transition is
\eqref{supp:eq:model} with $\sigma=0.3$, but the observation equation differs from
the LG model. This is a new spatial SV design motivated by the scalar SV
example.
In both models, the innovations $\epsilon_t^v$ and observation noises $e_t^v$
are independent standard normal variables across sites and times, independent
of one another and of $X_0$. Spatial coupling is retained in every simulation
and filtering transition.

The  \DIFaddbegin \DIFadd{simulation }\DIFaddend campaign uses only $d\in\{64,128,256\}$, with $N\in\{128,512\}$, nominal block sizes $b\in\{1,2,3,4\}$, $T=80$, and 32 filter replications per configuration on one fixed dataset per model and dimension. At every block size it compares BPF with TBPF at look-ahead horizons 1 and 5. The 4,608  block-filter runs are retained. PF and TPF at horizons 1 and 5 add 1,152 runs on exactly the same datasets, giving 5,760 unique runs and 180 method/configuration combinations. The design includes 48 model/dimension/particle-count/block-size settings, each evaluated at both guide horizons. Sharing each global baseline across block-size comparisons yields 96 horizon-matched comparisons without double-counting global runs. 
Contiguous blocks start at vertex 1. When $b=3$, the last block contains 1, 2, and 1 vertices at dimensions 64, 128, and 256, respectively. Thus $b$ denotes the maximum block size, and these final smaller blocks are included in every calculation. Block sizes 1, 2, and 4 divide all chosen dimensions exactly. Observations and latent truth are identical across particle counts, block sizes, and methods for a given model and dimension.

For LG models the full Kalman recursion supplies the exact likelihood and filtering mean. We record filtering RMSE against the Kalman mean, state RMSE, mean pre-resampling effective sample size (ESS) divided by $N$, log likelihood-functional estimates and their trajectories. In SV, state RMSE is a simulation diagnostic.  \DIFaddbegin \DIFadd{Log-estimate variance differs }\DIFaddend from ordinary relative variance . The empirical squared  \DIFaddbegin \DIFadd{coefficient of variation (CV) }\DIFaddend is
\begin{equation}
 \widehat{\mathrm{CV}}^2=\frac{n_{\mathrm{run}}}{n_{\mathrm{run}}-1}\left\{
 n_{\mathrm{run}}\frac{\sum_{r=1}^{n_{\mathrm{run}}}\exp(2L_{T,r})}{[\sum_{r=1}^{n_{\mathrm{run}}}\exp(L_{T,r})]^2}-1\right\},
\end{equation}
 \DIFaddbegin \DIFadd{where $n_{\mathrm{run}}=32$ is the replication count and $L_{T,r}$ the terminal log estimate in run $r$. With normalized contributions $p_r=e^{L_{T,r}}/\sum_s e^{L_{T,s}}$, $1/n_{\mathrm{run}}\le\sum_rp_r^2\le1$, hence $0\le\widehat{\mathrm{CV}}^2\le n_{\mathrm{run}}$.  }
\DIFaddbegin \DIFadd{Let $\widehat m_{t,r}^v$ be the weighted particle estimate of coordinate $v$ at time $t$ in run $r$, and let $m_{t,\mathrm{ref}}^v$ be its reference value. The reported error is $\mathrm{RMSE}_r=[(Td)^{-1}\sum_{t,v}(\widehat m_{t,r}^v-m_{t,\mathrm{ref}}^v)^2]^{1/2}$, averaged across runs after taking the square root. The reference is the exact Kalman mean for LG (filtering approximation RMSE) and latent truth for SV (latent-state reconstruction RMSE). Recorded block ESS is $[N\sum_i(w_t^{i,K})^2]^{-1}$ averaged over blocks and times; global ESS uses full-state weights and is a different diagnostic.
}

\DIFaddend \subsection{Guide design}\label{supp:subsec:guide-design}
The TBPF guide approximates the local backward likelihood in \eqref{eq:lookahead-guide}, using the next $\ell$ block observations, truncated at $T$. The current observation already enters the weights $w_t^{i,K}$ and is not reused in the guide. Although the filtering target conditions only on observations through $t$, guide construction uses future data and requires up to $\ell$ steps of look-ahead.

\DIFadd{We construct sitewise backward messages under independent scalar dynamics with autoregression $a=0.9$, approximating spatial interactions only when building the guide. For each site $v\in V$ and window endpoint $\tau$, let $h_{s,\tau}^v(x)$ approximate the likelihood of observations $y_{s+1:\tau}^v$ given scalar state $x$ at time $s$. Set $h_{\tau,\tau}^v(x)=1$. To construct the time-$t$ TBPF guide, use $\tau=\min(T,t+\ell)$ and recurse backward to $s=t$.

For LG, with unit process and observation variances, write $\log h_{s,\tau}^v(x)=-A_{s,\tau}^v x^2/2+B_{s,\tau}^v x+C_{s,\tau}^v$, where $x\in\mathbb R$ and $C_{s,\tau}^v$ is independent of $x$. Starting from $A_{\tau,\tau}^v=B_{\tau,\tau}^v=0$, the recursion for $s=\tau-1,\ldots,t$ is}
\begin{equation}\label{eq:lg-guide-recursion}
 \DIFadd{A_{s,\tau}^v=\frac{a^2(1+A_{s+1,\tau}^v)}{2+A_{s+1,\tau}^v}\qquad\text{\normalfont and}\qquad B_{s,\tau}^v=\frac{a(y_{s+1}^v+B_{s+1,\tau}^v)}{2+A_{s+1,\tau}^v}.}
\end{equation}
For SV,  \DIFaddbegin \DIFadd{retain the site likelihood $g_s^v$ and compute
}\[
 \DIFadd{h_{s,\tau}^v(x)=\int\mathcal N(dz;0.9x,0.3^2)g_{s+1}^v(z)h_{s+1,\tau}^v(z),\qquad v\in V,\quad s=\tau-1,\ldots,t.
}\]
\DIFadd{With }\DIFaddend 24-point Gauss--Hermite  \DIFaddbegin \DIFadd{nodes $z_j,w_j$ for the weight $e^{-z^2}$, the integral uses arguments $0.9x+\sqrt2(0.3)z_j$ and weights $w_j/\sqrt\pi$. The code interpolates log messages linearly on 401 points in }\DIFaddend $[-8,8]$,  \DIFaddbegin \DIFadd{extends endpoint values constantly, and subtracts each site's grid maximum after every backward step. Computation uses log-sum-exp. The effective horizon is $\min(\ell,T-t)$.
}\DIFaddend 

Block log guides are sums of the site log messages \DIFadd{$\sum_{v\in K}\log h_{t,\min(T,t+\ell)}^v(x^v)$}. Within each block proposal pool, we subtract the largest log-guide value and clip the result below at $-20$. The resulting guide values lie in $[e^{-20},1]$, giving a deterministic range ratio at most $e^{20}$. These same values enter both the tilted law $\nu_{t,\psi}^K=\psi_t^K\nu_t^K/c_t^K$ and the correction $c_t^K/H_t^K(X_t)$. This range bound meets the correction-stability assumption, although its worst-case bound does not predict the measured variance reductions. Particle propagation and observation weighting retain the coupled model.

\subsection{TPF comparator}\label{supp:subsec:tpf}
The TPF implements Algorithm 2 of \citet{whiteley2014}, twisting both ancestor selection and propagation for one uniformly selected offspring. Write $Z_t^i$ for predictive particles and $w_t^i\propto g_t(Z_t^i)$ for their normalized observation weights. For the arrival-time guide $\psi_{t+1}$, define $(f_{t+1}\psi_{t+1})(x)=\int f_{t+1}(x,dz)\psi_{t+1}(z)$. Then
 \DIFaddbegin \[
 \DIFadd{q_t^{\mathrm{TPF}}(dz)=\sum_i w_t^i f_{t+1}(Z_t^i,dz)\qquad\text{\normalfont and}\qquad c_{t+1}=\sum_i w_t^i(f_{t+1}\psi_{t+1})(Z_t^i).
}\]\DIFaddend 
The $N-1$ ordinary offspring follow  \DIFaddbegin \DIFadd{$q_t^{\mathrm{TPF}}$}\DIFaddend. For the distinguished offspring, the parent probability is proportional to $w_t^i(f_{t+1}\psi_{t+1})(Z_t^i)$ and its propagation density is proportional to $f_{t+1}(Z_t^i,z)\psi_{t+1}(z)$. The correction multiplying the usual bootstrap normalizer is
\[
 \frac{c_{t+1}}{N^{-1}\sum_j\psi_{t+1}(Z_{t+1}^j)}.
\]
Initialization is the ordinary first prediction from $\pi_0$ using $f_1$; no unused terminal propagation is performed. Filtering means use the observation-weighted predictive particles. PF is the same full-state bootstrap procedure without the twist.

For both models we use a Gaussian-shaped guide
\[
 \psi_t(z)=\exp\left\{-\frac{1}{2}\sum_v a_t^v(z^v-m_t^v)^2\right\},\qquad a_t^v>0.
\]
The guide at arrival $t$ uses observations $y_{t:t+\ell-1}$, so both TPF and TBPF use $\ell$ future observations when departing from the preceding time. For LG, its coefficients follow exact isolated-site Gaussian backward recursion including the arrival observation. This is not the exact coupled-model backward likelihood. For SV, we add the arrival log likelihood to the isolated-site grid message, then fit a local quadratic at its grid mode using centered first and second differences; precision is floored at $10^{-4}$. \DIFaddbegin \DIFadd{For each arrival time $t$ and site $v$, use the common grid $x_k=-8+k\Delta$, $k=0,\ldots,400$, with $\Delta=0.04$, and define
}\[
 \lambda_{t,k}^v=\log g_t^v(x_k)+\log h_{t,\min(T,t+\ell-1)}^v(x_k).
\]
\DIFadd{Let $k_t^v\in\{1,\ldots,399\}$ be the grid maximizer clamped to this interior index range. Define the slope, floored precision, and center by
}
 \begin{align*}
 a_t^v=\max\!\left\{-\frac{\lambda_{t,k_t^v+1}^v-2\lambda_{t,k_t^v}^v+\lambda_{t,k_t^v-1}^v}{\Delta^2},10^{-4}\right\},\\
  b_{t,\mathrm{quad}}^v=\frac{\lambda_{t,k_t^v+1}^v-\lambda_{t,k_t^v-1}^v}{2\Delta},\qquad m_t^v=x_{k_t^v}+\frac{b_{t,\mathrm{quad}}^v}{a_t^v}.
 \end{align*}
\DIFadd{The same precision $a_t^v$ is used in the center and the quadratic curvature. The }\DIFaddend original block-filter guide and its pool-dependent clipping are retained for TBPF.

Integration and tilted sampling are exact for the chosen quadratic guide under the actual coupled transition. For propagation from $t$ to $t+1$, consider a parent $x$ with \DIFadd{sitewise transition mean $\mu_{t+1}^v(x)=0.7x^v+0.1(x^{v-1}+x^{v+1})$} and let $\kappa_{t+1}^v=1+\sigma^2a_{t+1}^v$. Then
 \DIFaddbegin \begin{align*}
 \DIFadd{\log\!\int\psi_{t+1}(z)f_{t+1}(x,dz)}&\DIFadd{=-\frac{1}{2}\sum_v\left[\log \kappa_{t+1}^v+\frac{a_{t+1}^v(\mu_{t+1}^v(x)-m_{t+1}^v)^2}{\kappa_{t+1}^v}\right].}
\end{align*}
\DIFadd{For the distinguished offspring $j$, conditional on its parent state $x$, each coordinate $v\in V$ has distribution
}\[
 \DIFadd{Z_{t+1}^{j,v}\sim
 \mathcal N\left(\frac{\mu_{t+1}^v(x)+\sigma^2a_{t+1}^vm_{t+1}^v}{\kappa_{t+1}^v},\frac{\sigma^2}{\kappa_{t+1}^v}\right).
}\]\DIFaddend 
Thus guide approximation does not introduce a missing integration correction into the TPF. PF and this TPF have unbiased full-model likelihood estimates, while the blocked methods retain their block-functional targets. The procedures use different guide families, which is reported rather than attributing every difference solely to blocking.

\subsection{Simulation results and method comparisons}\label{supp:subsec:validation}

\subsubsection{ \DIFaddbegin \DIFadd{Retrospective minima over the tested configurations}\DIFaddend }\label{supp:nav:which-method-works-best}

Table~\ref{supp:tab:best_methods}  \DIFaddbegin \DIFadd{summarizes retrospective minima in }\DIFaddend the 5,760-run simulated-data comparison . Within each model, dimension, and particle count, we find the measured minimum across the 15 tested procedures (PF, two TPF horizons, four BPF block sizes, and eight TBPF combinations). The six settings per model receive equal weight; we do not average raw errors across dimensions.

\begin{table}[htbp]\color{black}\centering\small
\caption{\color{black}Observed minima over the tested configurations. Each model has six settings: three dimensions ($64,128,256$) and two particle counts ($128,512$). Choices are selected retrospectively from all tested block sizes and horizons. ``Best'' means the smallest observed metric. Lower variance or RMSE is better; higher best-setting counts favor the listed method.}\label{supp:tab:best_methods}
\setlength{\tabcolsep}{4pt}\renewcommand{\arraystretch}{1.2}
\begin{tabular}{@{}p{.28\linewidth}p{.25\linewidth}p{.43\linewidth}@{}}\toprule
Situation & Observed minimum & Evidence / qualification\\\midrule
LG: lowest log variance & \textcolor{tbgreen}{\textbf{TBPF}}, $b=1$ & Best in 5/6 settings; BPF wins at $d=256,N=128$.\\
SV: lowest log variance & \textcolor{tbgreen}{\textbf{TBPF}}, $\ell=5$ & Best in 6/6 settings; optimal $b$ ranges from 1 to 3.\\
LG: filtering accuracy & BPF or TBPF, $b=1$ & The best BPF and TBPF errors differ by less than 0.6\%.\\
SV: filtering accuracy & BPF or TBPF, $b=2$--$4$ & The best BPF and TBPF errors differ by less than 0.2\%.\\
\bottomrule\end{tabular}
\end{table}

The clearest pattern is that TBPF often minimizes log-estimate variance, while the best BPF and TBPF configurations have very similar filtering error.

\subsubsection{Graphical comparison}\label{supp:nav:graphical-comparison}
\begin{figure}[!htb]
	\centering
	 \DIFaddbeginFL \includegraphics[width=0.8\textwidth,height=.58\textheight,keepaspectratio]{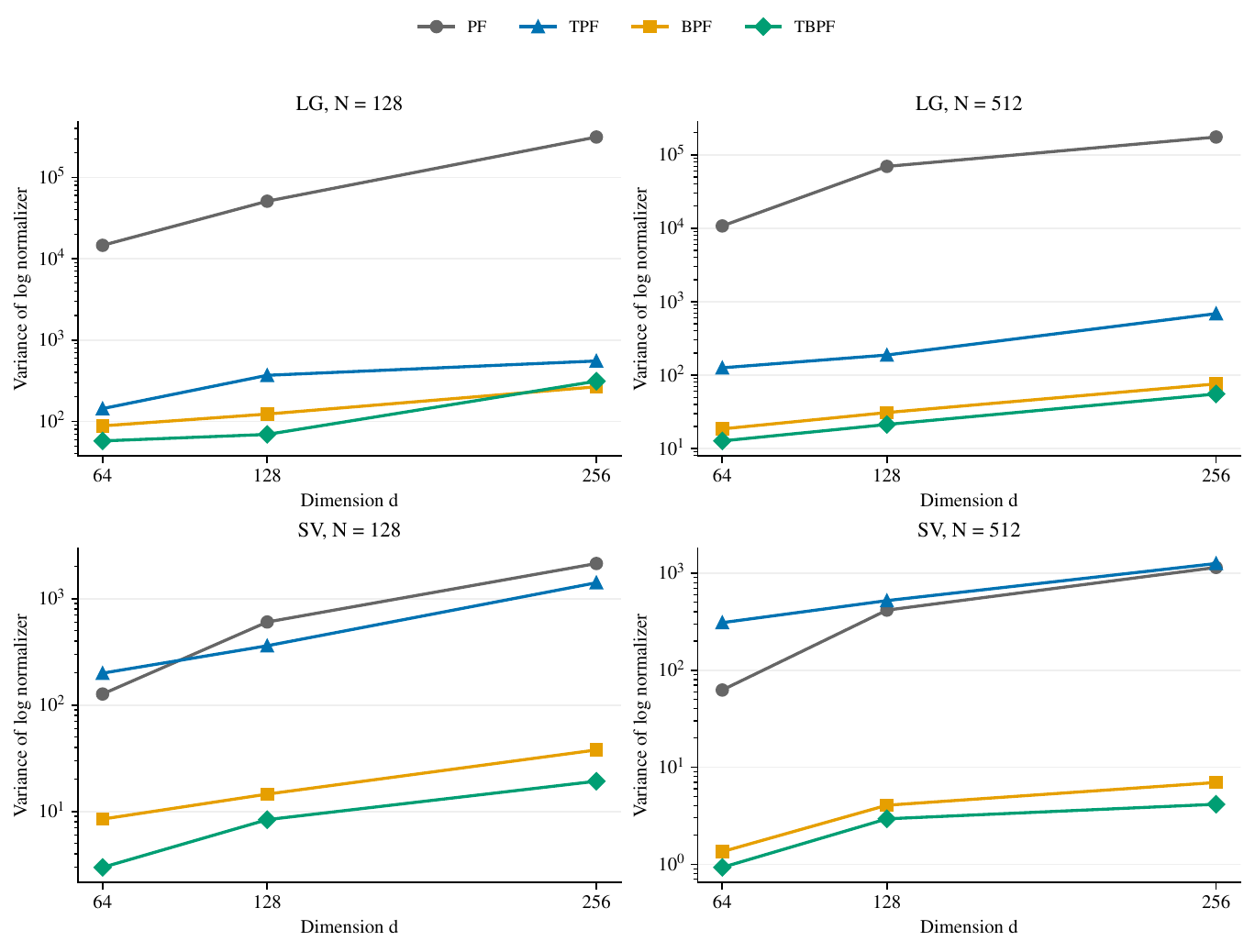}
	\DIFaddendFL \caption{ \DIFaddbeginFL \DIFaddFL{Retrospective log-normalizer }\DIFaddendFL variance \DIFaddbeginFL \DIFaddFL{envelopes }\DIFaddendFL versus dimension, with separate panels for model and particle count. The vertical axes are logarithmic. Each family is represented by its smallest observed variance over its tested block sizes and horizons: PF has one configuration, TPF two, BPF four, and TBPF eight. Lines connect these retrospective minima. PF and TPF estimate the full-model likelihood, while BPF and TBPF use block-functional targets. Lower variance is better.}
	\label{supp:fig:variance_dimension}
\end{figure}

\begin{figure}[!htb]
\centering
 \DIFaddbeginFL \includegraphics[width=\textwidth,height=.58\textheight,keepaspectratio]{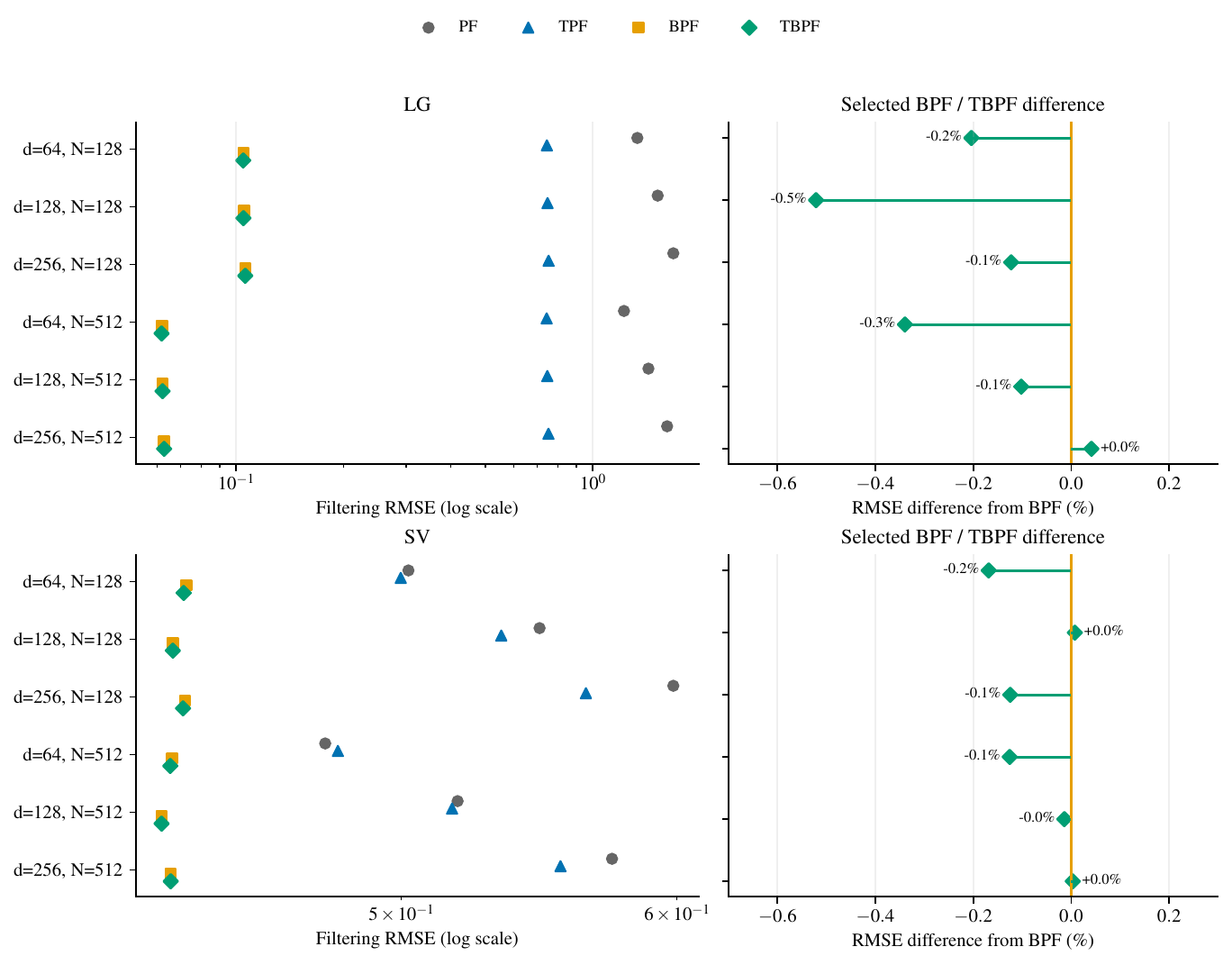}
\DIFaddendFL \caption{ \DIFaddbeginFL \DIFaddFL{Retrospective filtering-RMSE selection summaries }\DIFaddendFL in all twelve settings. Left: minimum mean filtering RMSE within each method family, using a logarithmic horizontal axis. LG errors are relative to the exact Kalman mean; SV errors are relative to the simulated latent state. Right: $100(\mathrm{RMSE}_{\mathrm{TBPF}}/\mathrm{RMSE}_{\mathrm{BPF}}-1)$ for the separately selected family minima. Lower values favor TBPF, with negative values indicating lower RMSE than BPF. The connecting segments show differences from the BPF reference. Lower RMSE is better in the left panel.}
\label{supp:fig:filtering_accuracy}
\end{figure}

Figures~\ref{supp:fig:variance_dimension} and~\ref{supp:fig:filtering_accuracy} give complementary views of the  \DIFaddbegin \DIFadd{simulation }\DIFaddend experiments. Each setting is a model, dimension, and particle count; there are six settings per model. The line and accuracy plots select each family's best tested configuration separately for the displayed metric, so the selected block size or horizon can change between points and figures. These retrospective minima describe the tested procedures rather than a single tuning rule. The main-text heatmap (Figure~\ref{fig:block_horizon_heatmap}) retains all block sizes and both horizons without selection. All plotted estimates use 32 replications on each fixed dataset.
 \DIFaddbegin \DIFadd{These plots are retrospective envelopes, not a fixed tuning rule. Table~\ref{supp:tab:selected-configurations} identifies every selected block size and horizon. }\DIFaddend

\par

\DIFaddbegin \begin{table}[ht]\color{black}\centering\small
\caption{\color{black}Selected configurations behind the retrospective envelopes in Figures~\ref{supp:fig:variance_dimension}--\ref{supp:fig:filtering_accuracy}. V denotes log variance and E denotes filtering RMSE. PF has one candidate; TPF, BPF, and TBPF have two, four, and eight. Each row selects within each family for the specified metric.}\label{supp:tab:selected-configurations}
\begin{tabular}{lrrlrrr}\toprule Model & $d$ & $N$ & Metric & TPF $\ell$ & BPF $b$ & TBPF $(b,\ell)$\\\midrule
LG & 64 & 128 & V & 5 & 2 & (1,1)\\
LG & 64 & 128 & E & 5 & 1 & (1,5)\\
LG & 128 & 128 & V & 5 & 1 & (1,1)\\
LG & 128 & 128 & E & 5 & 1 & (1,1)\\
LG & 256 & 128 & V & 5 & 1 & (2,1)\\
LG & 256 & 128 & E & 5 & 1 & (1,5)\\
LG & 64 & 512 & V & 5 & 1 & (1,1)\\
LG & 64 & 512 & E & 5 & 1 & (1,5)\\
LG & 128 & 512 & V & 5 & 1 & (1,5)\\
LG & 128 & 512 & E & 5 & 1 & (1,5)\\
LG & 256 & 512 & V & 5 & 1 & (1,1)\\
LG & 256 & 512 & E & 5 & 1 & (1,5)\\
SV & 64 & 128 & V & 5 & 2 & (1,5)\\
SV & 64 & 128 & E & 5 & 2 & (4,5)\\
SV & 128 & 128 & V & 5 & 1 & (2,5)\\
SV & 128 & 128 & E & 5 & 2 & (2,5)\\
SV & 256 & 128 & V & 5 & 1 & (1,5)\\
SV & 256 & 128 & E & 5 & 4 & (4,5)\\
SV & 64 & 512 & V & 5 & 4 & (3,5)\\
SV & 64 & 512 & E & 1 & 4 & (4,5)\\
SV & 128 & 512 & V & 5 & 2 & (2,5)\\
SV & 128 & 512 & E & 5 & 3 & (3,5)\\
SV & 256 & 512 & V & 5 & 1 & (2,5)\\
SV & 256 & 512 & E & 5 & 4 & (4,1)\\
\bottomrule\end{tabular}\end{table}
\DIFaddend \subsection{Real data: a 100-dimensional portfolio panel}\label{supp:sec:real_data}
We additionally evaluate all four procedures on the 100 value-weighted daily size/book-to-market portfolios from the Kenneth French Data Library \citep{FrenchData}. These portfolios are the intersections of ten size and ten book-to-market groups. We retain all 100 series in their published column order, using 503 trading days from 2 January 2018 to 31 December 2019 for preprocessing and the 253 trading days from 2 January to 31 December 2020 for evaluation. The largest cross-sectional median absolute standardized return occurs on 16 March 2020, reaching 11.04 training-period standard deviations. The selected panel contains no missing-value codes, and no dates or portfolios are dropped. The downloaded file identifies the August 2026 CRSP release.

Let $r_t^v$ be the published percentage return. Training-period means $\bar r^v$ and sample standard deviations $s^v$ define $u_t^v=(r_t^v-\bar r^v)/s^v$. The SV observation model takes $Y_t^v=u_t^v$ with the same latent ring dynamics and innovation scale $0.3$ as in the simulated SV experiment.  \DIFaddbegin \DIFadd{The self/neighbor coefficients $0.7,0.1,0.1$, }\DIFaddend the source-order ring, and the initial  \DIFaddbegin \DIFadd{$\mathcal N(0,I_{100})$ }\DIFaddend law are preset. The evaluation sequence starts from that common prior without a training-period filter warm-up.
The design has $d=100$, $N\in\{128,512\}$, $b\in\{1,2,3,4\}$ for blocked methods, and horizons $\ell\in\{1,5\}$ for twisted methods. There are 15 procedures per particle count and 32 independent random seeds, totaling 960 runs. All methods see identical transformed observations. The final block has size one when $b=3$. We report log-normalizer variance. No latent-state accuracy ranking is available for SV because the latent states are unobserved. Look-ahead guides use subsequent observations within the evaluation sequence. Differences between TPF and TBPF also include their guide constructions.

\begin{table}[ht]\color{black}\centering\small
\caption{\color{black}Observed best methods for the SV analysis of the real 100-portfolio panel. Configurations are selected separately for each objective from all tested block sizes and horizons. These are retrospective point-estimate rankings from 32 replications. Lower variance is better.}\label{supp:tab:real100_best}
\begin{tabular}{rl}\toprule $N$ & Lowest log variance\\\midrule
128 & TBPF ($b=1,\ell=5$)\\
512 & TBPF ($b=1,\ell=1$)\\
\bottomrule\end{tabular}\end{table}

TBPF has the smallest measured log-normalizer variance among all four methods at both particle counts. Its best tested variances are 837.53 and 259.42, versus the best BPF values 1048.52 and 593.66, at $N=128$ and $512$, respectively. These correspond to observed reductions of 20.1\% and 56.3\%. The comparisons select each family's best configuration separately and describe point estimates.

\par

\subsection{\DIFadd{Gaussian references, uncertainty, and implementation diagnostics}}\label{supp:revision-validation}
\paragraph{\DIFadd{Deterministic blocked Gaussian reference.}}\label{supp:gaussian-reference-notation}
\DIFadd{Let $A_{\mathrm{tr}}$ be the ring transition matrix, with diagonal $0.7$ and adjacent entries $0.1$, and let $m_t^{\mathrm{block}}$ and $\Sigma_t$ denote the blocked Gaussian filtering mean and covariance. Initialize $m_0^{\mathrm{block}}=0,\Sigma_0=I_d$. The predicted mean $m_{t,-}^{\mathrm{block}}$, predicted covariance $\Sigma_t^-$, block innovation covariance $S_t^K$, and block gain $G_t^K$ are given by
}\[
\DIFadd{m_{t,-}^{\mathrm{block}}=A_{\mathrm{tr}}m_{t-1}^{\mathrm{block}},\quad \Sigma_t^-=A_{\mathrm{tr}}\Sigma_{t-1}A_{\mathrm{tr}}^\mathsf T+I_d,
\quad S_t^K=(\Sigma_t^-)_{KK}+I_{|K|},\quad G_t^K=(\Sigma_t^-)_{KK}(S_t^K)^{-1}.
}\]
\DIFadd{Here $(\Sigma_t^-)_{KK}$ is the principal covariance submatrix indexed by block $K$, and $I_{|K|}$ is its identity matrix. Each block update is
}\[
\DIFadd{(m_t^{\mathrm{block}})^K=(m_{t,-}^{\mathrm{block}})^K+G_t^K[y_t^K-(m_{t,-}^{\mathrm{block}})^K],\qquad
\Sigma_t^K=(\Sigma_t^-)_{KK}-G_t^K(\Sigma_t^-)_{KK}.
}\]
\DIFadd{Recombine the means and set $\Sigma_t=\operatorname{blockdiag}_K\Sigma_t^K$. The product over $t,K$ of Gaussian densities with means $(m_{t,-}^{\mathrm{block}})^K$ and covariances $S_t^K$ gives $Z_T^{\mathrm{block}}$. A separate full Kalman recursion omits the covariance projection and gives the exact model likelihood $Z_T$. The implementation checks equality of both recursions when $b=d$.
}

\begin{table}[ht]\color{black}\centering\small
\caption{\color{black}Deterministic blocking discrepancy on the archived LG datasets. Mean RMSE compares block and full Kalman means; smaller values indicate less blocking error. Log differences compare $Z_T^{\mathrm{block}}$ with $Z_T$, not finite-particle bias.}\label{tab:blocked-reference}
\begin{tabular}{rrrr}\toprule $d$ & $b$ & Mean RMSE & $\log Z_T^{\mathrm{block}}-\log Z_T$\\\midrule
64 & 1 & 0.0376 & -3.427\\
64 & 2 & 0.0266 & -1.376\\
64 & 3 & 0.0220 & -2.403\\
64 & 4 & 0.0185 & 0.059\\
128 & 1 & 0.0378 & -8.004\\
128 & 2 & 0.0266 & -4.670\\
128 & 3 & 0.0218 & -4.865\\
128 & 4 & 0.0185 & -5.666\\
256 & 1 & 0.0382 & -32.666\\
256 & 2 & 0.0268 & -14.720\\
256 & 3 & 0.0220 & -14.830\\
256 & 4 & 0.0188 & -7.355\\
\bottomrule\end{tabular}\end{table}
\paragraph{\DIFadd{Retained-mean LG decomposition.}}\label{supp:decomposition-notation}
\DIFadd{A separate focused rerun retains every weighted particle mean for $d\in\{64,256\}$, $N\in\{128,512\}$ and $b\in\{1,4\}$, with BPF and TBPF at $\ell=5$, 16 common run seeds per pair, and the identical original observations. These 256 executions are distinct from the original campaign. Write $\widehat m_{t,r}$ for the vector of weighted particle means in run $r$, and $m_t^{\mathrm{block}}$ and $m_t^{\mathrm{full}}$ for the blocked and full Kalman means, respectively. For fixed observations and partition, these Gaussian reference means are deterministic. In particle-filter run $r$, define the particle error $e_{t,r}=\widehat m_{t,r}-m_t^{\mathrm{block}}$. The blocking error $b_t=m_t^{\mathrm{block}}-m_t^{\mathrm{full}}$ is common to all runs. The particle, blocking and total RMSEs are respectively
}\[
\DIFadd{\left(\frac{1}{Td}\sum_t\|e_{t,r}\|^2\right)^{1/2},\qquad
\left(\frac{1}{Td}\sum_t\|b_t\|^2\right)^{1/2},\qquad
\left(\frac{1}{Td}\sum_t\|e_{t,r}+b_t\|^2\right)^{1/2}.
}\]
\DIFadd{Table~\ref{supp:tab:lg-decomposition} averages each run's RMSE. In this subset, increasing $N$ reduces particle error, while larger blocks reduce deterministic blocking error but increase particle error. The same pattern appears at both dimensions. The two methods' particle and total RMSEs are close in these runs.
For each retained run, the exact squared-error identity is checked:
}\[
\DIFadd{\frac{1}{Td}\sum_t\|\widehat m_{t,r}-m_t^{\mathrm{full}}\|^2
=\frac{1}{Td}\sum_t\|e_{t,r}\|^2+\frac{1}{Td}\sum_t\|b_t\|^2
 +\frac{2}{Td}\sum_t\langle e_{t,r},b_t\rangle.
}\]
\DIFadd{The supplement includes all mean vectors and cross-term estimates with paired-bootstrap intervals. Table~\ref{tab:blocked-reference} is the deterministic blocking component. 
}\begin{table}[ht]\color{black}\centering\small
\caption{\color{black}Focused LG error decomposition. Each entry averages run-level RMSEs across 16 runs; brackets are individual paired-seed bootstrap 95\% intervals (4,000 resamples). The deterministic blocking component is not a Monte Carlo estimate. Smaller RMSE is better.}\label{supp:tab:lg-decomposition}
\begin{tabular}{rrrlrrr}\toprule $d$ & $N$ & $b$ & Method & Particle RMSE [interval] & Blocking RMSE & Total RMSE [interval]\\\midrule
64 & 128 & 1 & BPF & 0.0979 [0.0969,0.0989] & 0.0376 & 0.1051 [0.1042,0.1060]\\
64 & 128 & 1 & TBPF & 0.0984 [0.0975,0.0993] & 0.0376 & 0.1058 [0.1048,0.1067]\\
64 & 128 & 4 & BPF & 0.2320 [0.2304,0.2338] & 0.0185 & 0.2327 [0.2312,0.2344]\\
64 & 128 & 4 & TBPF & 0.2318 [0.2302,0.2333] & 0.0185 & 0.2326 [0.2310,0.2340]\\
64 & 512 & 1 & BPF & 0.0491 [0.0484,0.0498] & 0.0376 & 0.0619 [0.0613,0.0625]\\
64 & 512 & 1 & TBPF & 0.0495 [0.0490,0.0500] & 0.0376 & 0.0623 [0.0618,0.0628]\\
64 & 512 & 4 & BPF & 0.1270 [0.1260,0.1282] & 0.0185 & 0.1282 [0.1272,0.1294]\\
64 & 512 & 4 & TBPF & 0.1270 [0.1257,0.1283] & 0.0185 & 0.1282 [0.1269,0.1296]\\
256 & 128 & 1 & BPF & 0.0991 [0.0985,0.0996] & 0.0382 & 0.1067 [0.1061,0.1072]\\
256 & 128 & 1 & TBPF & 0.0985 [0.0981,0.0989] & 0.0382 & 0.1060 [0.1056,0.1065]\\
256 & 128 & 4 & BPF & 0.2319 [0.2309,0.2329] & 0.0188 & 0.2329 [0.2317,0.2338]\\
256 & 128 & 4 & TBPF & 0.2311 [0.2302,0.2321] & 0.0188 & 0.2321 [0.2312,0.2331]\\
256 & 512 & 1 & BPF & 0.0497 [0.0494,0.0500] & 0.0382 & 0.0628 [0.0626,0.0630]\\
256 & 512 & 1 & TBPF & 0.0494 [0.0491,0.0497] & 0.0382 & 0.0626 [0.0624,0.0629]\\
256 & 512 & 4 & BPF & 0.1272 [0.1267,0.1277] & 0.0188 & 0.1287 [0.1282,0.1291]\\
256 & 512 & 4 & TBPF & 0.1270 [0.1263,0.1277] & 0.0188 & 0.1285 [0.1278,0.1292]\\
\bottomrule\end{tabular}\end{table}

\paragraph{\DIFadd{Fresh conditional-precision check.}}
\DIFadd{Using the original 32-run slice as a precision pilot, we fixed 128 fresh seeds per procedure before generating additional outcomes. This count targets half the pilot interval scale under the approximate $n_{\mathrm{run}}^{-1/2}$ relation for $n_{\mathrm{run}}$ filter runs. We kept $d=64,N=512,b=1$ and both horizons fixed, reused the original datasets, and ran BPF and both TBPF configurations with common seed-indexed vectors. The 768 additional executions are separate from the original campaign and the independent-data check. Table~\ref{supp:tab:precision} reports all four estimates and interval widths. 
}\begin{table}[ht]\color{black}\centering\small
\caption{\color{black}Fresh fixed-count confirmation at $d=64,N=512,b=1$: 128 runs per procedure, with paired-seed 95\% percentile intervals (4,000 resamples). The count was fixed using the previous 32-run analysis as a precision pilot; the final runs use new seeds. All results are reported regardless of interval inclusion of one. Smaller ratios favor TBPF.}\label{supp:tab:precision}
\begin{tabular}{lrrrr}\toprule Model & $\ell$ & Ratio & 95\% interval & Interval width\\\midrule
LG & 1 & 0.759 & [0.510,1.116] & 0.606\\
LG & 5 & 0.779 & [0.547,1.117] & 0.570\\
SV & 1 & 0.594 & [0.416,0.849] & 0.434\\
SV & 5 & 0.545 & [0.384,0.801] & 0.417\\
\bottomrule\end{tabular}\end{table}

\FloatBarrier
\subsection{\DIFadd{Independent-data fixed-configuration check}}\label{supp:independent}
\DIFadd{We fixed $N=128$, $b=1$, $\ell=5$, $T=80$, and two data seeds (2026111801 and 2026111802) for each model/dimension. Each BPF/TBPF pair uses 16 seed-indexed runs, giving 384 additional filter executions. No configuration was tuned on these new data. These runs are separate from the original 5,760-run campaign and the 70/96 count. The two independent datasets per setting expose dataset variation. TBPF has lower observed log variance in 3/6 new LG cases and 6/6 new SV cases. 
}\begin{table}[ht]\color{black}\centering\small
\caption{\color{black}Independent-data check: observed TBPF/BPF log-variance ratios and individual paired-bootstrap 95\% intervals. Smaller ratios favor TBPF. Each interval conditions on its own dataset and uses 2,000 resamples of 16 run pairs.}\label{tab:independent}
\begin{tabular}{lrrr}\toprule Model & $d$ & Dataset 1 & Dataset 2\\\midrule
LG & 64 & 1.461 [0.521,3.840] & 0.516 [0.146,1.359]\\
LG & 128 & 0.431 [0.235,0.899] & 1.251 [0.571,2.710]\\
LG & 256 & 0.420 [0.163,1.090] & 1.659 [0.814,4.119]\\
SV & 64 & 0.540 [0.143,1.589] & 0.337 [0.137,0.862]\\
SV & 128 & 0.353 [0.119,0.798] & 0.861 [0.381,2.578]\\
SV & 256 & 0.627 [0.229,1.590] & 0.705 [0.150,1.686]\\
\bottomrule\end{tabular}\end{table}

\DIFaddend \appendixsection{Notation reference}\label{supp:app:notation}
The table collects the main notation used in  theory and experiments.
Symbols used only within a particular calculation retain their local definitions;
important reused symbols are distinguished below.

\begingroup
\small
\renewcommand{\arraystretch}{1.35}
\begin{longtable}{@{}p{0.27\textwidth}p{0.68\textwidth}@{}}
\caption{Notation used in the manuscript.}\label{supp:tab:notation}\\
\toprule
Symbol & Meaning or definition\\
\midrule
\endfirsthead
\multicolumn{2}{l}{\tablename\ \thetable\ (continued)}\\
\toprule
Symbol & Meaning or definition\\
\midrule
\endhead
\midrule
\multicolumn{2}{r}{Continued on next page}\\
\endfoot
\bottomrule
\endlastfoot
\multicolumn{2}{l}{\textbf{Model, spaces, and operators}}\\*
$V$, $d$ & Vertex set of the spatial graph and its size $d=|V|$.\\
$v$, $N(v)$ & A site and the neighborhood on which its transition depends. $N(v)$ is a set, whereas $N$ is the particle count.\\
$\cK$, $K$, $b$ & Block partition, one block, and maximum block size $b=\max_K|K|$.\\
\DIFadd{$J$, $q(J)$} & \DIFadd{Site subset and number of partition blocks meeting it: $q(J)=|\{K\in\cK:K\cap J\ne\varnothing\}|$; see Lemma~\ref{supp:lem:replacement}.}\\
$\cX$, $\cX^K$, $x^K$ & Full state space, block state space, and restriction of a state to block $K$.\\
$t$, $T$, $N$ & Time index, fixed time horizon, and number of particles.\\
$y_{1:T}$ & Observed data sequence, held fixed in  theoretical analysis.\\
$f_t(x,dz)$, $f_t^v$ & Transition kernel from time $t-1$ to $t$ and its sitewise factors.\\
$g_t^v$, $g_t^K$, $g_t$ & Site likelihood, block likelihood $\prod_{v\in K}g_t^v$, and full likelihood $\prod_{v\in V}g_t^v$.\\
$\mu(h)$, $\mu^K$ & Integral $\int h\,d\mu$ and block-$K$ marginal of $\mu$; see Section~\ref{sec:baseline}.\\
$P_t$ & Prediction operator $P_t\mu=\mu f_t$; see Section~\ref{sec:baseline}.\\
$C_t$, $B$ & Likelihood correction $C_t\mu=g_t\mu/\mu(g_t)$ and blocking $B\mu=\bigotimes_{K\in\cK}\mu^K$; see Section~\ref{sec:baseline}.\\
$\pi_t$, $\widetilde\pi_t$ & Exact filter and deterministic block filter, respectively: $\pi_t=C_tP_t\pi_{t-1}$ and $\widetilde\pi_t=C_tBP_t\widetilde\pi_{t-1}$. See \eqref{eq:block}.\\
$\widetilde\pi_t^K$, $p_K$ & Block marginal of the deterministic filter and coordinate projection $p_K(x)=x^K$; see Appendix~\ref{supp:subsec:consistency}.\\
$C_b(\cX)$, $\|u\|_\infty$ & Bounded continuous real functions and the supremum norm of a function.\\
$\delta_x$ & Unit point mass at $x$.\\
\addlinespace
\multicolumn{2}{l}{\textbf{Particles, weighted measures, and twisting}}\\*
$X_t^i$, $Z_t^i$ & Recombined particle after resampling and proposal before resampling, respectively. Superscript $K$ restricts either to a block.\\
$Z_t$, $X_t^{1:N}$ & Complete proposal pool $(Z_t^1,\ldots,Z_t^N)$ and complete recombined offspring array. In $H_t^K(X_t)$, $X_t$ denotes the offspring array.\\
$q_t^N$, $q_t$ & Empirical proposal measure $N^{-1}\sum_i\delta_{Z_t^i}$ and deterministic prediction $P_t\widetilde\pi_{t-1}$; see \hyperref[supp:prediction-notation]{Step 2 of the consistency proof}.\\
$a_{t,N}^{K}$, $w_t^{i,K}$ & Empirical block likelihood average for $N$ particles and normalized proposal weight; see \eqref{eq:weights}.\\
$a_{t,*}^K$ & Deterministic block likelihood increment $q_t^K(g_t^K)$ in the consistency proof; see Appendix~\ref{supp:subsec:consistency}.\\
$\nu_t^K$, \DIFadd{$\nu_{t,N}^K$} & Weighted block proposal law $\sum_iw_t^{i,K}\delta_{Z_t^{i,K}}$\DIFadd{; the second notation displays its dependence on $N$}.\\
$\widehat\pi_t^{TB}$ & Weighted, pre-resampling product measure $\bigotimes_K\nu_t^K=C_tBq_t^N$ along a TBPF history; see Algorithm~\ref{supp:alg:tbpf}.\\
$\eta_t^N$ & TBPF empirical measure $N^{-1}\sum_i\delta_{X_t^i}$ after resampling and recombination; see Theorem~\ref{thm:moments}.\\
$\psi_t^K$ & Positive finite block guide, fixed before current offspring are generated; it may depend on the available history and proposal pool.\\
$c_t^K$, $H_t^K(x)$ & Weighted-pool guide mean and candidate-array guide average: $c_t^K=\nu_t^K(\psi_t^K)$ and $H_t^K(x)=N^{-1}\sum_j\psi_t^K(x^{j,K})$; see Section~\ref{sec:algorithm}.\\
$\nu_{t,\psi}^K$, \DIFadd{$\nu_{t,N,\psi}^K$} & Tilted one-child law $\psi_t^K\nu_t^K/c_t^K$\DIFadd{, with particle-count dependence explicit in the second notation}.\\
\DIFadd{$s_i^K$, $G_i^K$, $\widetilde w_t^{i,K}$} & \DIFadd{Within the time-$t$ algorithm loop: frozen guide value, unnormalized observation weight, and normalized guide-tilted ancestor weight; see Algorithm~\ref{supp:alg:tbpf}.}\\
\DIFadd{$\operatorname{Cat}(w)$} & \DIFadd{Categorical law on proposal labels with probability vector $w$.}\\
$R_t$, $\widetilde R_t$ & Ordinary and twisted conditional resampling kernels on full offspring arrays; see \eqref{supp:eq:twist}. \\
$J_t^K$ & Uniformly selected distinguished child label at time $t$ in block $K$; see Algorithm~\ref{alg:tbpf}.\\
$A_t^{j,K}$ & Proposal ancestor label selected for child $j$ in block $K$. Distinct from the deterministic scalar $a_{t,*}^K$; see Algorithm~\ref{alg:tbpf}.\\
$V_t^K$ & Tilted replacement child at time $t$ in block $K$ in the distinguished-offspring construction; see Lemma~\ref{supp:lem:replacement}.\\
\DIFadd{$\eta_N^{\mathrm{ref}}$, $\eta_N^{\mathrm{rep}}$} & \DIFadd{Empirical measures of the reference and replaced arrays in Lemma~\ref{supp:lem:replacement}.}\\
$\bar\eta_t^{K,N}$ & Reference empirical block measure $N^{-1}\sum_i\delta_{Y_{t,N}^{i,K}}$, with $Y_{t,N}^{i,K}$ drawn conditionally independently from $\nu_{t,N}^K$, before replacing one child in the local-moments proof; see the proof of Theorem~\ref{supp:thm:moments}.\\
$\bar X_t^i$, $\bar\eta_t^N$ & Auxiliary ordinary reference row and its empirical measure in the replacement coupling; see the proof of Theorem~\ref{supp:thm:consistent}.\\
$\mathbb P_B$, $\mathbb P_{TB}$ & Laws of the augmented baseline extension and twisted histories in Theorem~\ref{supp:thm:rn}; $\E_B$ and $\E_{TB}$ are their expectations. The baseline particle-array marginal is BPF.\\
$\mathcal H_T^+$, $\Xi_t$, $A_t$ & Augmented history, frozen pre-offspring guide record, and optional retained resampling marks. The guide-generation and conditional resampling-mark kernels are $Q_t$ and $\widetilde K_t$, respectively; see \eqref{supp:eq:auxiliary-kernels}.\\
 \DIFaddbegin \DIFadd{$\mathcal F_{t-1}$, $\mathcal G_t$ }\DIFaddend & Information through the previous  \DIFaddbegin \DIFadd{resampling step and through current pool/}\DIFaddend guide construction,  \DIFaddbegin \DIFadd{respectively. $\mathcal G_{t,N}$ displays both the time and particle count for the latter when varying }\DIFaddend $N$ .\\
\addlinespace
\multicolumn{2}{l}{\textbf{Normalizers, variance, and consistency}}\\*
\DIFadd{$\alpha_t^K$, $\beta_t^K$, $r_t^K$} & \DIFadd{Minimum and maximum guide values on the proposal support and their ratio $r_t^K=\beta_t^K/\alpha_t^K$; see Proposition~\ref{supp:prop:second}.}\\
\DIFadd{$\rho_t^K$} & \DIFadd{Deterministic upper bound on $r_t^K$ across histories and pools; required to be uniform in $N$ for normalizer consistency; see Proposition~\ref{prop:correction} and Theorem~\ref{thm:consistency}.}\\
\DIFaddbegin \DIFadd{$Z_T$ }& \DIFadd{Exact model likelihood used in the Gaussian-reference comparison in Appendix~\ref{supp:revision-validation}.}\\
\DIFadd{$Z_T^{\mathrm{block}}$ }& \DIFadd{Deterministic block-recursion normalizer defined in Theorem~\ref{thm:consistency}.}\\
\DIFadd{$C_t^K$, $\mathcal C_t$ }& \DIFadd{Block correction $c_t^K/H_t^K(X_t)$ and its product over blocks, distinct from the observation operator $C_t$; see \hyperref[supp:correction-filtration]{the cumulative-correction proof}.}\\
\DIFadd{$\nu$, $R$} & \DIFadd{Fixed block-pool law and baseline offspring-array law $R=\nu^{\otimes N}$ in Proposition~\ref{prop:guide-improvement}.}\\
\DIFadd{$h$, $\bar h$, $\theta$} & \DIFadd{Bounded block perturbation, its offspring average $\bar h(x)=N^{-1}\sum_jh(x^j)$, and the parameter in $\psi_\theta=1+\theta h$; see Proposition~\ref{prop:guide-improvement}.}\\
\DIFadd{$Q_\theta$, $\mathcal V(\theta)$, $\mathcal V_{\log}(\theta)$ }& \DIFadd{Conditional tilted joint law and corrected ordinary/log variances at a fixed pool with unchanged continuation. See Proposition~\ref{prop:guide-improvement} and \eqref{eq:log-guide-improvement}.}\\
\DIFadd{$F$, $\mathcal Z_T$ }& \DIFadd{Generic history functional in Theorem~\ref{thm:rn}, and the specific normalizer functional $\mathcal Z_T=\prod_{t,K}a_{t,N}^{K}$ in \eqref{eq:variance-identity}.}\\
\DIFadd{$W(X,U)$, $\mathcal A(x)$ }& \DIFadd{Payoff of the offspring array $X$ and continuation variables $U$, and its conditional second moment $\mathcal A(x)=\int W(x,u)^2\mathsf M(x,du)$; see Proposition~\ref{prop:guide-improvement}.}\\
\DIFadd{$\mathsf M$, $\Lambda_\theta$, $m_h$ }& \DIFadd{Continuation kernel, array density, and guide-perturbation mean $m_h=\nu(h)$ in Proposition~\ref{prop:guide-improvement}.}\\
\DIFadd{$P_0$} & \DIFadd{Fixed-pool baseline joint law $P_0(dx,du)=R(dx)\mathsf M(x,du)$ in Proposition~\ref{prop:guide-improvement}.}\\
\DIFadd{$\alpha_\theta$, $s$} & \DIFadd{Normalized step $\alpha_\theta=\theta/(1+\theta m_h)$ and centered array statistic $s=\bar h-m_h$ in Proposition~\ref{supp:prop:guide-improvement}.}\\
\DIFadd{$\alpha$, $Q_\alpha^c$, $\Lambda_\alpha^c$ }& \DIFadd{Centered-guide parameter, joint law, and its density $1+\alpha s$, defined in Corollary~\ref{supp:cor:centered-step}.}\\
\DIFadd{$Y_{\log}$, $\mu_{\log}$ }& \DIFadd{Log payoff $\log W$ and its baseline expectation in the conditional log-variance criterion; see \eqref{eq:log-guide-improvement}.}\\
\DIFadd{$Y_{\log,\theta}$, $\mu_{\log,\theta}$, $\zeta_\theta$ }& \DIFadd{Corrected log payoff, its mean under $Q_\theta$, and log-density derivative $\partial_\theta\log\Lambda_\theta$ in Proposition~\ref{supp:prop:log-derivative}.}\\
\DIFadd{$h_0$, $\psi_\alpha^c$, $\mathcal V_c(\alpha)$} & \DIFadd{Centered perturbation $h_0=h-m_h$, guide $\psi_\alpha^c=1+\alpha h_0$, and corrected-payoff variance in Corollary~\ref{supp:cor:centered-step}.}\\
\DIFadd{$H_0$, $c_{\mathrm{step}}$, $b_{\mathrm{step}}$ }& \DIFadd{Centered perturbation bound $\|h_0\|_\infty$ and moments $\E_R[\mathcal A s]$, $\E_R[\mathcal A s^2]$ in Corollary~\ref{supp:cor:centered-step}.}\\
\DIFadd{$\alpha_{\mathrm{safe}}$ }& \DIFadd{Explicit centered guide step with the guaranteed variance decrease in \eqref{supp:eq:explicit-guide-step}.}\\
\DIFadd{$g_{\mathrm{ex}}$, $\psi_\beta$, $\mathcal V_{\mathrm{ex}}(\beta)$} & \DIFadd{Terminal likelihood, power guide $g_{\mathrm{ex}}^\beta$, and corrected-payoff variance in the main-text example and Appendix~\ref{supp:pool-example}.}\\
\DIFadd{$\sigma^2$ (local moments)} & \DIFadd{Deterministic limiting test variance in Theorem~\ref{thm:moments}; distinct from the experimental state-noise variance.}\\
\DIFaddend $D_T$ & Path correction $\prod_{t,K}c_t^K/H_t^K(X_t)=d\mathbb P_B/d\mathbb P_{TB}$; see \eqref{supp:eq:rn}.\\
$\widehat Z_T^B$ & BPF normalizer estimate $\prod_{t,K}a_{t,N}^{K}$, evaluated along a BPF history; see \eqref{eq:weights}.\\
$\widehat Z_T^{TB}$ & Corrected TBPF normalizer $D_T\prod_{t,K}a_{t,N}^{K}$, with every factor evaluated along the same TBPF history; see \eqref{supp:eq:tb-normalizer}.\\
$L_t$ (algorithm) & Accumulated log estimate $\log\widehat Z_t^{TB}$ in \eqref{supp:eq:logz}.\\

$d_*$, $u_m$ & Metric for weak convergence and its countable family of bounded continuous determining functions in the \hyperref[supp:weak-metric]{weak-topology construction in the consistency proof}.\\
 \DIFaddbegin \DIFadd{$\xrightarrow{\mathrm{weak},\mathbb P}$}\DIFaddend & \DIFadd{Convergence in probability in the weak topology on probability measures; see Theorem~\ref{thm:consistency}.}\\
\addlinespace
\multicolumn{2}{l}{\textbf{Guides, experiments, and diagnostics}}\\*
\DIFadd{$X_t^v$, $Y_t^v$} & \DIFadd{Latent coordinate and observation at site $v$ in the model equations; particle indices instead appear in $X_t^i$ and $Z_t^i$; see Appendix~\ref{supp:subsec:models}.}\\
\DIFadd{$\sigma$, $\epsilon_t^v$, $e_t^v$} & \DIFadd{State-noise scale and independent standard Gaussian state and observation noises in the experiments; see Appendix~\ref{supp:subsec:models}.}\\
\DIFadd{$\kappa_{\mathrm{sp}}$, $a$} & \DIFadd{Spatial coupling coefficient $0.1$ and isolated-guide autoregression coefficient $0.9$; see \eqref{supp:eq:model} and \eqref{eq:lg-guide-recursion}.}\\
\DIFadd{$\ell$} & \DIFadd{Guide horizon: TBPF uses $y_{t+1:\min(T,t+\ell)}$ from time $t$; the TPF arrival-time guide uses $y_{t:\min(T,t+\ell-1)}$; see \eqref{eq:lookahead-guide} and Appendix~\ref{supp:subsec:tpf}.}\\
\DIFadd{$\psi_t$, $w_t^i$, $c_{t+1}$} & \DIFadd{Full-state TPF guide, normalized observation weights, and integrated guide average $\sum_iw_t^i(f_{t+1}\psi_{t+1})(Z_t^i)$; see Appendix~\ref{supp:subsec:tpf}.}\\
\DIFadd{$q_t^{\mathrm{TPF}}$} & \DIFadd{Full-state TPF proposal mixture, distinct from deterministic block prediction $q_t$; see Appendix~\ref{supp:subsec:tpf}.}\\
\DIFadd{$\kappa_{t+1}^v$ }& \DIFadd{Gaussian integration denominator $1+\sigma^2a_{t+1}^v$ for the TPF guide; see Appendix~\ref{supp:subsec:tpf}.}\\
\DIFadd{$h_{s,\tau}^v$, $A_{s,\tau}^v$, $B_{s,\tau}^v$, $C_{s,\tau}^v$ }& \DIFadd{Site-$v$ backward message for $y_{s+1:\tau}^v$ given the state at time $s$, and its LG log-message coefficients; see Appendix~\ref{supp:subsec:guide-design}.}\\
\DIFadd{$a_t^v$, $m_t^v$, $b_{t,\mathrm{quad}}^v$ }& \DIFadd{TPF guide precision, center, and SV local slope at arrival time $t$ and site $v$; see Appendix~\ref{supp:subsec:tpf}.}\\
\DIFadd{$\lambda_{t,k}^v$, $k_t^v$, $x_k$, $\Delta$ }& \DIFadd{Sitewise log-guide grid values, clamped maximizing index, common grid locations, and grid spacing in Appendix~\ref{supp:subsec:tpf}.}\\
\DIFadd{$\mu_{t+1}^v(x)$ }& \DIFadd{Site-$v$ transition mean at time $t+1$ conditional on parent state $x$; see Appendix~\ref{supp:subsec:tpf}.}\\
\DIFadd{$r$, $n_{\mathrm{run}}$, $L_{T,r}$} & \DIFadd{Run index, number of independent filter runs, and terminal log-normalizer estimate in run $r$ for a fixed method and configuration; see Appendix~\ref{supp:subsec:models}.}\\
\DIFadd{$\widehat{\Var}_r$, $r_V$} & \DIFadd{Sample variance across runs and the matched TBPF/BPF log-normalizer variance ratio; see Section~\ref{sec:simulations}.}\\
\DIFadd{$\widehat m_{t,r}$, $\widehat m_{t,r}^v$} & \DIFadd{Weighted particle mean vector and its site-$v$ coordinate in run $r$; see \hyperref[supp:decomposition-notation]{the LG error decomposition}.}\\
\DIFadd{$m_t^{\mathrm{block}}$, $m_t^{\mathrm{full}}$} & \DIFadd{Blocked and full Kalman filtering means, distinct from the TPF guide centers $m_t^v$; see \hyperref[supp:decomposition-notation]{the LG error decomposition}.}\\
\DIFadd{$e_{t,r}$, $b_t$} & \DIFadd{Run-specific particle error $\widehat m_{t,r}-m_t^{\mathrm{block}}$ and common blocking error $m_t^{\mathrm{block}}-m_t^{\mathrm{full}}$; see the \hyperref[supp:decomposition-notation]{LG error decomposition}.}\\
\DIFadd{$m_{t,-}^{\mathrm{block}}$, $\Sigma_t$} & \DIFadd{Predicted blocked Gaussian mean and updated blocked Gaussian covariance in Appendix~\ref{supp:revision-validation}.}\\
\DIFadd{$A_{\mathrm{tr}}$, $S_t^K$, $G_t^K$} & \DIFadd{Ring transition matrix, block innovation covariance, and block Kalman gain; see \hyperref[supp:gaussian-reference-notation]{the blocked Gaussian recursion}.}\\
\DIFadd{$m_{t,\mathrm{ref}}^v$, $\mathrm{RMSE}_r$} & \DIFadd{Reference state coordinate (full Kalman mean for LG, latent truth for SV) and the resulting root mean squared error in run $r$; see Appendix~\ref{supp:subsec:models}.}\\
\DIFadd{$\widehat{\mathrm{CV}}^2$} & \DIFadd{Empirical squared coefficient of variation of the ordinary-scale normalizer estimates; see Appendix~\ref{supp:subsec:models}.}\\
\DIFadd{$\beta$} & \DIFadd{Power-guide exponent in $\psi_\beta=g_{\mathrm{ex}}^\beta$; see Appendix~\ref{supp:pool-example}. Distinct from the guide maximum $\beta_t^K$.}\\
\addlinespace
\end{longtable}
\endgroup
\end{document}